\documentclass[]{jfm}
\usepackage{subcaption}
\usepackage{graphicx}
\usepackage{newtxtext}
\usepackage{newtxmath}
\usepackage{natbib}
\usepackage{hyperref}
\usepackage{amsmath}
\allowdisplaybreaks

\hypersetup{
    colorlinks = true,
    urlcolor   = blue,
    citecolor  = black,
}

\newcommand{\RomanNumeralCaps}[1]
\linenumbers
\graphicspath{{.}} 

\title{Obliquely interacting solitary waves and wave wakes in free-surface flows}

\author{Lei Hu\aff{1}, Xu-Dan Luo\aff{2}  \and Zhan Wang\aff{1,3}\corresp{\email{zwang@imech.ac.cn}}}

\affiliation{\aff{1}Key Laboratory for Mechanics in Fluid Solid Coupling Systems, Institute of Mechanics, Chinese Academy of Sciences, Beijing 100190, PR China
\aff{2}Key Laboratory of Mathematics Mechanization, Academy of Mathematics and Systems Science, Chinese Academy of Sciences, Beijing 100190, PR China
\aff{3}School of Engineering Science, University of Chinese Academy of Sciences, Beijing 100049, PR China}

\begin{document}
\maketitle

\begin{abstract}
This paper investigates the weakly nonlinear isotropic bi-directional Benney--Luke (BL) equation, which is used to describe oceanic surface and internal waves in shallow water, with a particular focus on soliton dynamics. Using the Whitham modulation theory, we derive the modulation equations associated with the BL equation that describe the evolution of soliton amplitude and slope. By analyzing rarefaction waves and shock waves within these modulation equations, we derive the Riemann invariants and modified Rankine--Hugoniot conditions. These expressions help characterize the Mach expansion and Mach reflection phenomena of bent and reverse bent solitons. We also derive analytical formulas for the critical angle and the Mach stem amplitude, showing that as the soliton speed is in the vicinity of unity, the results from the BL equation align closely with those of the Kadomtsev--Petviashvili (KP) equation. Corresponding numerical results are obtained and show excellent agreement with theoretical predictions. Furthermore, as a far-field approximation for the forced BL equation -- which models wave and flow interactions with local topography -- the modulation equations yield a slowly varying similarity solution. This solution indicates that the precursor wavefronts created by topography moving at subcritical or critical speeds take the shape of a circular arc, in contrast to the parabolic wavefronts observed in the forced KP equation.
\end{abstract}

\begin{keywords}
solitary waves; surface gravity waves; isotropism
\end{keywords}


\section{Introduction}
\begin{figure}
\centering
\includegraphics[height=2.5in]{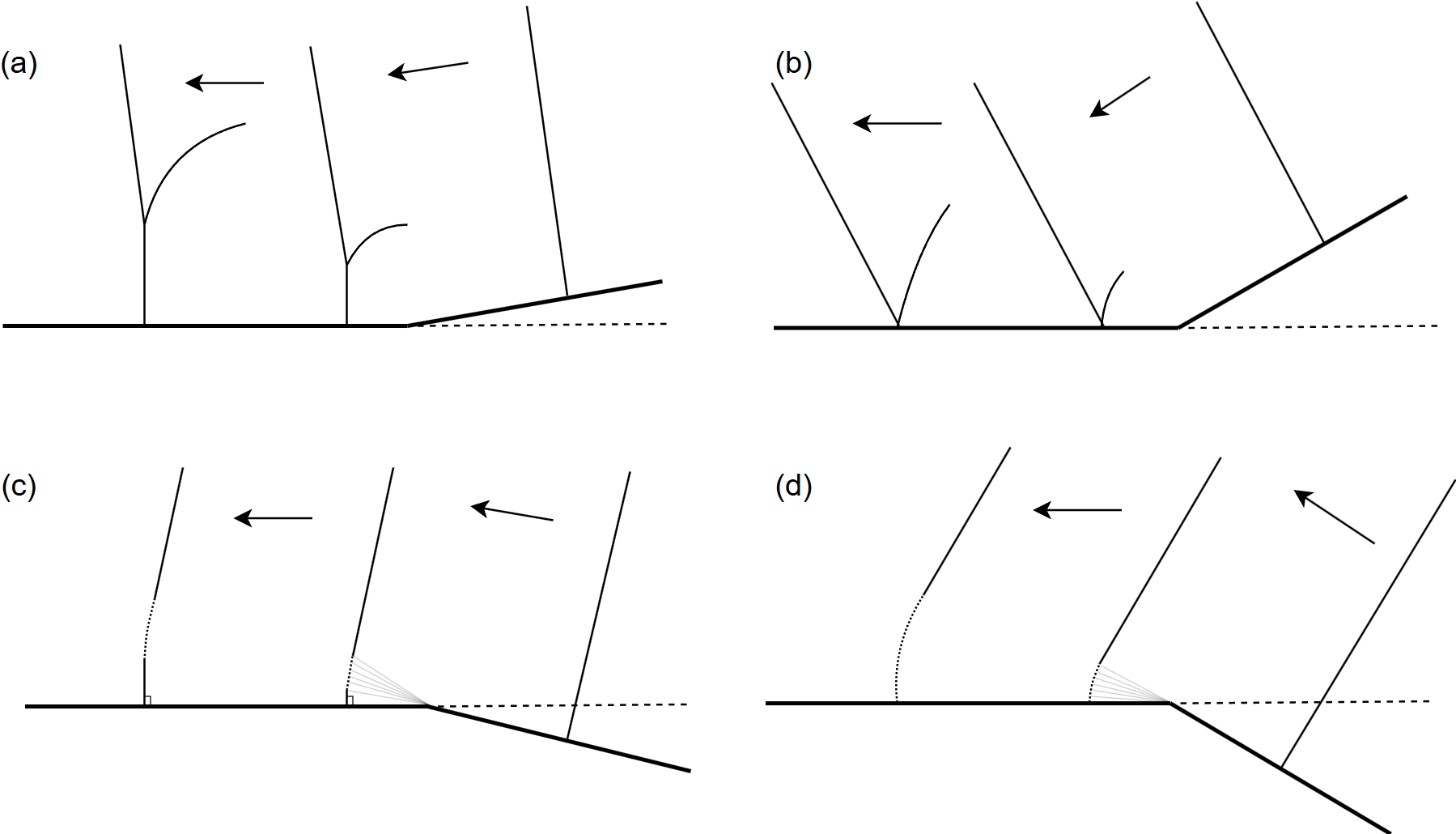}
\caption{Diagram of a shock wave encounters the compressive corner (a,b) and expansive corner (c,d): (a) Mach reflection; (b) regular reflection; (c) Mach expansion; (d) regular expansion.}
\label{fig:sketch_mach}
\end{figure}

Mach reflection is a phenomenon in fluid dynamics that has significant impacts on both theoretical research and practical applications. It was discovered and named by the physicist Ernst Mach, who studied shock wave interactions at the wavefront \citep{mach1875}. In his innovative experiment, meticulously reviewed by \citet{reichenbach1983} and replicated by \citet{krehl1991}, Mach reflection occurs when a shock wave encounters a compressive corner. Three shock waves typically emerge in the Mach reflection process: the incident and reflected waves and the `Mach stem' that bridges between them. In contrast, Mach expansion is observed when a shock wave interacts with an expansive corner. This interaction leads to the gradual divergence of expansion waves from the corner, resulting in the formation of a dynamically evolving fan-shaped region. This region serves as a bridge between the incident shock wave and the Mach stem \citep{whitham1957,whitham1959,bazhenova1975}, as illustrated in figure~\ref{fig:sketch_mach}. 

This phenomenon was also observed in oceanic surface waves \citep{cornish1910}. \citet{gilmore1950} elucidated the intersection and reflection phenomena of hydraulic jumps in liquids and shock waves in gases by drawing parallels between the two. Their discussion included analyses of regular and Mach reflections, comparing experimental observations with theoretical projections. Nevertheless, a comprehensive theoretical explanation of Mach reflection in water waves was not achieved until the pioneering contributions of \citet{miles1977,miles1977a}. Miles examined the interaction of two obliquely colliding solitons in surface gravity waves and proposed that the dynamics depend on their incident angles and amplitudes. A weakly interacting scenario occurs when the incident angles are sufficiently small, allowing the two solitons to approximately linearly superpose, resulting in regular reflection. As the incident angle increases, the solitons undergo Mach reflection, leading to the formation of the Miles resonant soliton. This phenomenon is characterized by the triple-point interaction among the incident, reflected, and fused (Mach/Miles) solitons, which has been validated by both experimental results \citep{perroud1957,melville1980} and numerical simulations \citep{funakoshi1980,tanaka1993}.

To describe the (2+1)-dimensional wave motion in a fluid of finite depth, \citet{kadomtsev1970} proposed a unidirectional model amenable to weak transverse variation, known as the Kadomtsev--Petviashvili (KP) equation:
\begin{equation}
\left( u_t+uu_x+u_{xxx} \right)_x+u_{yy}=0\,,\label{eq:KP_topo_flat}
\end{equation}
where $u$ is the surface elevation, $x$ and $y$ are the longitudinal and transverse spatial coordinates, respectively, and the subscripts denote partial derivatives. The complete integrability of the KP equation \citep{ablowitz1991} yields the analyticity of resonant multi-soliton solutions \citep{kodama2004,biondini2006,biondini2007,kodama2010}, which provides a foundation for the investigation of Miles' resonant soliton phenomena. Miles' theory experienced rapid development due to these explicit multi-soliton solutions and direct numerical simulations of the KP equation \citep{tsuji2007,biondini2009,kodama2009}. These new findings have also been confirmed by laboratory experiments \citep{li2011,kodama2016} and oceanic field observations \citep{ablowitz2012,wang2012a}.

The KP equation~\eqref{eq:KP_topo_flat}, also termed the KP-II equation, admits stable line solitons \citep{ablowitz1991}:
\begin{equation}
u=A\;\text{sech}^2 \left[ \frac{\sqrt{A}}{12} (x+q y-c t) \right] \,,\quad c=\frac{A}{3}+q^2\,,\label{eq:sol_KP_0}
\end{equation}
where $A$ is the amplitude, $c$ is the wave frequency, and $q$ denotes the slope of the line soliton. Thus, the propagation of line solitons can be characterized by determining $A=A(y,t)$ and $q=q(y,t)$. \citet{ryskamp2021} employed the Whitham modulation theory, with $A$ and $q$ varying slowly in time and space, to derive the modulation equations for the KP equation (see also \citealt{lee1990,neu2015}), which are
\begin{equation}
\begin{aligned}
&A_t+2q A_y+\frac{4}{3}A q_y=0\,,\\
&q_t+\frac{1}{3}A_y+2q q_y=0\,.
\end{aligned}\label{eq:modu-KP_yt}
\end{equation}
The above $2\times2$ quasi-linear hyperbolic system is a limiting case of the Whitham modulation equations for periodic waves, as discussed by \citet{ablowitz2017}. This scenario specifically addresses $x$-independent modulations of line solitons, as \citet{biondini2020} highlighted. 

Assisted by the exact solution \citep{kodama2018,biondini2006} for the Y-shaped soliton of equation~\eqref{eq:KP_topo_flat}, \citet{ryskamp2022}, through an analysis of equations~\eqref{eq:modu-KP_yt}, revised the classical Rankine--Hugoniot (RH) jump conditions and proposed the modified RH conditions. They accurately described the Miles resonant solitons and precisely characterized the Mach reflection phenomenon of line solitons within the KP framework. The amplitudes and slopes of the Mach stem and the reflected soliton were analytically determined by
\begin{equation}
A_i=q_0^2\,,\quad q_i = \sqrt{A_0}\,,\quad A_w=\left( q_0+\sqrt{A_0} \right)^2\,,\quad q_w=0\,,\quad q_0>0\,,\label{eq:res_kp_ref}
\end{equation}
where the subscripts $w$, $i$, and 0 represent the parameters of the Mach, reflected, and incident solitons, respectively. By applying the modified RH jump conditions, they derived the critical slope $q_{\text{cr}}=\sqrt{A_0}$ that distinguishes the region between regular and Mach reflections. These conclusions were further verified through numerical experiments conducted by \citet{ryskamp2022}.

As for the Mach expansion of line solitons, \citet{ryskamp2021} utilized the equivalent characteristic form of equations~\eqref{eq:modu-KP_yt} to perform eigenvalue analysis, leading to a precise description of line-soliton Mach expansion in the context of the KP equation. The study identified the critical angle for Mach expansion, along with the corresponding amplitude for forming a Mach stem, denoted as
\begin{equation}
A_w=\left(q_0+\sqrt{A_0}\right)^2\,,\quad q_w=0\,,\quad q_{\text{cr}}=-\sqrt{A_0}\,,\quad q_0<0\,,\label{eq:res_kp_exp}
\end{equation}
which can also be viewed as an extension of result~\eqref{eq:res_kp_ref} for the case $q_0<0$. Results presented by \citet{ryskamp2021} have been validated through numerical simulations, indicating that when the angle between the direction of propagation of the line soliton and the reflecting surface (referred to as the symmetry plane) is small, the line soliton undergoes Mach expansion. Starting from the point of incidence, they gradually expand to form an arc-shaped wave that connects the incident soliton and the Mach stem. As the angle increases, the Mach stem disappears, resulting in regular expansion. The planar depiction of the Mach expansion process for line solitons in the KP equation closely resembles the Mach expansion observed in aerodynamics, as illustrated in figures~\ref{fig:sketch_mach}(c,d). 

The previous discussions on Mach reflection/expansion of solitons are predominantly based on the KP equation. However, it is important to note that the KP equation is inherently anisotropic and describe unidirectional wave propagation. This limitation means it does not effectively capture wave phenomena with significant transverse variations in the ocean. In this paper, we focus on the weakly nonlinear isotropic bi-directional Benney--Luke (BL) equation:
\begin{equation}
\xi_{tt}-\Delta \xi_{tt}-\Delta\xi+\left(|\nabla \xi|^2\right)_t+\xi_t\Delta\xi=0\,,\label{eq:BL-topo-flat}
\end{equation}
where $\xi$ denotes the surface velocity potential, and $\Delta$ and $\nabla$ are the Laplacian and gradient operators, respectively. As shown by \cite{ablowitz2006}, the BL equation~\eqref{eq:BL-topo-flat} can be formally reduced to the KP equation~\eqref{eq:KP_topo_flat} in the appropriate limits.

The BL equation was initially proposed by \citet{benney1964} for studying free-surface shallow-water waves and was later extended to isotropic and two-way propagating internal waves by \citet{yuan2020} and \citet{yuan2022}. \citet{bokhove2016} used this equation to simulate a soliton splash occurring from a wave running through a confined channel, with the numerical results showing strong agreement with experimental data, which supports the notion that the Benney--Luke approximation serves as a `reasonable' reduced model for free-surface water waves. \citet{melville1980} utilized a finite difference method to investigate the initial value problem for the BL equation concerning two-dimensional soliton reflection, demonstrating a high degree of consistency with Miles' theoretical framework. \citet{ablowitz2011} employed both analytical and numerical techniques to explore how the BL equation can be utilized to modify the web solutions of the KP equation. Furthermore, \citet{ablowitz2013} developed a direct perturbation method, leveraging inherent conservation laws to ascertain the gradual progression of the parameters within non-decaying, two-dimensional solutions to the BL equation, with numerical simulations corroborating these findings. Building on the relationship between the KP and BL equations, \citet{gidel2017} extended the conclusions drawn from the Mach reflection phenomena of the KP equation to the BL equation. Using a variational Galerkin finite element method, numerical simulations conducted by \citet{gidel2017} aligned well with the theoretical predictions proposed by \cite{miles1977a}. However, only asymptotic solutions were applied to characterize the Mach reflection and Mach expansion phenomena within the framework of the BL equation. Consequently, a rigorous theory based on an exact solution remains absent, underscoring the primary concern addressed in the present paper.

\citet{milewski1998} incorporated the effects of bottom topography into the derivation of the BL equation, leading to the formulation of the forced Benney--Luke (fBL) equation that captures the influence of topography on surface waves \citep{curtis2015} and internal waves \citep{yuan2020,yuan2022}. The fBL equation can also be reduced to the forced Kadomtsev--Petviashvili (fKP) equation (see, for example, \citealt{lee1989}). \citet{lee1990} analyzed the numerical outcomes from the fKP equation, alongside the modulation equation~\eqref{eq:modu-KP_yt}, to scrutinize the evolution of upstream-advancing waves, which exhibit a parabolic progression. However, when considering the isotropic nature of the fBL equation, the precise depiction of its two-dimensional upstream-advancing wave dynamics is also a point of interest that merits attention.

In this work, we focus on the oblique interactions of line solitons based on the BL equation, as well as the wave wakes generated by moving topography, governed by the fBL equation. Section~\ref{sec:rela} presents the relationship between the fBL and fKP equations, while section~\ref{sec:num} outlines the numerical schemes we implemented. In section~\ref{sec:modu}, we begin our exploration of line soliton dynamics by deriving the modulation equations from the BL equation. Using the bent soliton as our initial condition, we investigate the Mach expansion phenomenon for oblique solitons in section~\ref{sec:mach_exp}. Section~\ref{sec:mach_ref} concentrates on the Mach reflection phenomenon in the BL equation, utilizing reverse bent soliton initial data. We examine the shock wave issue within the modulation system, applying both classical and modified Rankine--Hugoniot conditions. As the modulation system serves as a far-field approximation to the fBL equation, in section~\ref{sec:wake}, we utilize it to delineate the evolution characteristics of waves generated by moving topography. We provide an approximate description of the evolution patterns for upstream-advancing waves and wave wakes, validated through numerical simulations of the primitive equation. Section~\ref{sec:con} provides a summary of the key findings and suggests potential extensions. Additionally, Appendix~\ref{appA} includes a detailed discussion of a more general Whitham modulation theory for the BL equation.

\section{Mathematical Formulation and Numerical Schemes}\label{sec:formulation}
\subsection{Relationship between the fBL equation and the fKP equation}\label{sec:rela}
Consider a three-dimensional incompressible and inviscid fluid flowing irrotationally, bounded above by a free surface $z=\eta(x,y,t)$ and below by a varying bottom topography $b(x,y,t)$, where $x$ and $y$ are the horizontal coordinates, and the $z$-axis points upwards with $z=0$ the undisturbed free surface. The velocity potential on the free surface, $\xi(x,y,t)$, is governed by the fBL equation \citep{milewski1998,curtis2015}
\begin{equation}
\left(1-\frac{\mu^2}{3}\Delta \right)\xi_{tt}-\Delta\xi+\epsilon\left[\left(|\nabla \xi|^2\right)_t+\xi_t\Delta\xi \right]+\epsilon b_t=0\,.\label{eq:BL-topo-ori}
\end{equation}
Here, $\mu$ and $\epsilon$ are small parameters measuring the nonlinear and dispersive effects, respectively, and we assume $\epsilon=O(\mu^2)$ to balance these two effects. When the bottom topography is flat, the fBL equation~\eqref{eq:BL-topo-ori} reduces to the classic BL equation. It is noted that equation~\eqref{eq:BL-topo-ori} can also describe nonlinear internal waves propagating over variable bottom topography, with variations primarily manifesting as differences in the coefficients of the constituent terms \citep{yuan2020,yuan2022}.

Equation~\eqref{eq:BL-topo-ori} can be reduced to the fKP equation by the following change of variables. Assuming a primarily left-propagating wave with weak transverse variation, we substitute
\begin{equation}
X=x+t\,,\quad Y=\mu y\,,\quad T=\epsilon t\,, 
\end{equation}
into equation~\eqref{eq:BL-topo-ori} and neglect terms of order $\textit{O}(\epsilon^2,\epsilon\mu)$, leading to 
\begin{equation}
\xi_{XT}+\frac{3}{2}\xi_{X}\xi_{XX}-\frac{\mu^2}{\epsilon}\left(\frac{1}{6}\xi_{XXXX}+\frac{1}{2}\xi_{YY} \right)+\frac{1}{2}b_{X}=0\,.
\label{eq:KP_temp}
\end{equation}
Returning to the original independent variables and letting $\zeta=-\xi_x$, equation~\eqref{eq:KP_temp} becomes the fKP equation:
\begin{equation}
\left(\zeta_t-\zeta_x-\frac{3\epsilon}{2}\zeta\zeta_x-\frac{\mu^2}{6}\zeta_{xxx} \right)_x-\frac{1}{2}\zeta_{yy}=\frac{\epsilon}{2}b_{xx}\,,\label{eq:KP_topo_ori}
\end{equation}
which covers the KP equation under flat topography $b\equiv0$. \citet{lee1990} conducted a detailed investigation into the solution behavior of equation~\eqref{eq:KP_topo_ori} when a localized forcing disturbance moves with a near-critical speed in horizontally unbounded shallow water. 

To normalize the fBL equation, we introduce new independent and dependent variables 
\begin{equation}
\tilde{x}=\frac{\sqrt{3}}{\mu}x\,,\quad\tilde{y}=\frac{\sqrt{3}}{\mu}y\,,\quad\tilde{t}=\frac{\sqrt{3}}{\mu}t\,,\quad\tilde{\xi}=\frac{\sqrt{3}\epsilon}{\mu}\xi\,,\quad\tilde{b}=\frac{\epsilon\mu}{\sqrt{3}}b\,,\label{eq:uni_trans}
\end{equation}
so that equation~\eqref{eq:BL-topo-ori} can be transformed to the standard form
\begin{equation}
\xi_{tt}-\Delta\xi_{tt} -\Delta\xi+\left(|\nabla\xi|^2\right)_t+\xi_t\Delta\xi+b_t=0\,,\label{eq:BL-topo}
\end{equation}
where all the tilde superscripts have been dropped for ease of notation. In the subsequent sections, our analyses associated with the BL equation will be based on equation~\eqref{eq:BL-topo}. Considering the flat topography situation, equation~\eqref{eq:BL-topo} admits the exact solution
\begin{equation}
\xi=\frac{a}{\delta\omega}\tanh\left[ \delta (kx+ly-\omega t) \right]\,,\quad\delta=\frac{\sqrt{a}}{2|\omega|}\,,\label{eq:sol_BL_0}
\end{equation}
where $k$ and $l$ denote the wavenumbers in the $x$ and $y$ directions, respectively, $\omega$ is the wave frequency, and $a$ is a free parameter measuring the amplitude. These parameters are connected through the dispersion relation $a=\omega^2/(k^2+l^2)-1$. Correspondingly, under the scales \eqref{eq:uni_trans}, the fKP equation reads 
\begin{equation}
\left(\zeta_t-\zeta_x-\frac{3}{2}\zeta\zeta_x-\frac{1}{2}\zeta_{xxx} \right)_x-\frac{1}{2}\zeta_{yy}=\frac{1}{2}b_{xx}\,.\label{eq:KP_topo}
\end{equation}
For the flat bottom scenario, it admits the line soliton solution
\begin{equation}
\zeta^{(\text{KP})}=A^{(\text{KP})}\mathrm{sech}^2\left[\frac{\sqrt{A^{(\text{KP})}}}{2k} (kx+ly-\omega t) \right]\,,\label{eq:sol_KP}
\end{equation}
with the dispersion relation $A^{(\text{KP})}=-2-l^2/k^2-2\omega/k$. Finally, a simple rescaling
\begin{equation}
\tilde{t}=t\,,\quad \tilde{x}=-2^{1/3}x-2^{1/3}t\,,\quad \tilde{y}=2^{2/3}y\,,\quad \tilde{\zeta}=\frac{2^{2/3}}{3}\zeta\,,\label{eq:KPts}
\end{equation}
shifts the fKP equation~\eqref{eq:KP_topo} with a flat bottom to the standard form (equation~\ref{eq:KP_topo_flat}), which serves as a bridge for the subsequent comparisons between the conclusions drawn from the KP equation, namely expressions~\eqref{eq:res_kp_ref} and~\eqref{eq:res_kp_exp}, and the BL equation. 

\subsection{Numerical Schemes}\label{sec:num}
Here, we briefly describe the numerical methods for dealing with the fBL equation. The numerical algorithm is based on the pseudo-spectral method, and all the derivatives and pseudo-differential operators are computed with the aid of their Fourier symbols. The time integration of the equations is carried out using the classic fourth-order Runge--Kutta scheme with the method of integrating factors (see \citealt{milewski1999a} for details). Firstly, to fit these waves in a periodic domain of computation, we need to make $\xi$ periodic, which can be achieved by writing
\begin{equation}
\Psi=\xi(\boldsymbol{x},t)+\boldsymbol{C}\cdot\boldsymbol{x}\,,
\end{equation}
where the constant vector $\boldsymbol{C}=(C_1,C_2)$ is chosen to ensure that $\Psi$ is periodic. Thus, the fBL equation~\eqref{eq:BL-topo} can be rewritten as 
\begin{equation}
\left( 1-\Delta \right)\Psi_{tt} -2\boldsymbol{C}\cdot\nabla\Psi_t-\Delta\Psi=\mathcal{N}(\Psi,b)\,, \label{eq:BL1}
\end{equation}
where $\mathcal{N}(\Psi,b)$ includes all the nonlinear and topography terms, shown as 
\begin{equation}
\mathcal{N}(\Psi,b)=-|\nabla\Psi|^2_t-\Psi_t\Delta\Psi-b_t\,.
\end{equation}

Secondly, by performing a Fourier transform of equation~\eqref{eq:BL1} with respect to the spatial variables, we can reformulate it into a first-order single complex evolution equation through the introduction of
\begin{equation}
\widehat{u}=\left( \partial_t+\mathcal{S}+\mathrm{i} \mathcal{L} \right)\widehat{\Psi}\,,\quad \widehat{v}=\left( \partial_t+\mathcal{S}-\mathrm{i} \mathcal{L} \right)\widehat{\Psi}\,,
\end{equation}
with
\begin{equation}
\mathcal{S}=-\frac{\boldsymbol{C}\cdot\mathrm{i}\boldsymbol{k}}{1+|\boldsymbol{k}|^{2}}\,,\quad \mathcal{L}=\sqrt{\frac{|\boldsymbol{k}|^{2}}{1+|\boldsymbol{k}|^{2}}+\frac{\left(\boldsymbol{C}\cdot\boldsymbol{k}\right)^2}{\left(1+|\boldsymbol{k}|^{2}\right)^{2}}}\,,
\end{equation}
where the hat denotes Fourier-transformed variables and $\boldsymbol{k}=\left( k, l\right)$ is the wavenumber vector. Therefore, equation~\eqref{eq:BL1} can be recast as
\begin{equation}
\left(\partial_t+\mathcal{S}-\mathrm{i}\mathcal{L}\right)\widehat{u}=\frac{1}{1+|\boldsymbol{k}|^{2}}\widehat{\mathcal{N}}\,,\quad\left(\partial_t+\mathcal{S}+\mathrm{i}\mathcal{L}\right)\widehat{v}=\frac{1}{1+|\boldsymbol{k}|^{2}}\widehat{\mathcal{N}}\,.
\label{eq:BL2}
\end{equation}
Upon noticing that $\Psi$ is real so that $\widehat{\Psi}^*(\boldsymbol{k})=\widehat{\Psi}(-\boldsymbol{k})$, the two equations in \eqref{eq:BL2} are equivalent, and $\Psi$ and $\Psi_t$ can be recovered from $u$ alone with 
\begin{subequations}  
\begin{align}
&\widehat{\Psi}(\boldsymbol{k})=\frac{1}{2\mathrm{i}\mathcal{L}}\left[\widehat{u}(\boldsymbol{k})-\widehat{u}^*(-\boldsymbol{k})\right]\,,\label{eq:BL3-1}\\
&\widehat{\Psi}_t(\boldsymbol{k})=\frac{1}{2}\left[ \widehat{u}(\boldsymbol{k})+\widehat{u}^*(-\boldsymbol{k}) \right]-\mathcal{S}(\boldsymbol{k})\widehat{\Psi}(\boldsymbol{k})\,,\label{eq:BL3-2}
\end{align}
\end{subequations}
where the asterisk represents the complex conjugation. When $\boldsymbol{k}=(0,0)$, it is noted that $\mathcal{L}(\boldsymbol{0})=0$, which leads to the singularity of $\widehat{\Psi}(\boldsymbol{k})$ in equation~\eqref{eq:BL3-1}. Indeed, the value of $\widehat{\Psi}(\boldsymbol{0})$ can be obtained by directly solving equation~\eqref{eq:BL3-2}. The approach of integrating factors, in conjunction with the fourth-order Runge--Kutta scheme and the pseudo-spectral method, is utilized for the time stepping of the first equation in \eqref{eq:BL2}. Interested readers are referred to \citet{milewski1999a} for more details.  

To investigate the non-periodic and non-decaying solutions to the fBL equation, a windowing method, as proposed by \citet{schlatter2005}, is implemented to facilitate successful numerical computations. Indeed, this method has been widely employed to study nonlinear interactions among solitary waves by \citet{kao2012,ablowitz2013,yuan2018l,yuan2022}. For line solitons, we extend the computational domain sufficiently long in the $x$-direction to ensure that all solitons remain spatially distant from the boundary lines. Consequently, this expansion necessitates that these waves intersect with the boundary lines in the $y$-direction. To retain the effectiveness of the Fourier transform in the $y$-direction, a fundamental approach involves introducing a smooth function within the windowing method. This function exhibits compact support, with values that divide the solution into the interior and boundary-layer regions. The smooth function $W(y)$ can be simply defined as
\begin{equation}
W(y)=\exp\left(-a\Big|\frac{y}{L_y}\Big|^n\right)\,,
\end{equation}
where $=\pm L_y$ denotes the boundary lines in the $y$-direction with the parameters $n=27$ and $a=(1.111)^n\ln10$ throughout this paper. Correspondingly, at each time step, the solution within the computational domain is redefined as
\begin{equation}
\Psi=W\Psi+(1-W)\widetilde{\Psi}\,,\label{eq:wm_trans}
\end{equation}
where $\Psi$ is the solution to the fBL equation and $\widetilde{\Psi}$ is assumed to satisfy the fBL equation in the boundary layers. Applying the windowing method obviates the requirement for the constant vector $\boldsymbol{C}$ to guarantee the periodicity of the boundaries in the $y$-direction. Thus, we take $C_2=0$ in~\eqref{eq:BL1}. Submitting the transformation~\eqref{eq:wm_trans} into the fBL equation~\eqref{eq:BL1} and denoting $\psi(x,y,t)=W(y)\Psi(x,y,t)$, an alternative equation for $\psi$ can be derived as follows:
\begin{equation}
(1-\Delta)\psi_{tt}-2C_1\psi_{xt}-\Delta\psi=\mathcal{N}(\psi,b)+G_1(W,\widetilde{\Psi},b)+G_{2}(W,\psi, \widetilde{\Psi},b)\,,
\end{equation}
where
\begin{align*}
G_1(W,\widetilde{\Psi},b)=&-\left(1-W\right)\left[ (1-\Delta)\widetilde{\Psi}_{tt}-2C_1\widetilde{\Psi}_{xt}-\Delta\widetilde{\Psi}-\mathcal{N}(\widetilde{\Psi},b)\right]\,,\\
G_{2}(W,\psi, \widetilde{\Psi},b)=&-\left( W_{yy}\widetilde{\Psi}_{tt}+2W_y\widetilde{\Psi}_{ytt}+W_{yy}\widetilde{\Psi}+2W_y\widetilde{\Psi}_y \right)-(1-W)\left[ 2\left(\astrut \nabla\psi\cdot\nabla \widetilde{\Psi} \right)_t\right.\\
&\left.-W\left( |\nabla \widetilde{\Psi}|^2_t+\widetilde{\Psi}_t\Delta \widetilde{\Psi} \right) -2W_y\left( \widetilde{\Psi}_y \widetilde{\Psi}_t+\widetilde{\Psi} \widetilde{\Psi}_{yt} \right)+\psi_t\Delta \widetilde{\Psi}-W_{yy}\widetilde{\Psi}_t \widetilde{\Psi}\astrut\right]\\
&-2W_y^2\widetilde{\Psi} \widetilde{\Psi}_t+2W_y\left( \psi_y \widetilde{\Psi}_t+\psi_{yt}\widetilde{\Psi}+\psi_t \widetilde{\Psi}_y \right)+W_{yy}\psi_t \widetilde{\Psi}+(1-W)b_t\,,
\end{align*}
subject to the initial condition
\begin{equation*}
\psi(x,y,0)=\Psi(x,y,0)-(1-W)\widetilde{\Psi}(x,y,0)\,.
\end{equation*}

Given the assumption that $\widehat{\Psi}$ complies with the fBL equation~\eqref{eq:BL1} within the boundary layers, we can deduce that $G_1=0$, leaving only the term $G_2$ to be influential. To ensure the spectral accuracy of the results, we analytically calculate the terms related to $\widehat{\Psi}$ and $W$ in $G_2$.

Based on the above numerical method, computations for equation~\eqref{eq:BL-topo} are performed with a computational domain of $x\times y=[-500,500]\times[-1000,1000]$, discretized with $1024\times2048$ Fourier modes along the propagating and transverse directions, respectively. The pseudo-spectral scheme is coupled with the 2/3-rule dealiasing technique (\textit{i.e.}, using a buffer of half the number of modes in each horizontal direction for each time step).

\section{Modulation Theory}\label{sec:modu}
In the past several years, research on the modulation theory of the KP equation has been extensively conducted. The modulation system of the KP equation~\eqref{eq:KP_topo} without forcing reveals 
\begin{equation}
\begin{aligned}
&k_y-(k q)_x=0\,,\\
&k_t-\left(\frac{a k}{2}+\frac{k q^2}{2}+k\right)_x=0\,,\\
&q_t+q \left(\frac{a}{2}+\frac{q^2}{2}\right)_x-q_x\left(\frac{a}{2}+\frac{q^2}{2}+1\right)-\left(\frac{a}{2}+\frac{q^2}{2}\right)_y=0\,,\\
&\left( a^{3/2} \right)_t-\left(a^{3/2}-\frac{1}{2} a^{3/2} q^2+\frac{3 a^{5/2}}{10}\right)_x-\left(a^{3/2} q\right)_y=0\,,
\end{aligned}\label{eq:modu-KP}
\end{equation}
where $a=A^{(\text{KP})}$ is the wave amplitude, and $k$ and $l=qk$ denote the wavenumbers in the $x$- and $y$-directions, respectively (\citealt{grava2018,biondini2020,ryskamp2022}). Although system \eqref{eq:modu-KP} contains four equations with three unknowns to solve, the first equation holds automatically under a smoothness assumption (see \citealt{ablowitz2017} for details).

Next, we derive the modulation equations for the BL equation. The complete Whitham modulation theory for equation \eqref{eq:BL-topo-flat} is thoroughly presented in Appendix~\ref{appA}; in the following, we briefly outline the derivation focusing on soliton dynamics, a limiting case of periodic waves. To adapt the modulation theory, we rescale the independent variables as follows: 
\begin{equation}
X=\varepsilon x\,,\quad Y=\varepsilon y\,,\quad T=\varepsilon t\,,
\end{equation}
where $\varepsilon$ is a small parameter; thus, equation~\eqref{eq:BL-topo-flat} can be rewritten as 
\begin{equation}
\xi_{TT}-\left(1+\varepsilon^2\partial_{TT}\right)\left(\partial_{XX}+\partial_{YY}\right)\xi+\varepsilon\left[\left(\xi_X^2+\xi_Y^2\right)_T+\xi_T\left(\xi_{XX}+\xi_{YY}\right)\right]=0\,.\label{eq:e_BL}
\end{equation}
To apply the method of multiple scales, we search for a solution in the form of $\xi(\eta,X,Y,T)$, where $\eta(X,Y,T)$ is a fast variable satisfying 
\begin{equation}
\eta_X=k(X,Y,T)/\varepsilon\,,\quad\eta_Y=l(X,Y,T)/\varepsilon\,,\quad\eta_T=-\omega(X,Y,T)/\varepsilon\,.
\end{equation}
Here, $k$ and $l$ are the local wavenumbers in the $X$- and $Y$-directions, respectively, and $\omega$ is the wave frequency, all of which are assumed to vary slowly with respect to $X$, $Y$, and $T$. Equating the mixed second derivatives of $\eta$ yields the following compatibility conditions:
\begin{equation}
k_T+\omega_X=0\,,\quad l_T+\omega_Y=0\,,\quad k_Y-l_X=0\,.
\label{eq:whitham1}
\end{equation}
The first two relations in conditions~\eqref{eq:whitham1} constitute the first and second modulation equations, which automatically imply the third one for $T>0$ if it is satisfied at $T=0$. In addition, these two modulation equations are also called the equations of conservations of waves. We then seek an asymptotic solution of $\xi$ expanding in powers of $\varepsilon$, namely,
\begin{equation}
\xi=\xi^{(0)}(\eta,X,Y,T)+\varepsilon\xi^{(1)}(\eta,X,Y,T)+O(\varepsilon^{2})\,.
\label{E: q expansion}
\end{equation}
Substituting~\eqref{E: q expansion} into equation~\eqref{eq:e_BL} and collecting terms in the same power of $\varepsilon$, the leading order equation can be satisfied if we take
\begin{equation}
\xi^{(0)}(X,Y,T;\varepsilon)\sim\frac{a}{\delta\omega}\tanh\left[\frac{\delta}{\varepsilon}(k X+lY-\omega t)\right]\,,
\end{equation}
where $\delta$ is a scaling factor, and $a$ is a slowly varying function, shown as
\begin{equation}
a=\frac{\omega^2-k^2-l^2}{k^2+l^2}\,,\quad \delta=\frac{1}{2}\frac{\sqrt{a}}{|\omega|}\,.\label{eq:a_delta}
\end{equation}
Proceeding to the next-to-leading order equation, it yields that $\xi^{(1)}$ satisfies a forced problem. The standard solvability argument implies the forcing terms must be orthogonal to $\xi^{(0)}$. Returning to the original independent variables, upon simplification, we can transform the solvability condition into 
\begin{equation}
\text{sgn}(\omega)\left(a\sqrt{a}\frac{5+4a}{1+a}\right)_t+\left( a\sqrt{a}\frac{5+3a}{\sqrt{1+a}}\frac{k}{\sqrt{k^2+l^2}}\right)_x+\left(a\sqrt{a}\frac{5+3a}{\sqrt{1+a}}\frac{l}{\sqrt{k^2+l^2}}\right)_y=0\,,
\label{eq:modu1}
\end{equation}
where sgn$(\omega)$ denotes the direction of wave propagation if, without loss of generality, we take the wavenumber $k>0$. For convenience, we assume the wave propagates along the negative $x$-direction (\textit{i.e.,} $\omega<0$), and more discussions about the sign of $\omega$ are given in Appendix \ref{appA}. We introduce the dependent variable $q=l/k$ as the slope of soliton~\eqref{eq:sol_BL_0}. Subsequently, the compatibility conditions~\eqref{eq:whitham1} and the modulation equations~\eqref{eq:modu1} lead to
\begin{subequations}\label{eq:modu4}
\begin{align}
&k_y-(qk)_x=0\,,\label{eq:modu4_1}\\
&k_t-\left( k\sqrt{1+q^2}\sqrt{1+a}\right)_x=0\,,\label{eq:modu4_2}\\
&q_t+q\left(\sqrt{1+q^2}\sqrt{1+a}\right)_x-q_x\left(\sqrt{1+q^2}\sqrt{1+a}\right)-\left(\sqrt{1+q^2}\sqrt{1+a}\right)_y=0\,,\label{eq:modu4_3}\\
&\left(a\sqrt{a}\frac{5+4a}{1+a}\right)_t-\left(a\sqrt{a}\frac{5+3a}{\sqrt{1+a}}\frac{1}{\sqrt{1+q^2}}\right)_x-\left(a\sqrt{a}\frac{5+3a}{\sqrt{1+a}}\frac{q}{\sqrt{1+q^2}}\right)_y=0\,.\label{eq:modu4_4}
  \end{align}
\end{subequations}

In fact, equations~\eqref{eq:modu4_1}--\eqref{eq:modu4_3} arise from the consistency equations~\eqref{eq:whitham1}, and equation~\eqref{eq:modu4_4} is related to a certain conservation law inherent in the BL equation. In particular, the constraint $k_y-(qk)_x=0$ can be shown to be invariant with respect to time by the method discussed by \cite{ablowitz2017}. Thus, if the phase $\eta(X,Y,T)$ is smooth, then equation~\eqref{eq:modu4_1} is automatically satisfied.

It is observed that equations~\eqref{eq:modu4_3} and~\eqref{eq:modu4_4} only involve two physical variables, $a$ and $q$, giving rise to a closed system. Equations~\eqref{eq:modu4_3} and~\eqref{eq:modu4_4} will be employed in the subsequent sections to capture soliton amplitude and slope evolutions.

\section{Bent Solitons -- Mach Expansion}\label{sec:mach_exp}
\begin{figure}
\centering
\begin{minipage}{1.7in}
\centering
\includegraphics[height=1.5in]{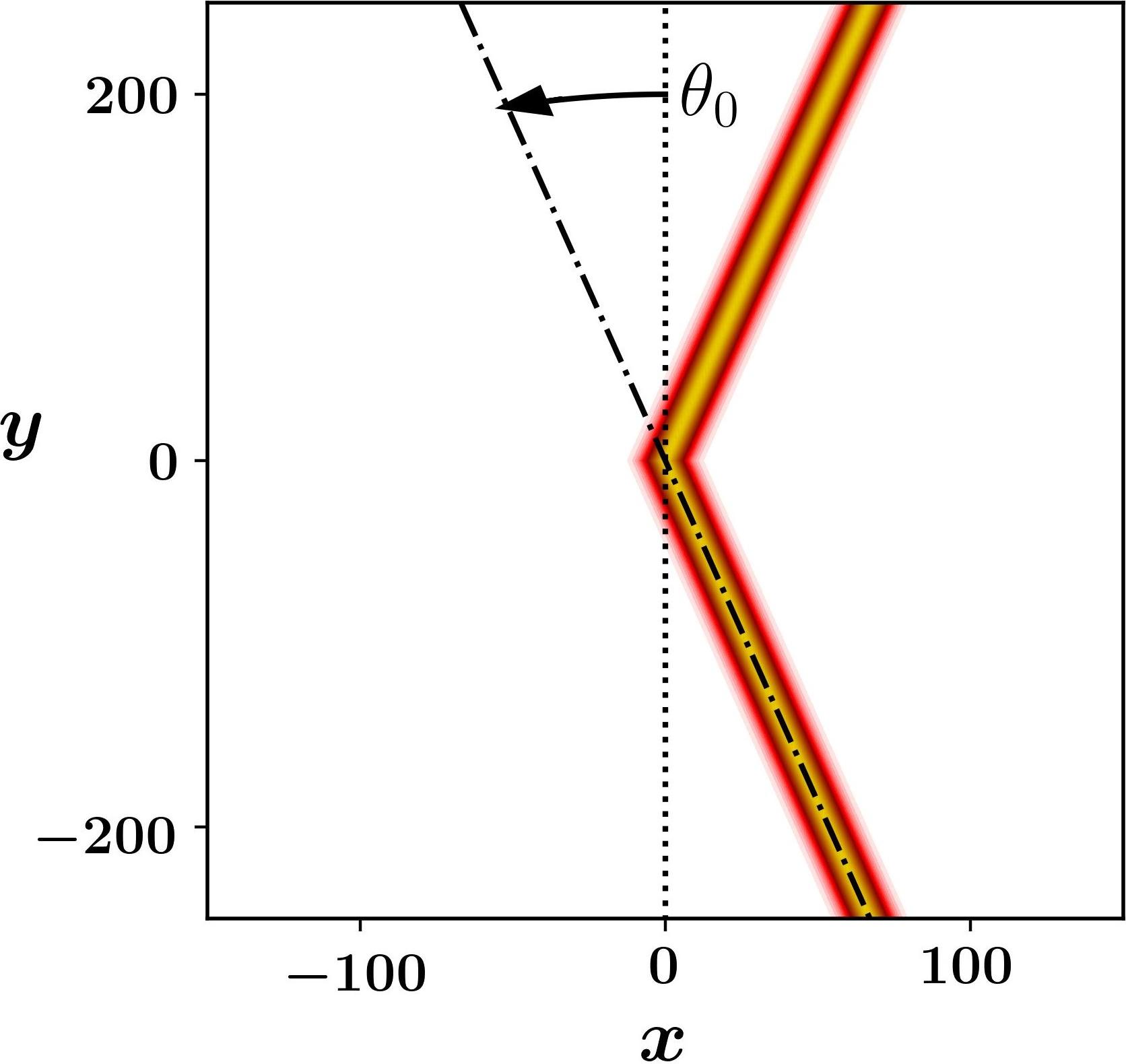}
\subcaption{$t=0$}
\label{fig:V15-a}
\end{minipage}
\centering
\begin{minipage}{1.7in}
\centering
\includegraphics[height=1.5in]{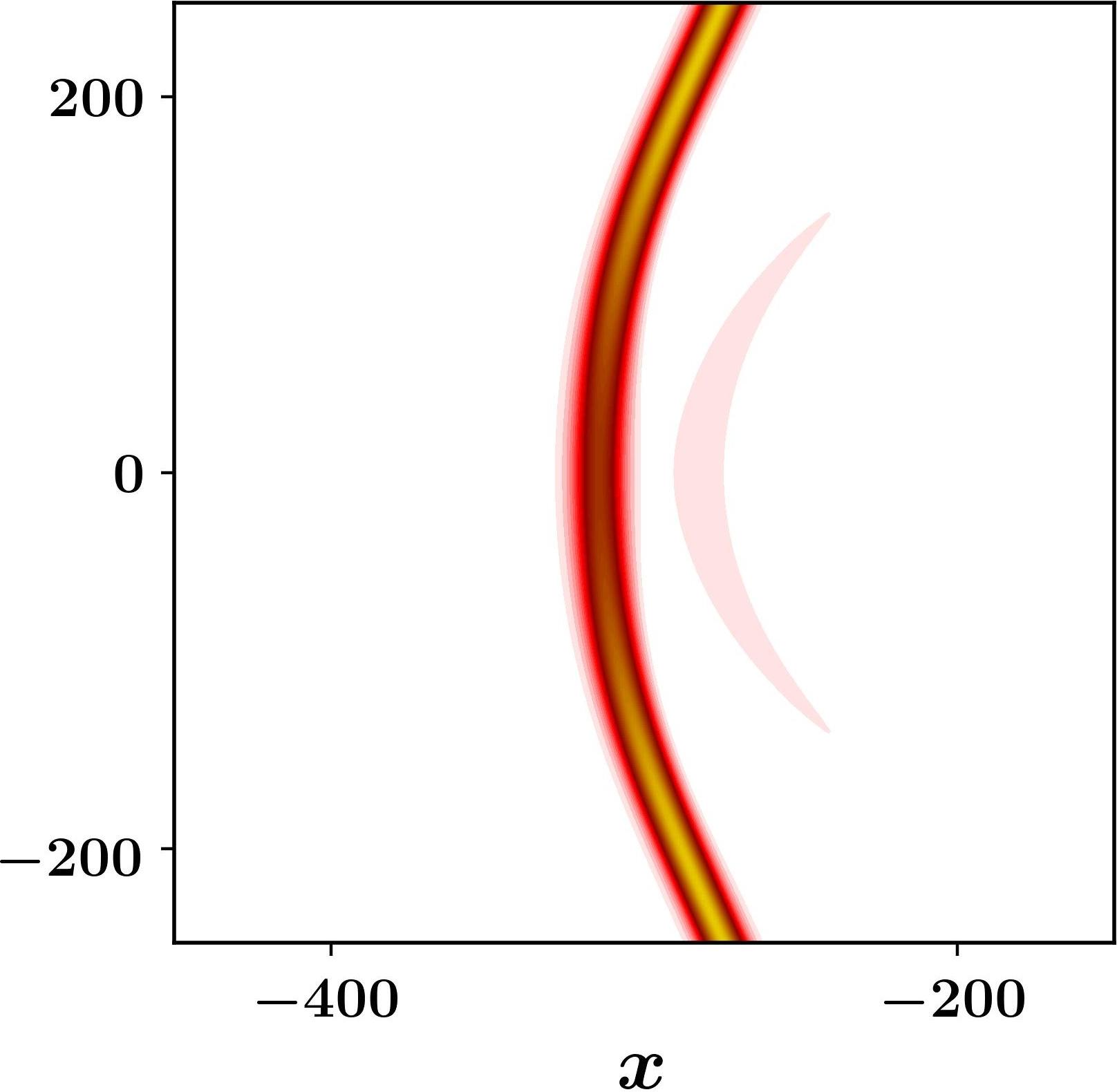}
\subcaption{$t=300$}
\label{fig:V15-b}
\end{minipage}
\centering
\begin{minipage}{1.7in}
\centering
\includegraphics[height=1.5in]{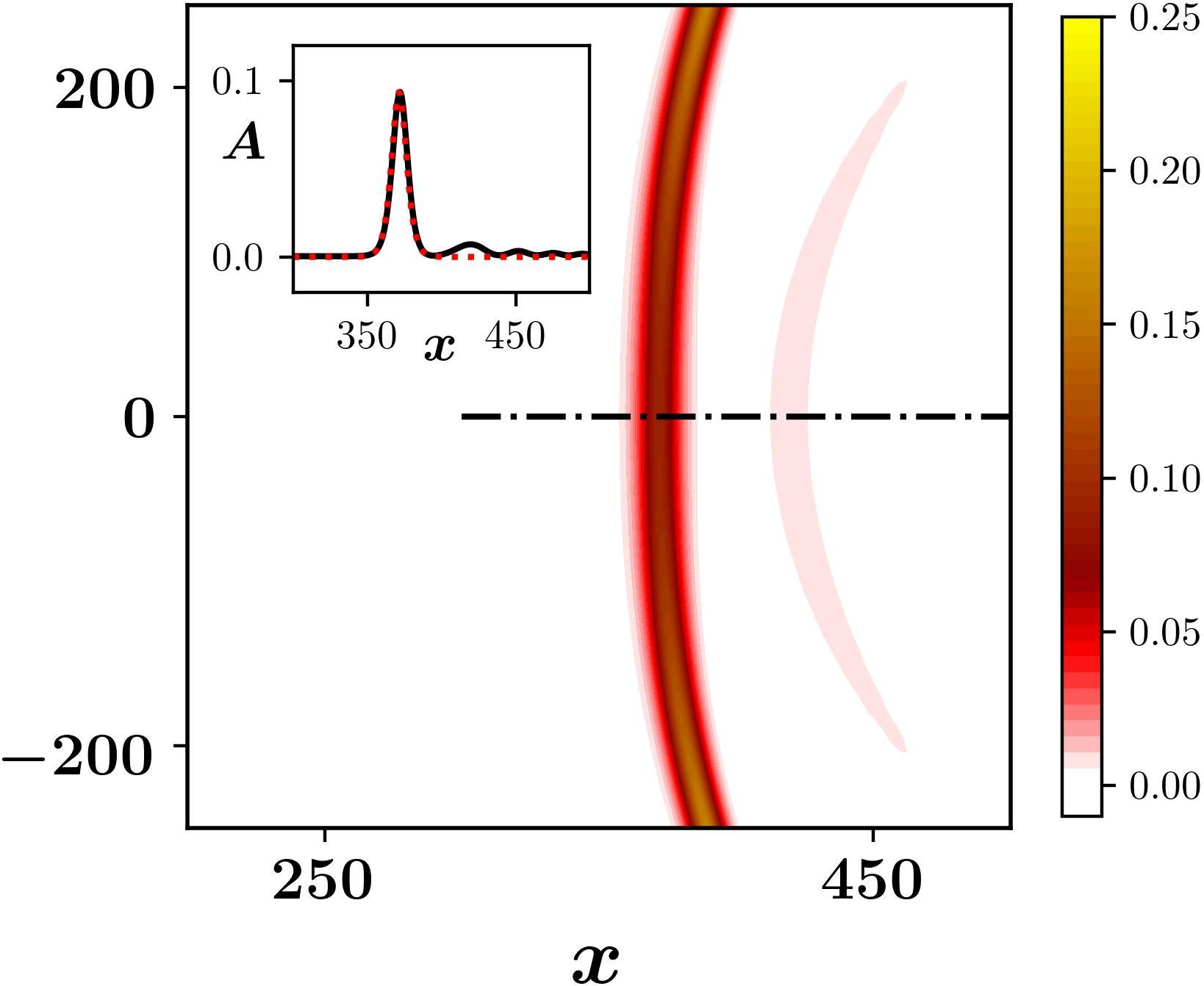}
\subcaption{$t=600$}
\label{fig:V15-c}
\end{minipage}\\
\caption{Time evolution of the bent-stem soliton initial data~\eqref{eq:ini_bs} with $\theta_0=-15^\circ$ and $a_0=0.21$, obtained by numerically solving the BL equation: (a) $t=0$, (b) $t=300$, and (c) $t=600$. The black solid curve in the subplot of (c) represents the cross-sectional profile at $y=0$ (corresponding to the dash-dotted line in the main plot), and the red dashed curve illustrates the theoretical prediction of the waveform given by~\eqref{eq:sol_mach_stem}.}\label{fig:V15}
\end{figure}

A bent soliton in water waves refers to a type of soliton propagating over the water surface while following a curved or bent trajectory. The evolution of the KP equation with a bent soliton initial condition displays smoothing and expanding dynamics akin to a Mach expansion in supersonic flow (\citealt{ryskamp2021}). To investigate the underlying phenomenon in the BL equation and compare it with the KP equation, we adopt the analogous bent soliton initial condition:
\begin{equation}
a(y,0)=a_0\,,\quad q(y,0)=\left\{
\begin{array}{ll}
q_0\,, & y\geq0\,,\\
-q_0\,, & y<0\,,
\end{array} \right.\quad k=\cos\theta_0\,,\quad q_0=\tan\theta_0\,,
\label{eq:ini_bs}
\end{equation}
where $\theta_0$ denotes the initial inclined angle of the soliton (see figure~\ref{fig:V15-a}). In this scenario, we are only concerned with variations of amplitude and slope along the $y$-direction. Consequently, the modulation equations~\eqref{eq:modu4_3} and~\eqref{eq:modu4_4} lead to a $2\times2$ quasi-linear hyperbolic system, given by
\begin{equation}
\begin{aligned}
&a_t-\frac{q\sqrt{1+a}}{\sqrt{1+q^2}}a_y-\frac{2a\left(5+3a\right)(1+a)^{3/2}}{\left(12a^2+25a+15\right) \left(1+q^2\right)^{3/2}}q_y=0\,,\\
&q_t-\frac{\sqrt{1+q^2}}{2\sqrt{1+a}}a_y-\frac{q\sqrt{1+a}}{\sqrt{1+q^2}}q_y=0\,,
\end{aligned}\label{eq:modu_yt}
\end{equation}
with the characteristic speeds 
\begin{equation}
\begin{aligned}
&U(a,q)=-\frac{\sqrt{1+a}}{\sqrt{1+q^2}}q-\frac{\sqrt{a}\sqrt{1+a}\sqrt{5+3a}}{\sqrt{12a^2+25a+15}\sqrt{1+q^2}}\,,\\
&V(a,q)=-\frac{\sqrt{1+a}}{\sqrt{1+q^2}}q+\frac{\sqrt{a}\sqrt{1+a}\sqrt{5+3a}}{\sqrt{12a^2+25a+15}\sqrt{1+q^2}}\,.
\end{aligned}
\end{equation}
Integrating along each characteristic direction unveils that
\begin{equation}
r(a,q)=\arctan q+I_0^a\,,\quad s(a,q)=\arctan q-I_0^a\,,\label{eq:rs}
\end{equation}
are the Riemann invariants for system~\eqref{eq:modu_yt}, where $I_m^n$ denotes the definite integral with the lower limit $m$ and the upper limit $n$, shown as
\begin{equation}
I_m^n=\int_m^n \frac{\sqrt{12z^2+25z+15}}{2(1+z)\sqrt{z}\sqrt{5+3z}}\text{d}z\,.
\end{equation}
Based on the Riemann variables, $r$ and $s$, the modulation equations~\eqref{eq:modu_yt} can be equivalently transformed to the diagonal form
\begin{equation}
r_t+U r_y=0\,,\qquad s_t+Vs_y=0\,.
\label{eq:rs_eq}
\end{equation}
Subsequently, theoretical analyses based on the method of characteristics will be performed for the modulation system~\eqref{eq:rs_eq}. Meanwhile, the fifth-order weighted essentially non-oscillatory (WENO5) scheme (see \citealt{shu1998} for details of the method) will be applied to numerically solve some initial-value problems for the quasi-linear hyperbolic system~\eqref{eq:modu_yt}.

\subsection{Mach expansion}
The occurrence of Mach expansion originates from the bent soliton initial condition~\eqref{eq:ini_bs} with $\theta_0<0$, depicted in figure~\ref{fig:V15-a}, and the negative sign of $\theta$ yields two rarefaction waves emanating from the point $(0,0)$. Figures~\ref{fig:V15-a}--\ref{fig:V15-c} show the evolution of the initial data~\eqref{eq:ini_bs} with $\omega_0=-1.1$ and $\theta_0=-15^\circ$. In figures~\ref{fig:V15-b} and \ref{fig:V15-c}, a vertical soliton is observed in conjunction with the evolution of the bent soliton, identified as the Mach stem. 

The specific characteristics of the Mach stem can be described by the hyperbolic system~\eqref{eq:modu_yt} associated with the initial data~\eqref{eq:ini_bs}. In this case, one wave propagates along the positive direction of the $y$-axis with the Riemann variable $r$ held constant, while the other wave propagates along the negative direction of the $y$-axis with the other Riemann variable $s$ held constant. Therefore, the slope $q_w$ and the amplitude $a_w$ of the Mach stem are subject to the constraint
\begin{equation}
r(a_0,\tan \theta_0)=r(a_w,q_w)\,,\quad s(a_0,-\tan \theta_0)=s(a_w,q_w)\,,
\end{equation}
which yields the expressions
\begin{equation}
q_w=0\,,\quad\theta_0=I_{a_0}^{a_w}=\int_{a_0}^{a_w} \frac{\sqrt{12a^2+25a+15}}{2(1+a)\sqrt{a}\sqrt{5+3a}}\text{d}a\,.\label{eq:aqw}
\end{equation}
Here, the amplitude $a_w$ of the Mach stem can be obtained by numerically solving the following boundary value problem:
\begin{equation}
\frac{\text{d}a(\theta)}{\text{d}\theta}=-\frac{2(1+a)\sqrt{a}\sqrt{5+3a}}{\sqrt{12a^2+25a+15}}\,,\quad a(\theta_0)=a_0\,,\quad a(0)=a_w\,.\label{eq:ode_for_aw}
\end{equation}

\begin{figure}
\centering
\begin{minipage}{1.7in}
\centering
\includegraphics[height=1.5in]{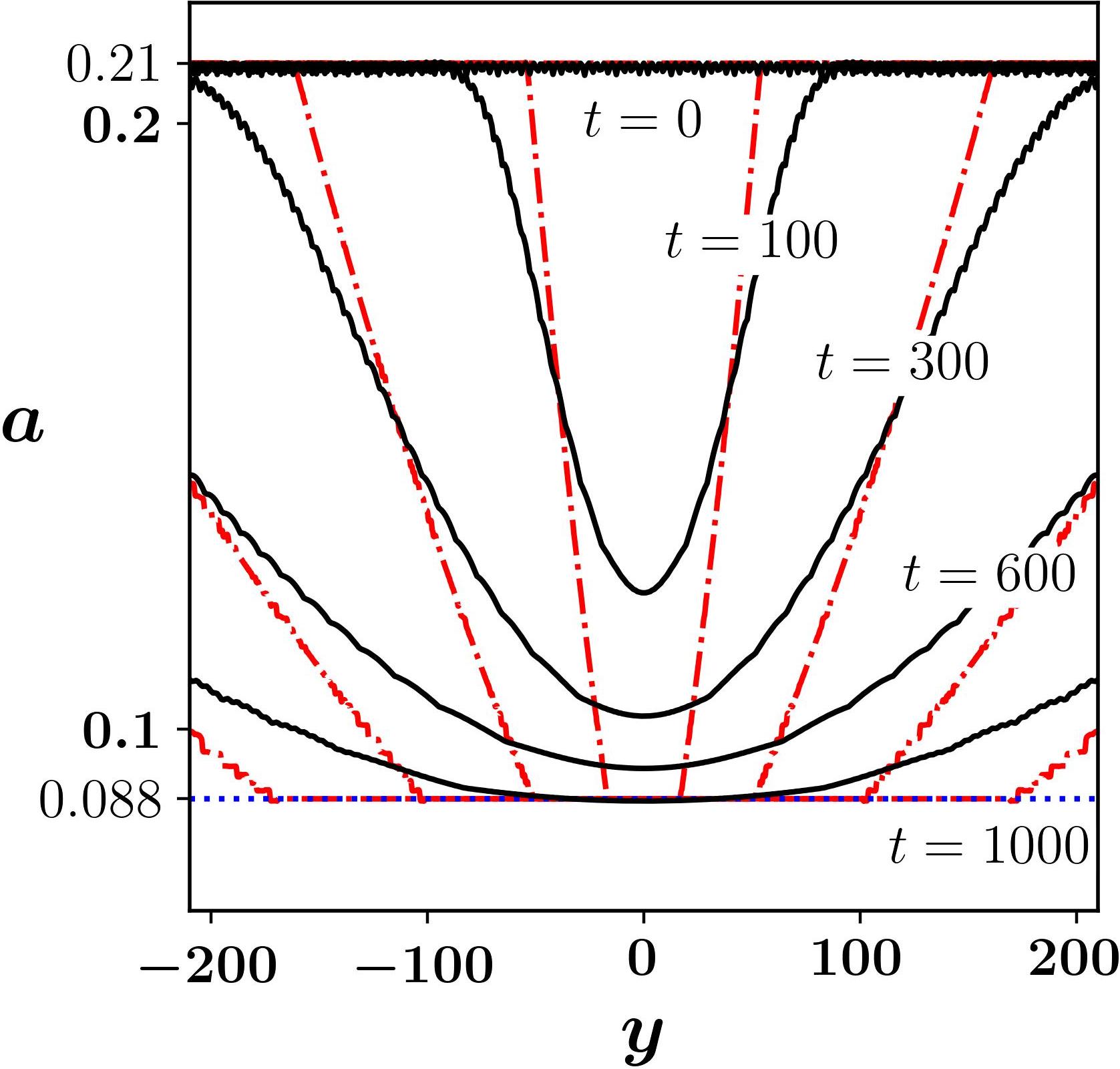}
\subcaption{}
\label{fig:V15-d}
\end{minipage}
\centering
\begin{minipage}{1.7in}
\centering
\includegraphics[height=1.5in]{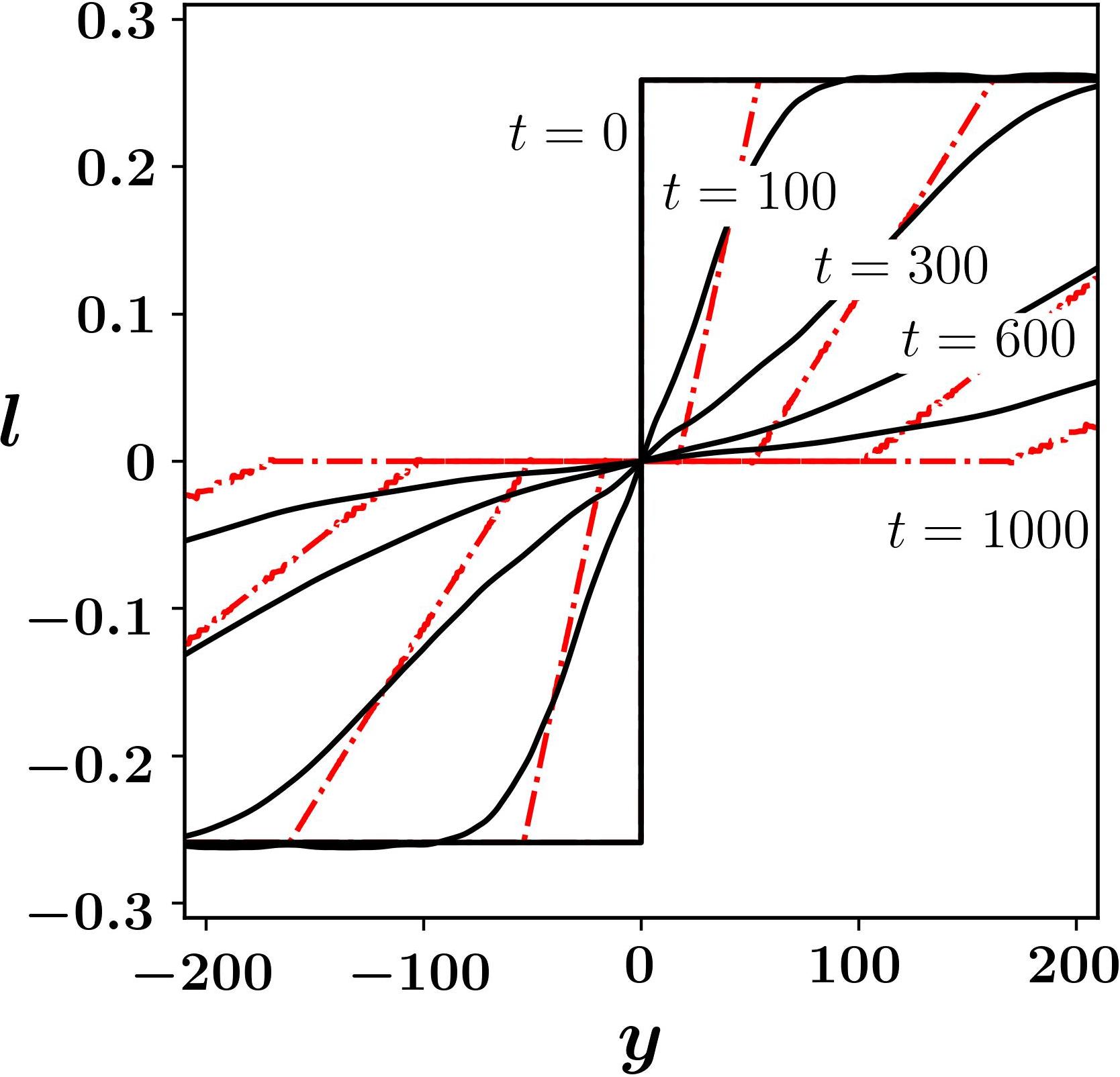}
\subcaption{}
\label{fig:V15-e}
\end{minipage}
\centering
\begin{minipage}{1.7in}
\centering
\includegraphics[height=1.5in]{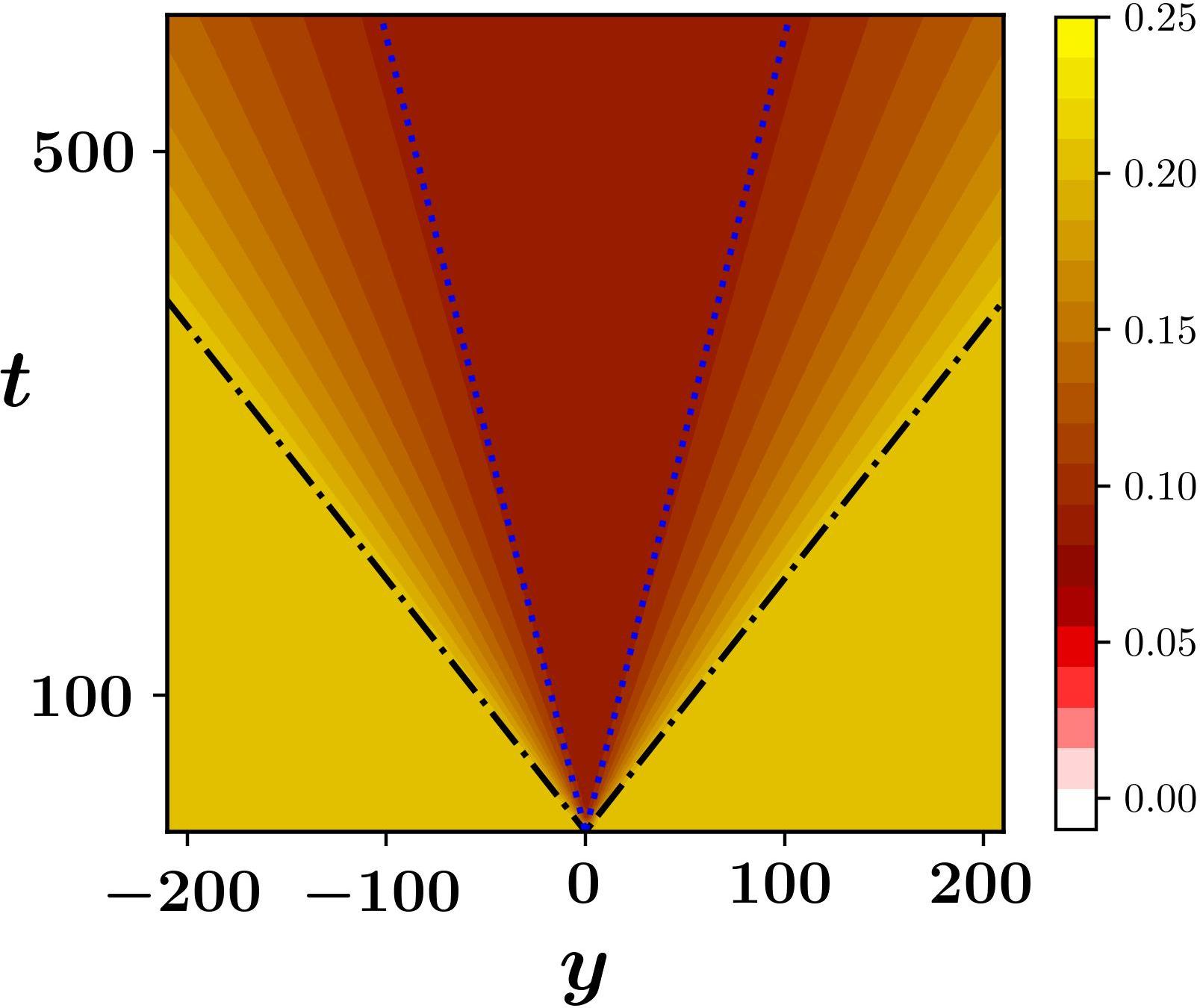}
\subcaption{}
\label{fig:V15-f}
\end{minipage}\\
\caption{(a)--(b) evolutions of the amplitude $a$ and slope $q$ for the bent soliton initial data~\eqref{eq:ini_bs} with $\theta_0=-15^\circ$ and $a_0=0.21$ at different times. They are obtained from the numerical simulations for the BL equation~\eqref{eq:BL-topo-flat} (dark solid curves) and the modulation equations~\eqref{eq:modu_yt} (red dashed curves). (c) space-time contour plots of the amplitude of the Mach stem with the same initial data, where the blue dashed lines and black dash-dotted lines characterize the edges of the evolution, predicted by the solutions~\eqref{eq:sol_modu_1}.}\label{fig:V15-aq} 
\end{figure}

Figures~\ref{fig:V15-d} and~\ref{fig:V15-e} show the evolution of the amplitude $a$ and the slope $q$ for the initial data~\eqref{eq:ini_bs} with $\theta_0=-15^\circ$ and $a_0=0.21$. Results depicted by the dark solid curves are derived from numerical simulations for the BL equation~\eqref{eq:BL-topo-flat}, whereas the outcomes illustrated by the red dashed curves are obtained from numerical simulations for the modulation equations~\eqref{eq:modu_yt}. It can be observed from figures~\ref{fig:V15-d} and~\ref{fig:V15-e} that the amplitude and the slope are well predicted by the solutions of the modulation equations~\eqref{eq:modu_yt}. Furthermore, after a long period of evolution, the amplitude and the slope of the Mach stem converge to $a_w\approx 0.088$ and $q_w=0$, respectively, as derived from \eqref{eq:aqw}. Interestingly, we have observed that the Mach wave takes the form of a soliton, accompanied by a series of arc-shaped oscillations trailing behind the wavefront, which will be explained in section~\ref{sec:wake}. The waveform of the Mach stem can be expressed as
\begin{equation}
a=a_w \text{sech}^2\left[ \frac{\sqrt{a_w}}{2|\omega_0|}(x-\omega_0t) \right]\,,\quad \omega_0=-\sqrt{a_0+1}\,.\label{eq:sol_mach_stem}
\end{equation}

Furthermore, the simple wave solution to the modulation equations~\eqref{eq:modu_yt} with the bent soliton initial data~\eqref{eq:ini_bs} reveals
\begin{equation}
\begin{aligned}
I_0^a&=\left\{
\begin{array}{ll}
I_0^{a_0}\,, & |y|>V_0t\,,\\[2pt]
\frac{\theta_0}{V_w-V_0}\frac{|y|}{t}-\frac{V_0I_0^{a_w}-V_wI_0^{a_0}}{V_w-V_0}\,, & V_wt\le|y|\le V_0 t\,,\\[8pt]
I_0^{a_w}\,, & |y|<V_wt\,,
\end{array}\right.\\[3pt]
q&=\text{sgn}(y)\left\{
\begin{array}{ll}
\tan\theta_0\,, & |y| > V_0 t\,,\\[3pt]
\tan(\theta_0+I_{a}^{a_0})\,, & V_w t\le |y|\le V_0 t\,,\\[3pt]
0\,, & |y|<V_wt\,,
\end{array}\right.
\end{aligned}
\label{eq:sol_modu_1}
\end{equation}
with 
\begin{equation}
V_w=V(a_w,0)\,,\quad V_0=V(a_0,\tan\theta_0)\,.
\end{equation}
The amplitude of the Mach stem, as determined by the modulation equations~\eqref{eq:modu_yt}, is illustrated in figure~\ref{fig:V15-f} through the contour plot. In the plot, the black dash-dotted lines and blue dashed lines represent the edges of the rarefaction wave, with the slopes given by $\pm V_0$ and $\pm V_w$.

The appearance of the Mach stem is due to $a_w>0$. When $a_w=0$, we can determine the critical angle of the slope for the occurrence of the Mach stem, which is given by
\begin{equation}
\theta_{\text{cr}}=\int_{a_0}^{0}\frac{\sqrt{12a^2+25a+15}}{2(1+a)\sqrt{a}\sqrt{5+3a}}\;\text{d}a\,.\label{eq:theta_cr_exp}
\end{equation}
Thus, in the case of $\theta_{\text{cr}}\le\theta_0\le0$, the bent soliton will evolve into a Mach stem perpendicular to the propagation direction, exhibiting a phenomenon similar to Mach expansion in gas dynamics. 

\subsection{Regular expansion}
When $\theta_0<\theta_{\text{cr}}$ (corresponding to the negative value of $a_w$), the Mach stem disappears; instead, the regular expansion phenomenon occurs. Similar to the analysis in \citet{ryskamp2021}, we let $a_w=0$, resulting in $q_w\neq 0$. By symmetry, we can obtain $q$ locally based on the simple wave criterion,
\begin{equation}
q_{w+}=\lim_{y\rightarrow0^+}q(y,t)=-\lim_{y\rightarrow 0^-}q(y,t)=q_{w-}\,,
\end{equation}
with the initial discontinuity at $y=0$. Then, with the aid of the Riemann invariants, the solution for the rarefaction wave is given by
\begin{equation}
\begin{aligned}
I_0^a&=\left\{
\begin{array}{ll}
I_0^{a_0}\,, & |y| > V_0t\,,\\[3pt]
\frac{I_0^{a_0}}{V_0-V_w}\frac{|y|}{t}-\frac{V_wI_0^{a_0}}{V_0-V_w}\,, & V_w t\le |y|\le V_0t\,,\\[6pt]
0\,, & |y|<V_w t\,,
\end{array} \right.\\[2pt]
q&=\text{sgn}(y)\left\{
\begin{array}{ll}
\tan \theta_0\,, & |y|>V_0t\,,\\[2pt]
\tan (\theta_0+I_{a}^{a_0}) & V_wt\le |y|\le V_0t\,,\\[2pt]
q_{w+}\,, & |y|<V_wt\,,
\end{array}\right.
\end{aligned}
\end{equation}
with 
\begin{equation}
V_w=V(0,q_{w+})\,,\quad V_0=V(a_0,\tan \theta_0)\,. 
\end{equation}
\begin{figure}
\centering
\begin{minipage}{1.7in}
\centering
\includegraphics[height=1.5in]{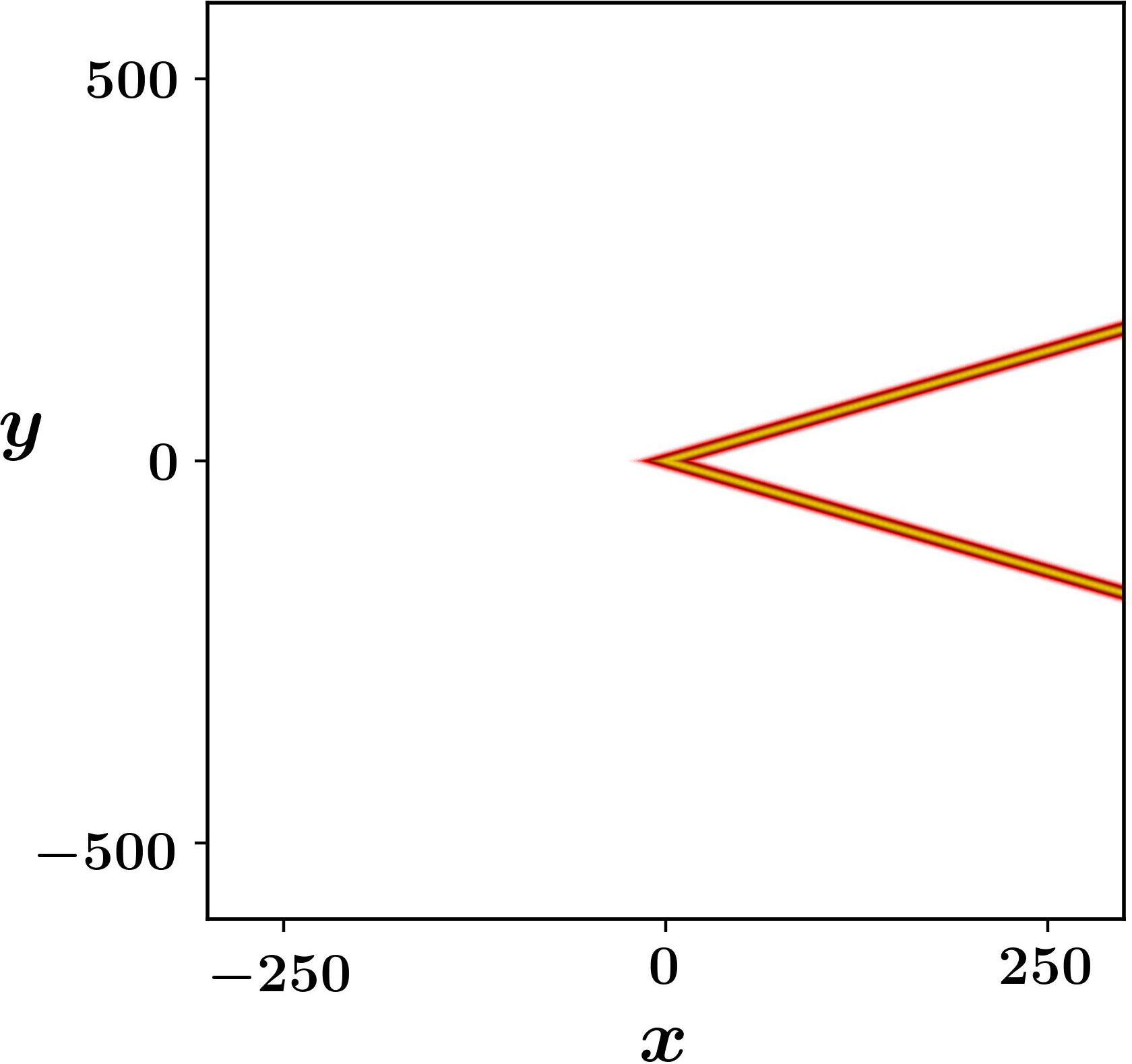}
\subcaption{$t=0$}
\label{fig:V60-a}
\end{minipage}
\centering
\begin{minipage}{1.7in}
\centering
\includegraphics[height=1.5in]{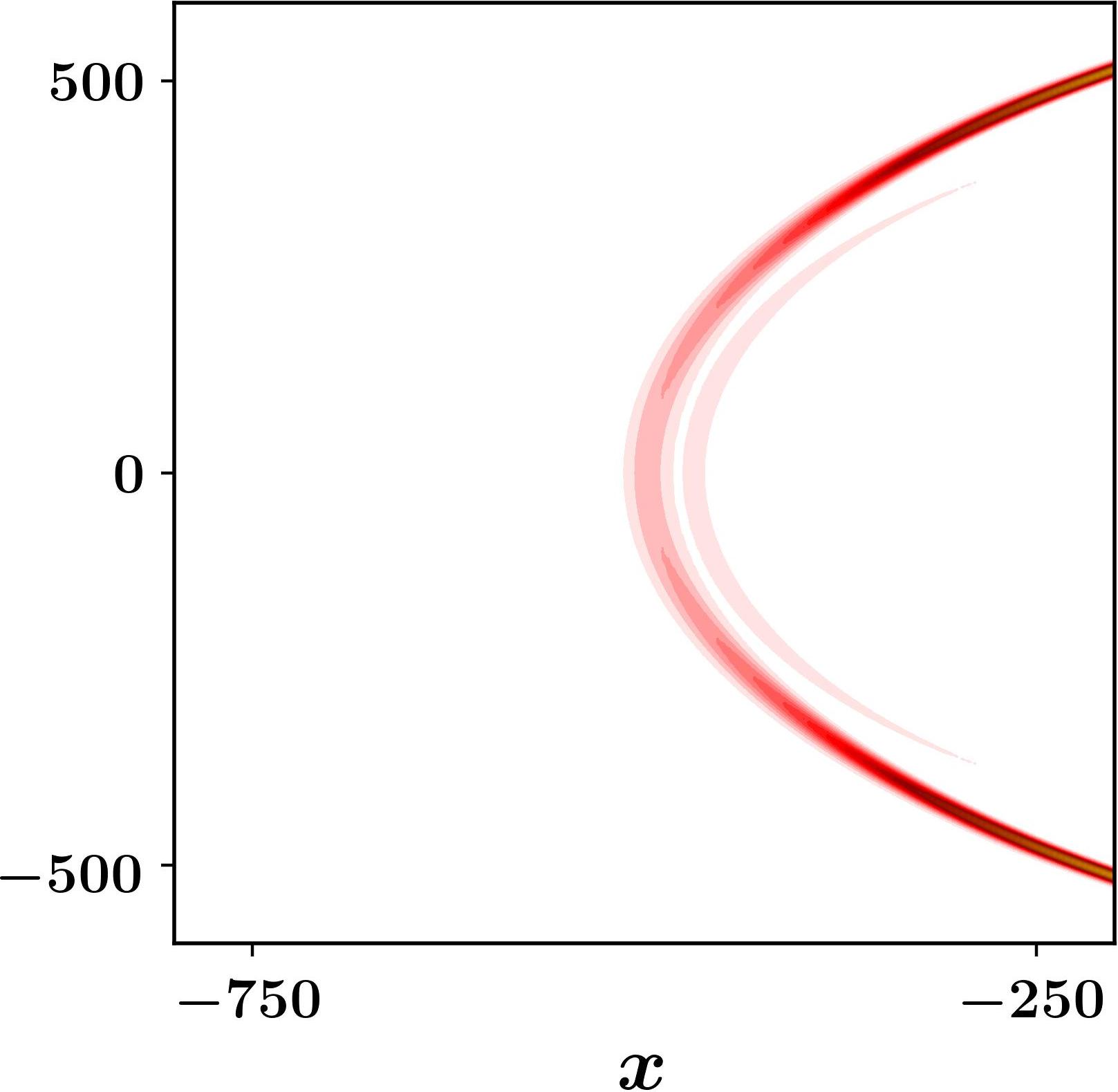}
\subcaption{$t=500$}
\label{fig:V60-b}
\end{minipage}
\centering
\begin{minipage}{1.7in}
\centering
\includegraphics[height=1.5in]{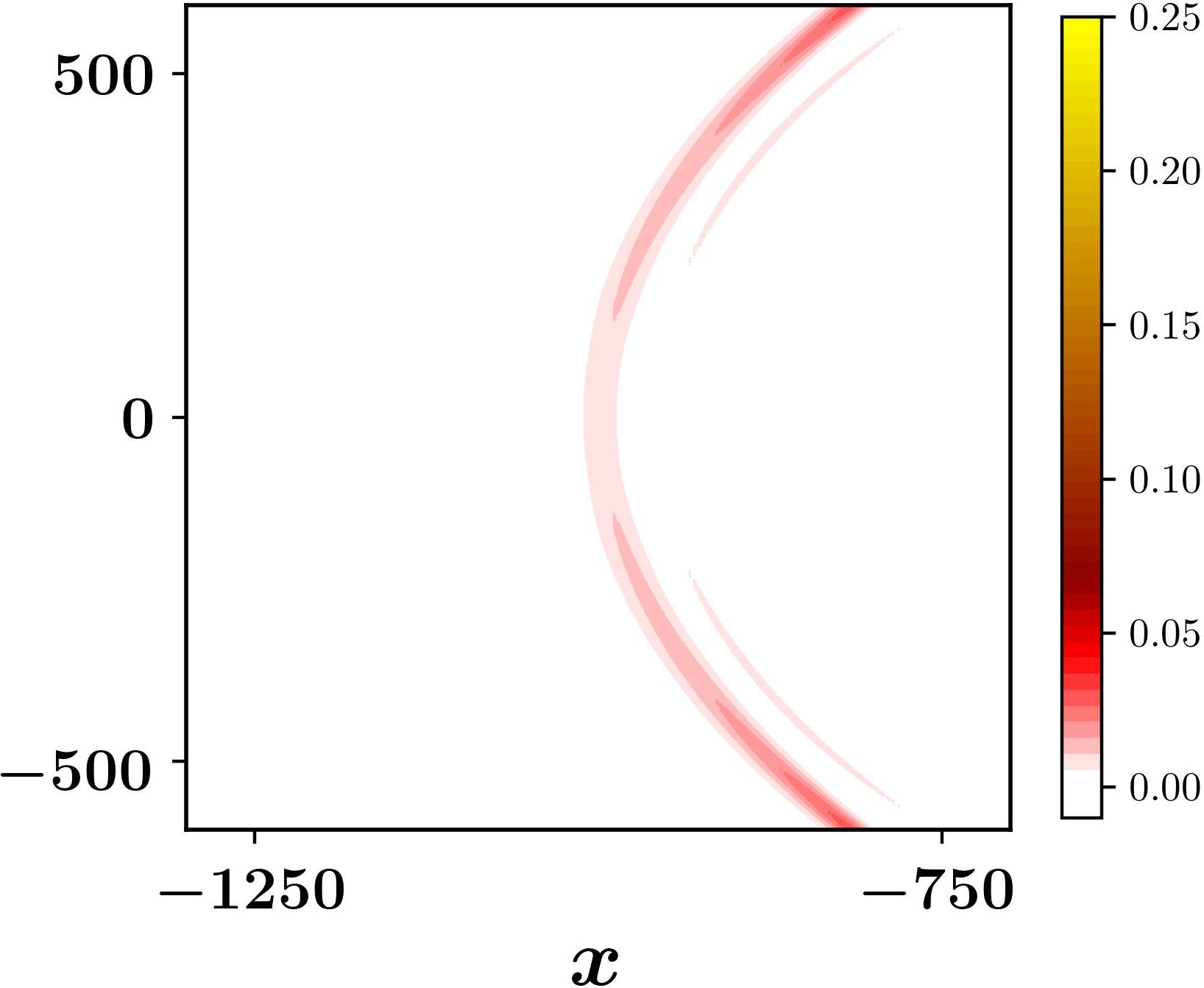}
\subcaption{$t=1000$}
\label{fig:V60-c}
\end{minipage}\\
\caption{Time evolution of the bent-stem soliton initial data~\eqref{eq:ini_bs} with $\theta_0=-60^\circ$ and $a_0=0.21$: (a) $t=0$, (b) $t=500$, and (c) $t=1000$.}\label{fig:V60}
\end{figure}

Figure~\ref{fig:V60} shows the regular reflection of the bent soliton with $\theta=-60^\circ$ and $a_0=0.21$ for the initial condition~\eqref{eq:ini_bs}. To mitigate the interference of trailing waves from the cyclic boundary on the outcomes, we extend the computational domain to $x\times y=2000\times 1000$ in this scenario while maintaining Fourier modes unchanged. Unlike the Mach expansion shown in figure~\ref{fig:V15}, a stable Mach stem formation does not emerge in this situation. What is noted instead is the presence of an arc-shaped wavefront, which exhibits a gradual reduction in amplitude attenuating with time and serves to connect two bent solitons.

\subsection{Comparison between the BL and KP equations}
\begin{figure}
\centering
\begin{minipage}{2in}
\centering
\includegraphics[height=1.7in]{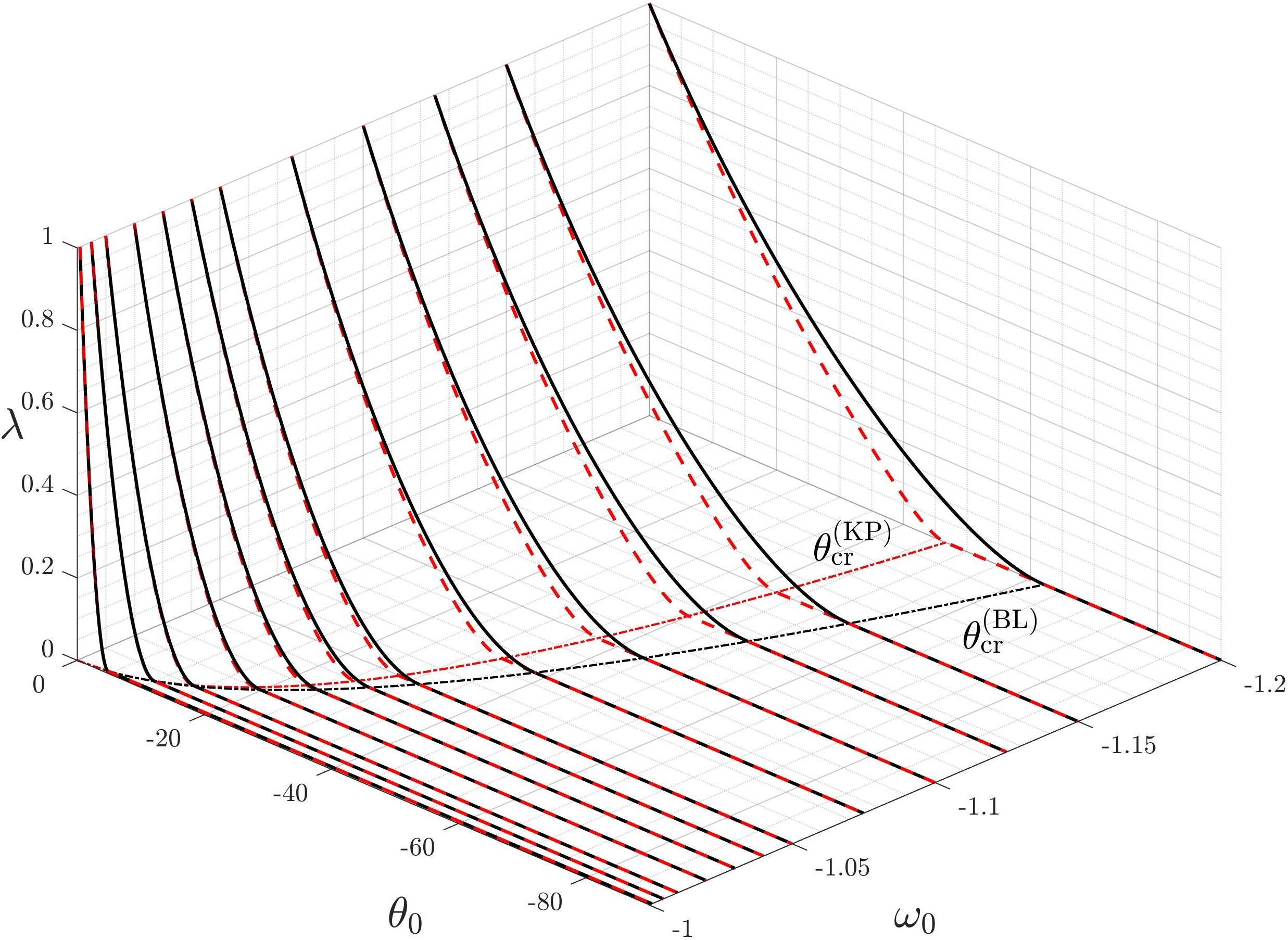}
\subcaption{}\label{fig:same_c-A}
\end{minipage}
\hspace{1cm}
\begin{minipage}{2in}
\centering
\includegraphics[height=1.7in]{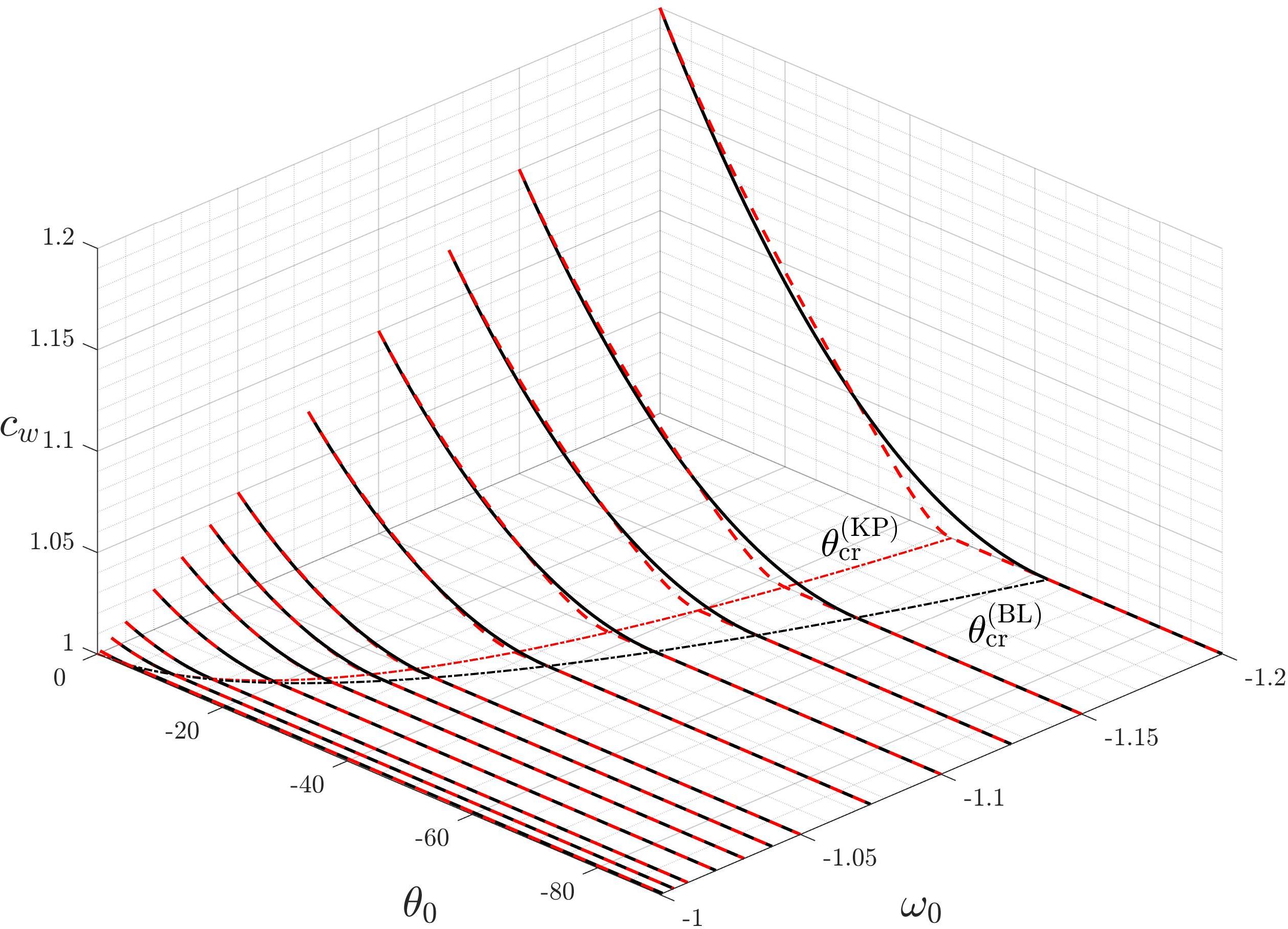}
\subcaption{}
\label{fig:same_c-c}
\end{minipage}
\caption{Comparative plot of the wall reduction factor (a) and the speed (b) of the Mach stem in the KP (red dashed curves) and BL (black solid curves) equations to the initial angle $\theta_0$ and speed $\omega_0$ of soliton. The dash-dotted curves denote the critical angle of the Mach expansion versus the initial speed $\omega_0$ for the KP (red) and BL (black) equations given by expressions~\eqref{eq:kp_mach_cr} and~\eqref{eq:bl_mach_cr}, respectively.}\label{fig:same_c}
\end{figure}

To capture the effects of isotropism, we investigate wave propagation in both the BL and KP equations, and the main results are shown in figure~\ref{fig:same_c}, where we take the bent soliton in the form of \eqref{eq:ini_bs} as the initial data. In fact, due to its reduction of the BL equation, the KP equation shares characteristics similar to those of the BL equation to some extent, especially for the central part of the soliton evolution \citep{ablowitz2011,yuan2022,gidel2017}. Thus, we mainly focus on the differences between these two equations for the Mach expansion of the oblique soliton, namely, the wall reduction factor and the speed of the Mach stem. 

According to the modulation equations~\eqref{eq:modu-KP}, the corresponding amplitude and critical slope of the Mach stem in the KP equation can be obtained as follows:
\begin{equation}
A^{(\text{KP})}_w=\left( \sqrt{A^{(\text{KP})}_0}+\frac{q_0}{\sqrt{3}}\right)^2\,,\quad q^{(\text{KP})}_{\text{cr}}=-\sqrt{3A^{(\text{KP})}_0}\,, 
\end{equation}
where the initial amplitude $A^{(\text{KP})}_0$ is derived based on the dispersion relation
\begin{equation}
A^{(\text{KP})}_0=-2-q_0^2-2\omega_0/k=-2-\tan^2 \theta_0-2\frac{\omega_0}{\cos \theta_0}\,.
\end{equation}
Thus, one obtains the wall reduction factor, speed, and critical angle with respect to the initial angel $\theta_0$ and speed $\omega_0$, shown as
\begin{equation}
\begin{aligned}
&\lambda^{(\text{KP})}=\frac{A^{(\text{KP})}_w}{A^{(\text{KP})}_0}=\left(\frac{\sqrt{2}\sin\theta_0}{\sqrt{3}\sqrt{-3-\cos 2\theta_0-4\omega_0\cos\theta_0}}+1\right)^2\,,\\ 
&c_w^{(\text{KP})}=-\frac{A^{(\text{KP})}_w}{2}-1\,,\\
&\theta_{\text{cr}}^{(\text{KP})}=\arctan q^{(\text{KP})}_{\text{cr}}=-\text{arcsec}\left(\frac{\sqrt{9\omega_0^2-8}-3\omega_0}{4}\right)\,,
\end{aligned}\label{eq:kp_mach_cr}
\end{equation}
where the wall reduction factor refers to the ratio of the amplitude of the Mach stem to that of the incident soliton \citep[see][]{ryskamp2021}.

To facilitate a comparative analysis between the BL and KP equations, based on the relationship $\zeta=-\xi_x$, we define the soliton amplitude in the BL equation as $A^{(\text{BL})}=-ka/\omega$, with the dispersion relation $a=\omega^2/( k^2+l^2 )-1$. Thus, we can reveal the wall reduction factor, speed, and critical angle for the BL equation, 
\begin{equation}
\begin{aligned}
&\lambda^{(\text{BL})}=\frac{A^{(\text{BL})}_w}{A^{(\text{BL})}_0}=\frac{a_w\sqrt{a_0+1}}{k_0a_0\sqrt{a_w+1}}\,,\\ 
&c_w^{(\text{BL})}=-\sqrt{A^{(\text{BL})}_w+1}\,,\\
&\theta_{\text{cr}}^{(\text{BL})}=\int_{a_0}^{0} \frac{\sqrt{12a^2+25a+15}}{2(1+a)\sqrt{a}\sqrt{5+3a}}\text{d}a\sim \sqrt{6\varepsilon}\left( 1-\frac{11}{180}\varepsilon+\frac{2123}{36000}\varepsilon^2 \right)+O(\varepsilon^3)\,,
\end{aligned}\label{eq:bl_mach_cr}
\end{equation}
with $a_0=\omega_0^2-1$, $\varepsilon=-\omega_0-1$, and $a_w$ being determined through numerical solutions of equation~\eqref{eq:ode_for_aw}.

Figure~\ref{fig:same_c} presents a comparison of the wall reduction factor and the speed of the Mach stem between the KP and BL equations as functions of the initial angle $\theta_0$ and speed $\omega_0$. The wall reduction factor and the Mach stem speed for the KP and BL equations are depicted by the red dashed and black solid curves, respectively. The dash-dotted curves correspond to the critical angle of the Mach wave fan for both the KP and BL equations, derived from expressions~\eqref{eq:kp_mach_cr} and~\eqref{eq:bl_mach_cr}. Clearly, compared with the KP equation, the decrease of the critical angle for the BL equation is more rapid as $|\omega_0|$ is far away from the unity. Upon visual inspection of figure~\ref{fig:same_c}, it is discernible that when the speed $|\omega_0|$ is close to 1, there is a pronounced correspondence between the two datasets. Furthermore, as $|\omega_0|$ increases, the divergence between the datasets becomes progressively more pronounced. Examination of figure~\ref{fig:same_c-A} reveals that, for each speed, the wall reduction factor diminishes as the initial angle approaches the critical angle. This signifies that the amplitude of the Mach stem formed by the Mach expansion relative to the incident wave amplitude decreases. Additionally, it is observed that the wall reduction factor for the KP equation is consistently lower than that for the BL equation. This discrepancy becomes increasingly evident as the initial speed increases. While in figure~\ref{fig:same_c-c}, for the speed of the Mach stem $c_w$, regardless of the initial speed, the differences in $c_w$ between the two models are insignificant as the initial angle surpasses the criticality. However, as $\theta_0$ approaches the critical angle $\theta_{\text{cr}}$, a distinct divergence emerges in the values of $c_w$.

\section{Reverse Bent Solitons -- Mach Reflection}\label{sec:mach_ref}
\subsection{Mach reflection}
\begin{figure}
\centering
\begin{minipage}{1.7in}
\centering
\includegraphics[height=1.5in]{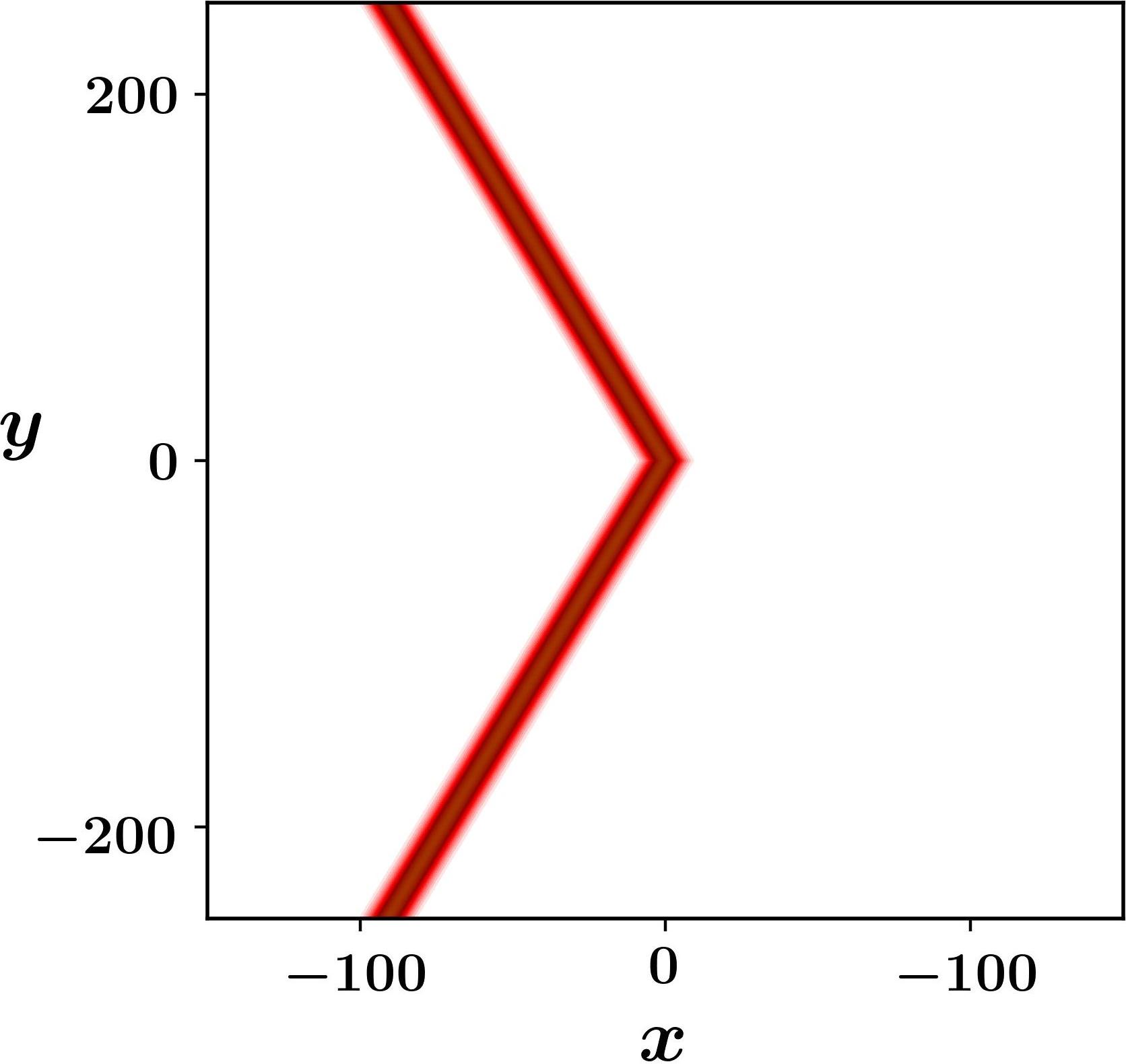}
\subcaption{$t=0$}\label{fig:RV20-a}
\end{minipage}
\centering
\begin{minipage}{1.7in}
\centering
\includegraphics[height=1.5in]{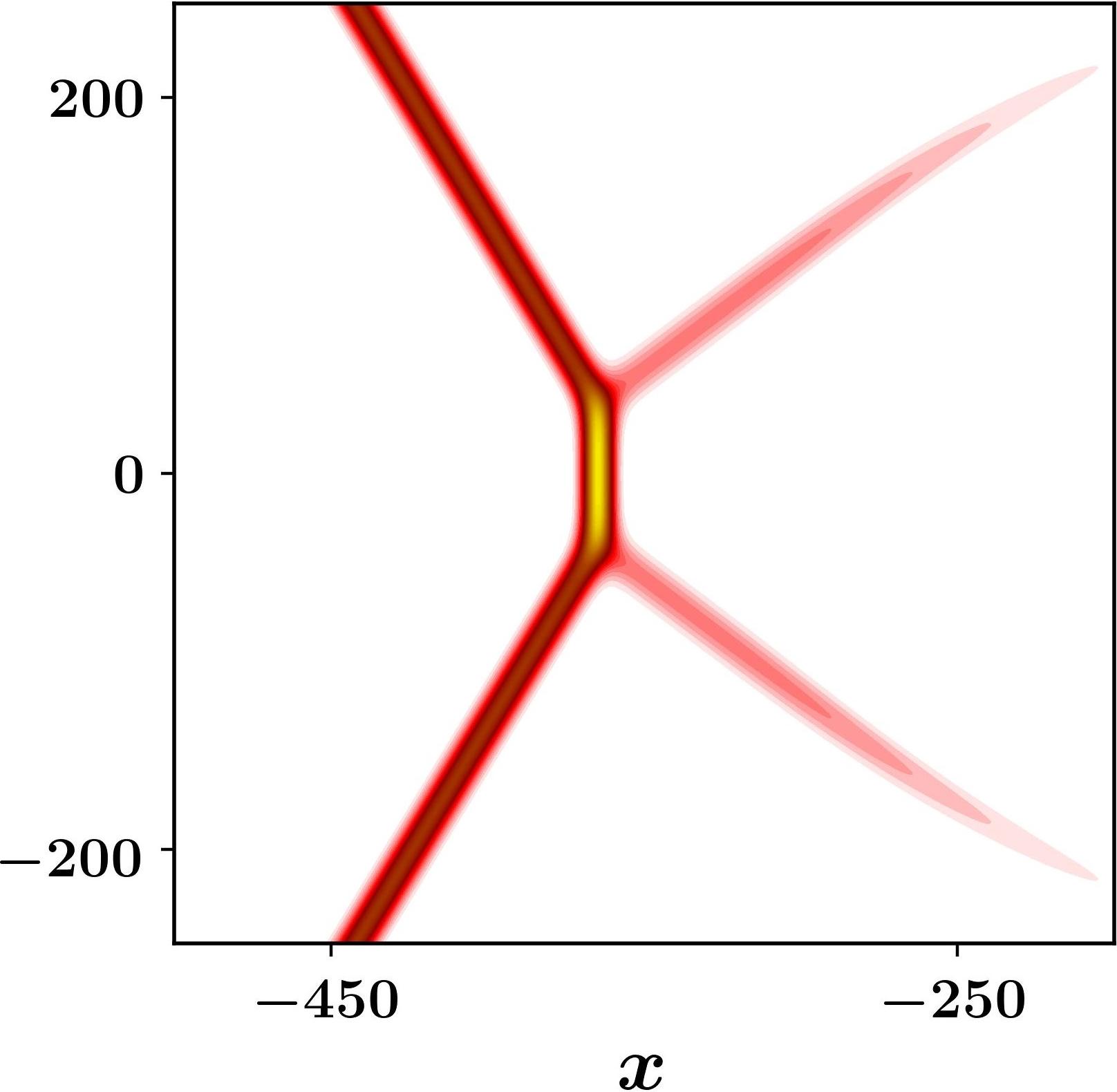}
\subcaption{$t=300$}
\label{fig:RV20-b}
\end{minipage}
\centering
\begin{minipage}{1.7in}
\centering
\includegraphics[height=1.5in]{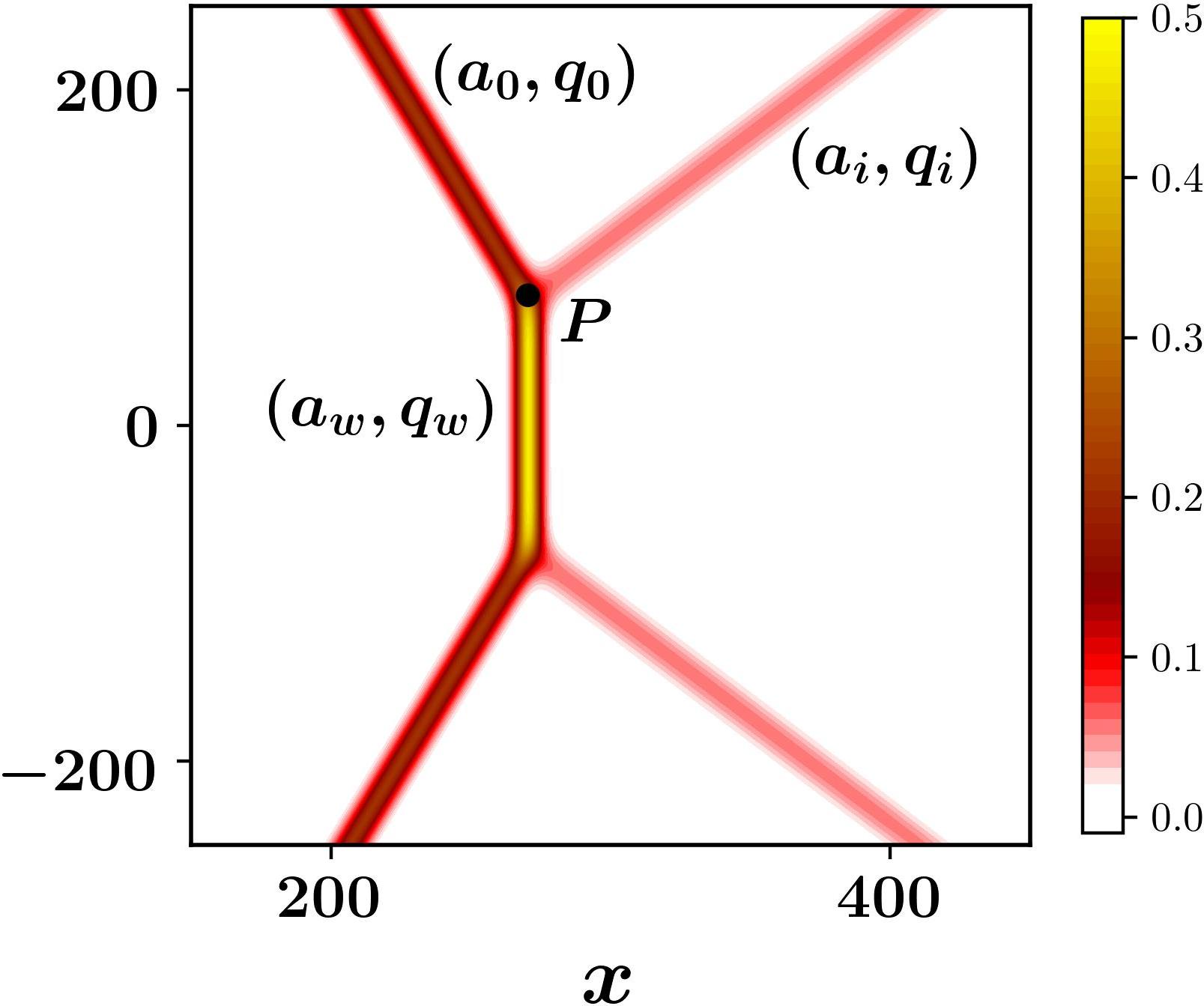}
\subcaption{$t=600$}\label{fig:RV20-c}
\end{minipage}
\caption{Snapshot during the evolution of the bent-stem soliton initial data~\eqref{eq:ini_bs} with $\theta_0=20^\circ$ and $a_0=0.21$: (a) $t=0$, (b) $t=300$, and (c) $t=600$.}\label{fig:RV20-3D}
\end{figure}

When $\theta_0>0$ in the initial data for the bent soliton, not only does the Mach stem occur, but two solitons also arise symmetrically with the slope sgn$(y)q<0$. Figure~\ref{fig:RV20-3D} illustrates the evolution of the reverse bent soliton with $a_0=0.21$ and $\theta_0=20^\circ$ in the initial condition~\eqref{eq:ini_bs}. In figure~\ref{fig:RV20-3D}, one can observe that, as the reverse bent soliton $(a_0,q_0)$ propagates, the Mach stem $(a_w,q_w)$ and the soliton $(a_i,q_i)$ are generated at the bend point, and as time evolves, these three solitons maintain their intersection at the same point $P$, which is identified as the Miles resonant solitons \citep{miles1977a}. 

In this case, two oblique solitons (analogous to oblique shock waves in gas dynamics) symmetrically emanate from the bend point. \citet{ryskamp2022} investigated a similar initial condition for the KP equation. Building upon the methodology outlined in \cite{ryskamp2022}, we employ the modified RH conditions to scrutinize the dynamics of the reverse bent soliton, shown as
\begin{equation}
s_y\llbracket f \rrbracket_{m,n}=\llbracket h\rrbracket_{m,n}\,,
\end{equation}
where $s_y$ denotes the velocity of the discontinuity in the $y$-direction, the multivalued functions $f$ and $h$ satisfy the conservation law $f_t+h_y=0$ and, $m$ as well as $n$, denote the count of values of the multivalued functions on either side of the discontinuity point. 

Therefore, the modulation equation~\eqref{eq:modu4_4} can be represented in the conservation form as demonstrated below:
\begin{equation}
f_t+g_x+h_y=0\,,
\end{equation}
with
\begin{equation}
f=a\sqrt{a}\frac{5+4a}{1+a}\,,\quad g=-a\sqrt{a}\frac{5+3a}{\sqrt{1+a}}\frac{1}{\sqrt{1+q^2}}\,,\quad h=-a\sqrt{a}\frac{5+3a}{\sqrt{1+a}}\frac{q}{\sqrt{1+q^2}}\,.
\end{equation}
Thus, the velocity $(s_x,s_y)$ at the discontinuity point $P$ satisfies the modified RH conditions, shown as
\begin{equation}
s_x=\frac{g_0+g_i-g_w}{f_0+f_i-f_w}\,,\quad s_y=\frac{h_0+h_i-h_w}{f_0+f_i-f_w}\,,\label{eq:v1}
\end{equation}
where the subscripts $w$, $i$, and 0 represent the parameters of the Mach, reflected, and incident solitons, respectively, as shown in figure~\ref{fig:RV20-c}.

It is noted that the KP equation is an integrable system with an elegant mathematical structure and infinite conservation laws. However, the BL equation is non-integrable; thus, some conclusions in \cite{ryskamp2022} drawn from the KP equation may not directly apply to the BL equation. The following modifications are applied to address the difficulty of non-integrability. Considering the consistency equation~\eqref{eq:modu4_3}, by alternately neglecting the $x$- and $y$-directions, we can decompose it into two separate equations in the conservation form
\begin{equation}
F^{(1)}_t+G^{(1)}_x=0\,,\quad F^{(2)}_t+H^{(2)}_y=0\,,
\end{equation}
with
\begin{equation}
\begin{aligned}
F^{(1)}&=\frac{1}{q}\,,\quad G^{(1)}=-\frac{\sqrt{1+q^2}\sqrt{1+a}}{q}\,,\\
F^{(2)}&=q\,,\quad H^{(2)}=-\sqrt{1+q^2}\sqrt{1+a}\,.
\end{aligned}
\end{equation}
Moreover, the velocity $(s_x,s_y)$ at the point $P$ satisfies the classical RH conditions
\begin{equation}
\begin{aligned}
s_x=\frac{G^{(1)}_0-G^{(1)}_w}{F^{(1)}_0-F^{(1)}_w}=\frac{G^{(1)}_i-G^{(1)}_w}{F^{(1)}_i-F^{(1)}_w}=\frac{G^{(1)}_0-G^{(1)}_i}{F^{(1)}_0-F^{(1)}_i}\,,\\
s_y=\frac{H^{(2)}_0-H^{(2)}_w}{F^{(2)}_0-F^{(2)}_w}=\frac{H^{(2)}_i-H^{(2)}_w}{F^{(2)}_i-F^{(2)}_w}=\frac{H^{(2)}_0-H^{(2)}_i}{F^{(2)}_0-F^{(2)}_i}\,,
\end{aligned}\label{eq:v2}
\end{equation}
where the fourth term of the continued equality in each expression can be derived from the second and third terms. 
\begin{figure}
\centering
\begin{minipage}{2in}
\centering
\includegraphics[height=1.7in]{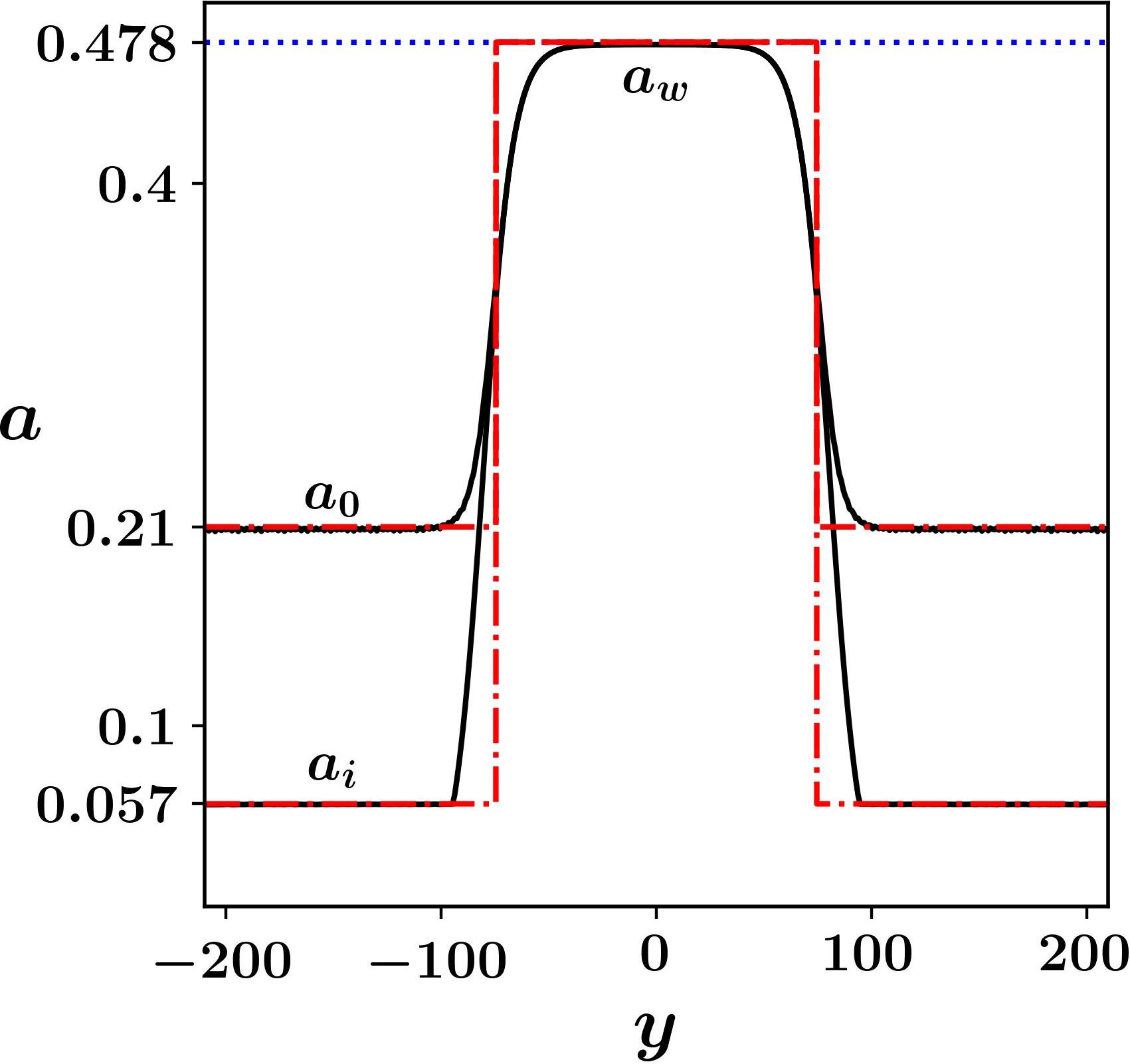}
\subcaption{}\label{fig:RV20-A}
\end{minipage}
\hspace{0.5cm}
\begin{minipage}{2in}
\centering
\includegraphics[height=1.7in]{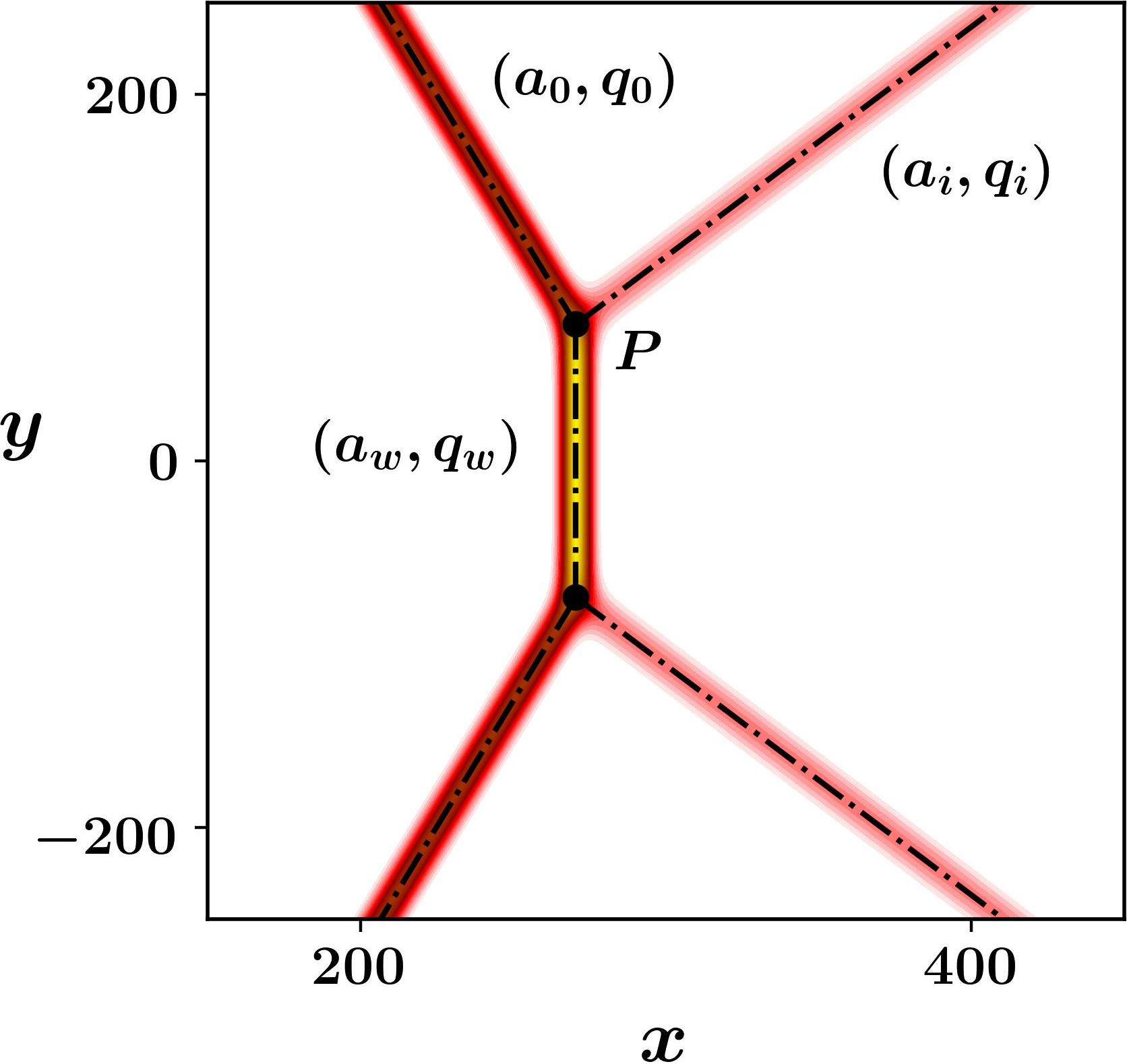}
\subcaption{}
\label{fig:RV20-Q}
\end{minipage}
\caption{(a) evolutions of the amplitude $a$ for the bent soliton initial data~\eqref{eq:ini_bs} with $\theta_0=20^\circ$ and $a_0=0.21$ at $t=600$. They are obtained from the direct numerical simulations of the BL equation~\eqref{eq:BL-topo-flat} (dark solid curves) and the theoretical prediction based on the algebraic equations~\eqref{eq:v1} and~\eqref{eq:v2} (red dashed curves) with $a_w\approx 0.478149$ and $a_i\approx 0.0568022$. (b) the corresponding contour plot of (a): the black dash-dotted lines represent the theoretical prediction obtained through the algebraic equations~\eqref{eq:v1} and~\eqref{eq:v2} with $q_i\approx 0.797368$.}\label{fig:RV20-AQ}
\end{figure}

By solving the system of equations~\eqref{eq:v1} and~\eqref{eq:v2} and considering the initial data $(a_0,q_0)$, we can derive the numerical solution $(a_w,q_w)$ and $(a_i,q_i)$. Additionally, the velocity $(s_x,s_y)$ can also be accurately determined. The comparative illustration between the results of theoretical predictions and numerical simulations is depicted in figure~\ref{fig:RV20-AQ}. In figure~\ref{fig:RV20-A}, the dark solid lines correspond to numerical simulations of the BL equation~\eqref{eq:BL-topo-flat}, while the red dashed lines represent theoretical predictions derived from the algebraic expressions~\eqref{eq:v1} and~\eqref{eq:v2}. The numerically determined values for the amplitude $a$ and the slope $q$ exhibit a remarkable concordance with the theoretical expectations. The contour plot, as depicted in figure~\ref{fig:RV20-Q}, further substantiates the precision of the theoretical forecasts; the black dash-dotted lines therein correspond to the theoretical predictions culled from the aforementioned algebraic relations. The concurrence between the outcomes of the numerical simulations and the theoretical prognoses is striking, with the modulation theory adeptly delineating the evolution of reverse bent solitons, known as Mach reflection.

\subsection{Regular reflection}
When the velocity $s_y\leq0$, the Mach stem will not appear. In this case, by substituting $s_y=0$ into equations~\eqref{eq:v1} and~\eqref{eq:v2}, the critical slope $q_{\text{cr}}$ is attained, which only depends on the initial amplitude $a_0$. The corresponding parameters for the reflected soliton are as follows: 
\begin{equation}
a_i=a_0\,,\quad q_i=-q_0\,.\label{eq:cr1}
\end{equation}
Moreover, based on the expressions of $s_x$ in equations~\eqref{eq:v1} and~\eqref{eq:v2}, one obtains the relation between $a_w$ and $a_0$ as
\begin{equation}
2\left(4a_0+5\right)a_0^{3/2}a_w-\left(a_0+1\right)a_w^{5/2}-2\left(3 a_0+4\right)a_0^{5/2}=0\,,\quad a_w\geq a_0\,.\label{eq:cr2}
\end{equation}
Considering equation~\eqref{eq:cr2}, the numerical solution for $a_w$ in terms of $a_0$ can be derived, leading to the determination of the critical slope
\begin{equation}
q_{\text{cr}}=\frac{\sqrt{a_w-a_0}}{\sqrt{1+a_0}}\sim\sqrt{6\varepsilon}\left(1-\frac{19}{20}\varepsilon+\frac{911}{800}\varepsilon^2\right)+O(\varepsilon^3)\,.
\end{equation}

\begin{figure}
\centering
\begin{minipage}{1.7in}
\centering
\includegraphics[height=1.5in]{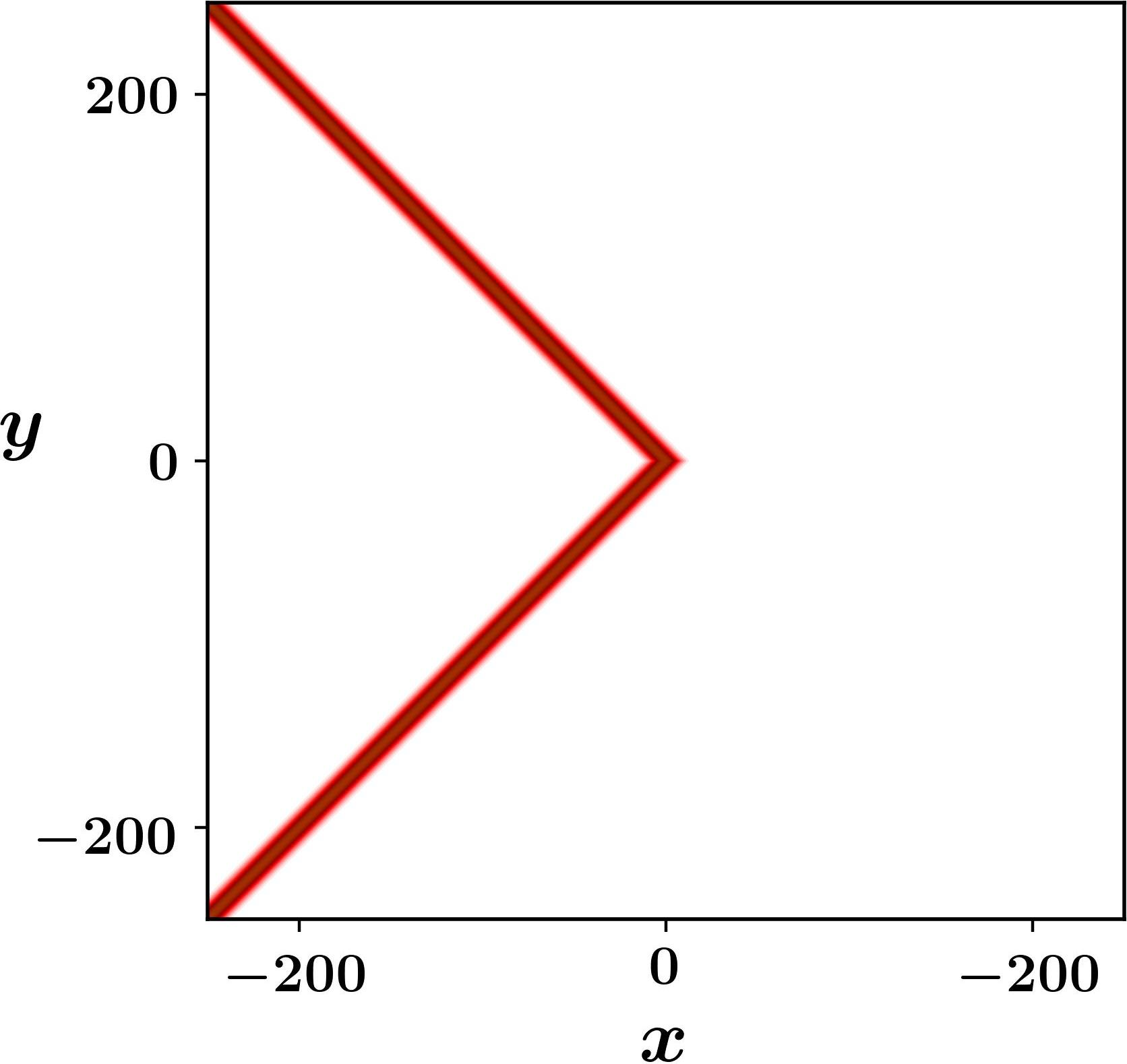}
\subcaption{$t=0$}\label{fig:RV45-a}
\end{minipage}
\centering
\begin{minipage}{1.7in}
\centering
\includegraphics[height=1.5in]{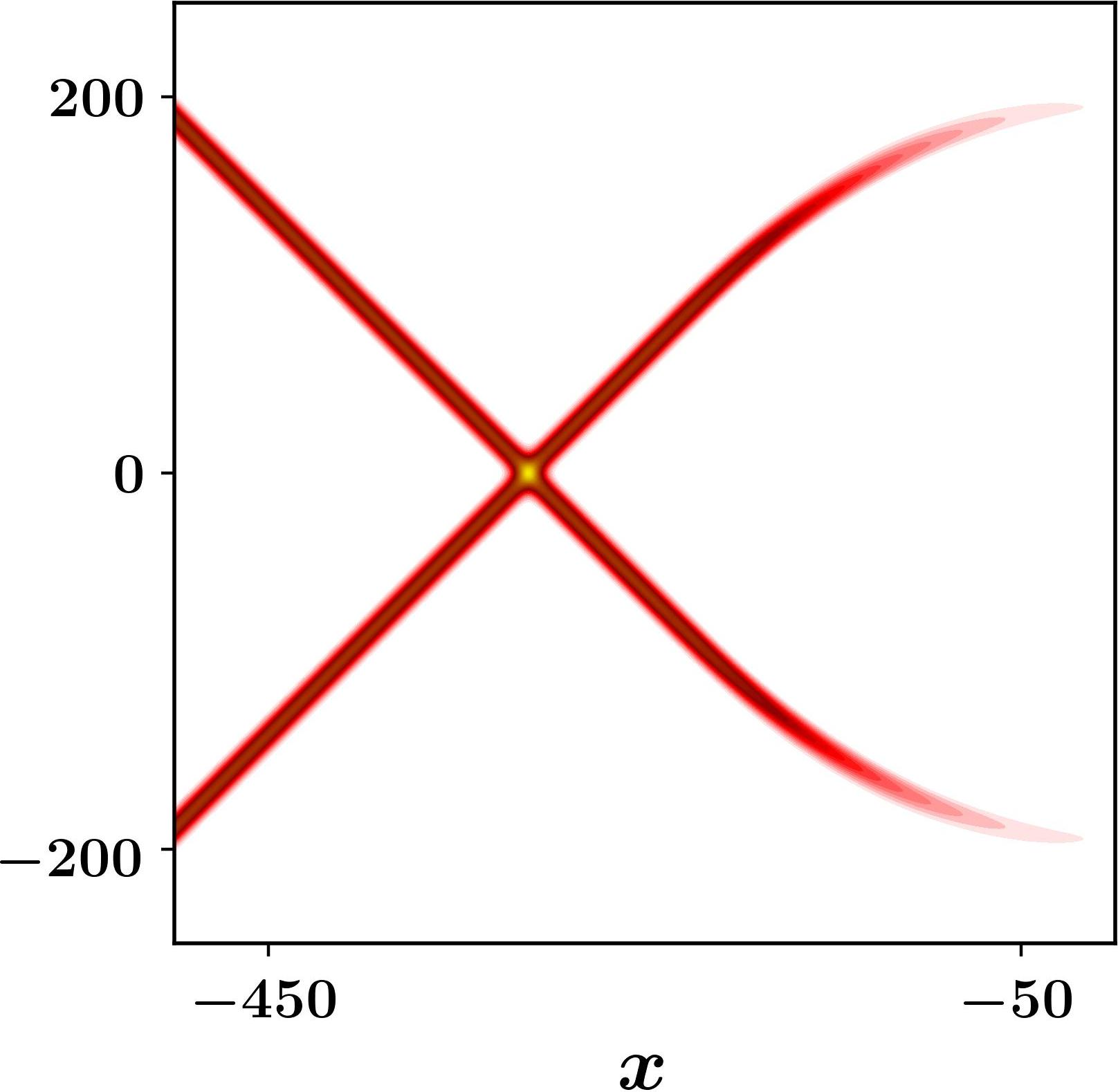}
\subcaption{$t=200$}
\label{fig:RV45-b}
\end{minipage}
\centering
\begin{minipage}{1.7in}
\centering
\includegraphics[height=1.5in]{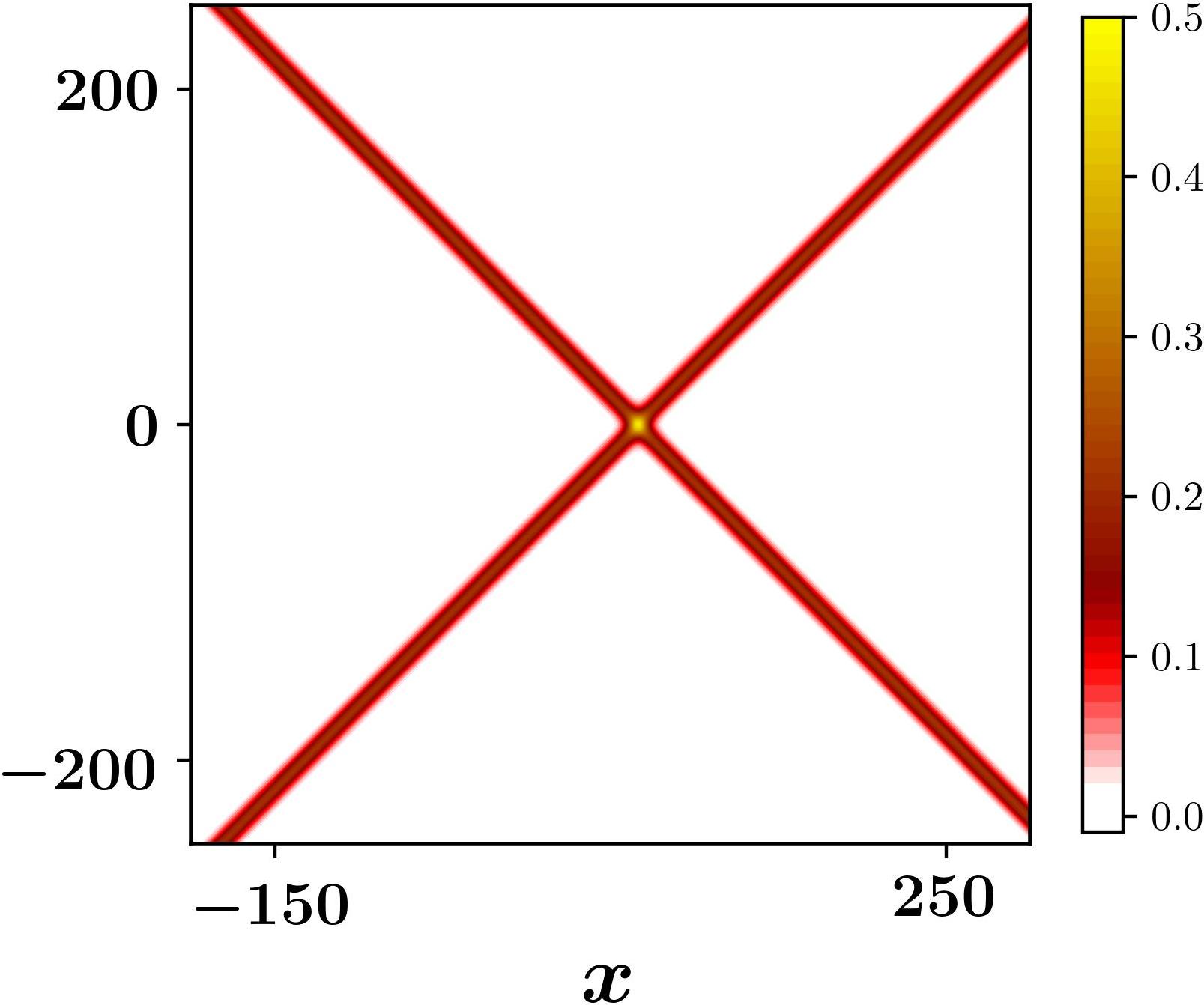}
\subcaption{$t=450$}\label{fig:RV45-c}
\end{minipage}
\caption{Time evolution of the bent-stem soliton initial data~\eqref{eq:ini_bs} with $\theta_0=45^\circ$ and $a_0=0.21$: (a) $t=0$, (b) $t=200$, and (c) $t=450$.}\label{fig:RV45-3D}
\end{figure}

Thus, when $\theta_0>\arctan q_{\text{cr}}$, the soliton dynamics demonstrates the X-shaped pattern rather than the Mach stem, as shown in figure~\ref{fig:RV45-3D}, termed the regular reflection. The generation mechanism of the X-shaped pattern is the interaction of two line solitons with an ignorable phase shift. The numerical results for the regular reflection of the reverse bent soliton with $\theta_0=45^\circ$ and $a_0=0.21$ in the initial data~\eqref{eq:ini_bs} are depicted in figure~\ref{fig:RV45-3D}. The X-shaped soliton pattern is observed to evolve symmetrically, with the two solitons intersecting at the same point as the reverse bent soliton propagates. 

\subsection{Comparison between the BL and KP equations}
Through the above results, based on the modulation equations~\eqref{eq:modu-KP} for the KP equation, we can conclude similar results for the Mach stem in the reverse bent soliton case. Specifically, the velocity $(s_x,s_y)$ at the discontinuity point $P$ is shown as
\begin{equation}
\begin{aligned}
s_x=\frac{g_0+g_i-g_w}{f_0+f_i-f_w}=\frac{G^{(1)}_0-G^{(1)}_w}{F^{(1)}_0-F^{(1)}_w}=\frac{G^{(1)}_i-G^{(1)}_w}{F^{(1)}_i-F^{(1)}_w}=\frac{G^{(1)}_0-G^{(1)}_i}{F^{(1)}_0-F^{(1)}_i}\,,\\
s_y=\frac{h_0+h_i-h_w}{f_0+f_i-f_w}=\frac{H^{(2)}_0-H^{(2)}_w}{F^{(2)}_0-F^{(2)}_w}=\frac{H^{(2)}_i-H^{(2)}_w}{F^{(2)}_i-F^{(2)}_w}=\frac{H^{(2)}_0-H^{(2)}_i}{F^{(2)}_0-F^{(2)}_i}\,,
\end{aligned}\label{eq:rv_KP}
\end{equation}
with
\begin{equation}
\begin{gathered}
f=a^{3/2}\,,\quad g=-a^{3/2}+\frac{1}{2}a^{3/2}q^2-\frac{3}{10}a^{5/2}\,,\quad h=-a^{3/2}q\,,\\
F^{(1)}=\frac{1}{q}\,,\quad G^{(1)}=-\frac{a}{2q}-\frac{1}{q}-\frac{q}{2}\,,\\
F^{(2)}=q\,,\quad H^{(2)}=-1-\frac{a}{2}-\frac{q^2}{2}\,.
\end{gathered}
\end{equation}
The integrability of the KP equation implies the existence of exact Y-shaped soliton solutions, and the amplitude and the slope of each leg are uniquely determined by the phase parameters $\kappa_1$, $\kappa_2$, and $\kappa_3$ (see \citealt{biondini2007}). Without loss of generality, we assume $\kappa_1<\kappa_2<\kappa_3$. Thus, we have
\begin{equation}
\begin{gathered}
A_0^{(\text{KP})}=\frac{(\kappa_2-\kappa_1)^2}{6\times 2^{1/3}}\,,\quad A_w^{(\text{KP})}=\frac{(\kappa_3-\kappa_1)^2}{6\times 2^{1/3}}\,,\quad A_i^{(\text{KP})}=\frac{(\kappa_3-\kappa_2)^2}{6\times 2^{1/3}}\,,\\
q_0=-\frac{\kappa_2+\kappa_1}{2^{2/3}}\,,\quad q_w=-\frac{\kappa_3+\kappa_1}{2^{2/3}}\,,\quad q_i=-\frac{\kappa_3+\kappa_2}{2^{2/3}}\,.
\end{gathered}
\label{eq:aq_for_KP_1}
\end{equation}
We let $\kappa_3=-\kappa_1$ to ensure $q_w=0$ for the symmetry of the reverse bent soliton case. Replacing $\kappa_1$ and $\kappa_2$ with $A_0^{(\text{KP})}$ and $q_0$ makes it straightforward to infer that
\begin{equation}
\begin{gathered}
A_w^{(\text{KP})}=\left( \frac{q_0}{\sqrt{3}}+\sqrt{A_0^{(\text{KP})}} \right)^2\,,\quad A_i^{(\text{KP})}=\frac{q_0^2}{3}\,,\quad q_i=-\sqrt{3}\sqrt{A_0^{(\text{KP})}}\,,\\
s_x=-1-\frac{A_0^{(\text{KP})}}{2}=c_w^{(\text{KP})}\,,\quad s_y=\frac{1}{3}\left( \sqrt{3A_0^{(\text{KP})}}-q_0 \right)\,.
\end{gathered}\label{eq:aq_for_KP_2}
\end{equation}
Thus, $s_y=0$ leads to the critical slope $q_{\text{cr}}^{(\text{KP})}=\sqrt{3A_0^{(\text{KP})}}$. The scaling~\eqref{eq:KPts} converts~\eqref{eq:aq_for_KP_2} to the data for equation~\eqref{eq:KP_topo_flat}, which agree well with the findings presented in \citet{ryskamp2022}. 

Finally, we can obtain the reduction factor and the critical angle in terms of the initial angle $\theta_0$ and speed $\omega_0$, \textit{i.e.},
\begin{equation}
\begin{gathered}
\lambda^{(\text{KP})}=\frac{A^{(\text{KP})}_w}{A^{(\text{KP})}_0}=\left(\frac{\sqrt{2}\sin\theta_0}{\sqrt{3}\sqrt{-3-\cos 2\theta_0-4\omega_0\cos\theta_0}}+1\right)^2\,,\\
\theta_{\text{cr}}^{(\text{KP})}=\arctan q^{(\text{KP})}_{\text{cr}}=\text{arcsec}\left( \frac{\sqrt{9\omega_0^2-8}-3\omega_0}{4}\right)\,.
\end{gathered}
\end{equation}
\begin{figure}
\centering
\begin{minipage}{2in}
\centering
\includegraphics[height=1.8in]{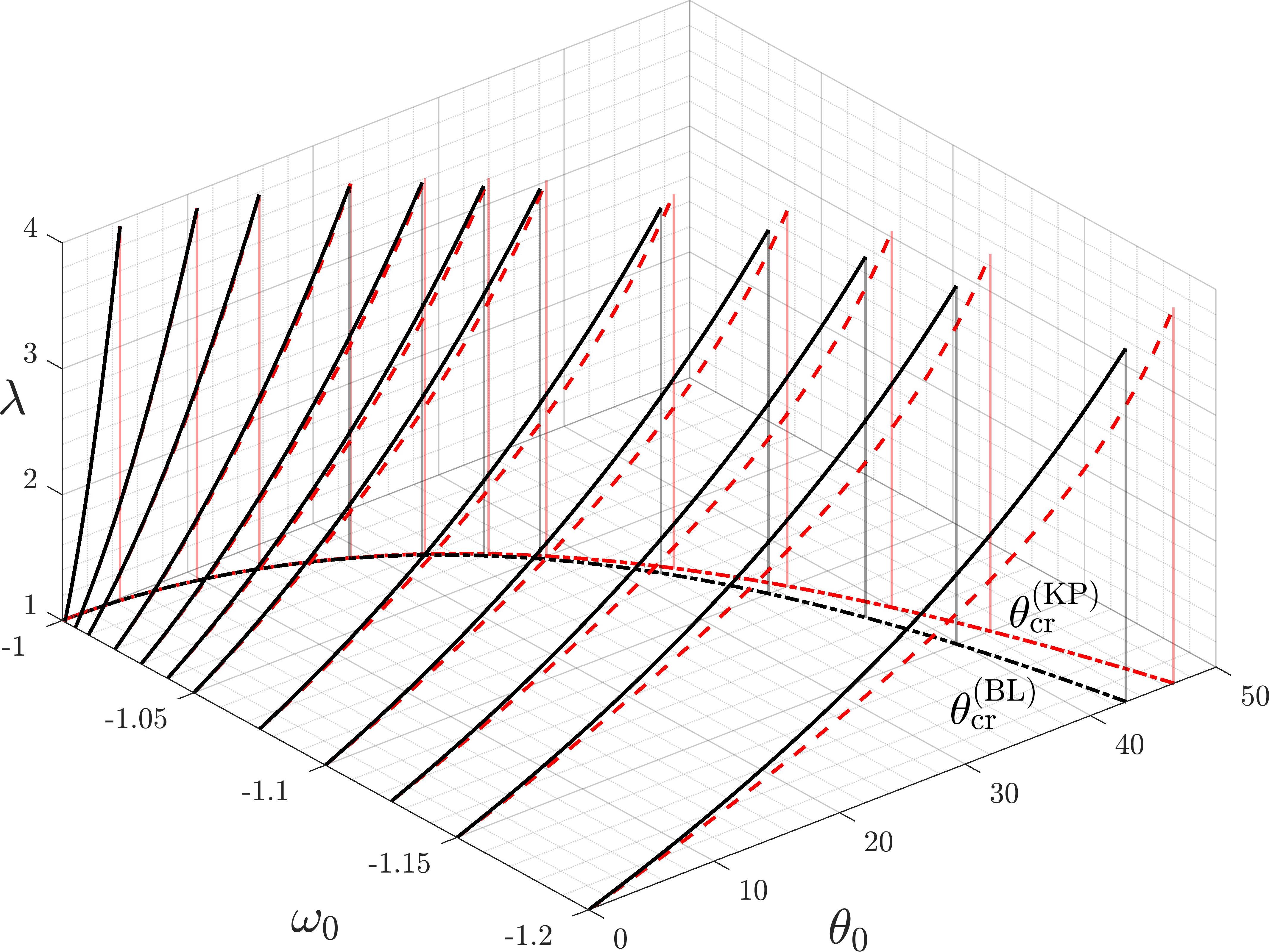}
\subcaption{}\label{fig:RV_same_c-A}
\end{minipage}
\hspace{1cm}
\begin{minipage}{2in}
\centering
\includegraphics[height=1.8in]{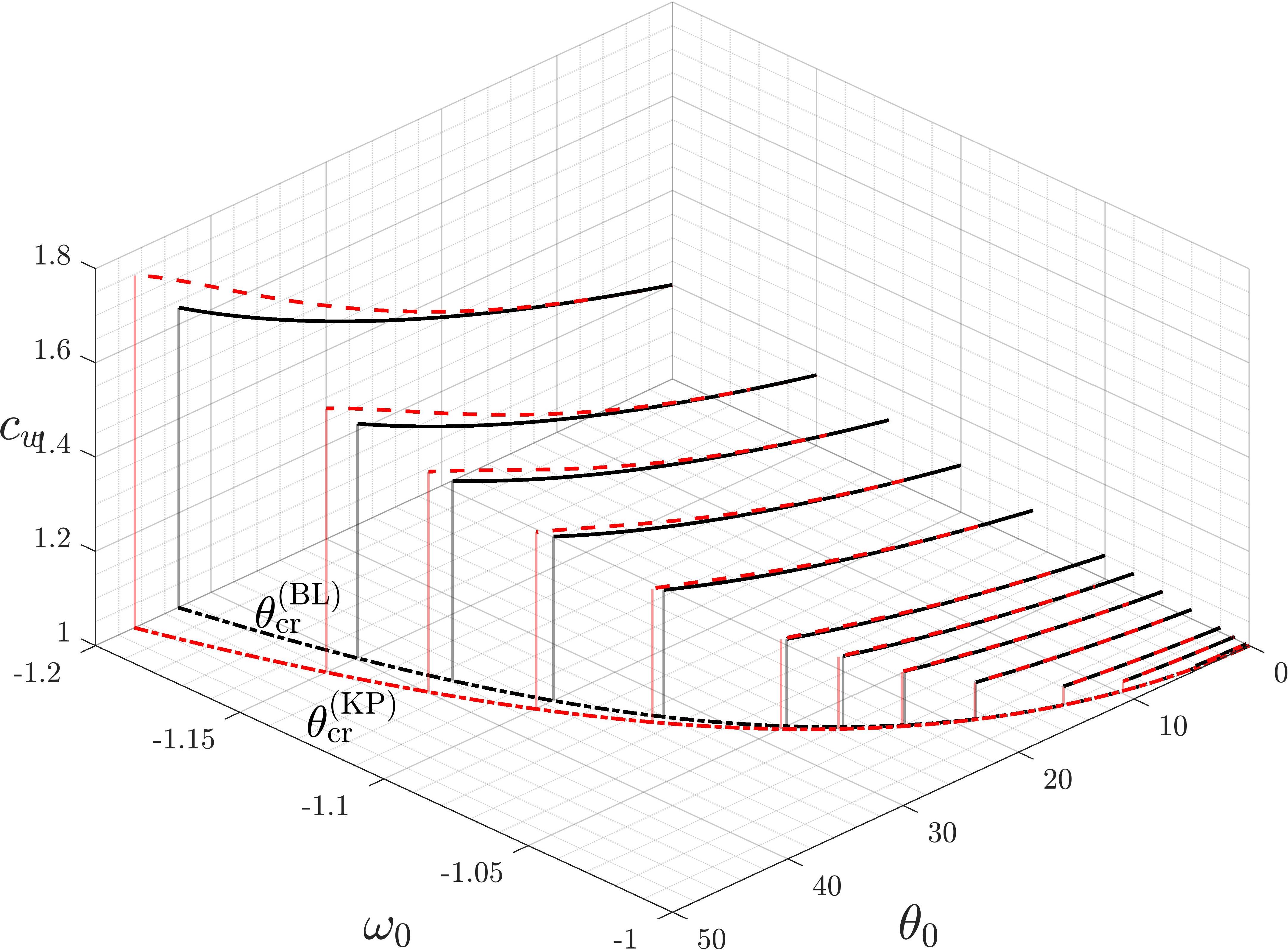}
\subcaption{}
\label{fig:RV_same_c-c}
\end{minipage}
\caption{The same as figure~\ref{fig:same_c} except that $\theta_0>0$ for the Mach reflection of the line solitons.}\label{fig:RV_same_c}
\end{figure}

Figure~\ref{fig:RV_same_c} illustrates the comparison of the wall reduction factor and the Mach stem speed for the KP and BL equations as functions of the initial angle $\theta_0$ and speed $\omega_0$. This figure deviates from the previous one, denoted as figure~\ref{fig:same_c}, in that it specifically considers scenarios where the initial angle $\theta_0>0$, thus focusing on the Mach reflection of line solitons. In figure~\ref{fig:RV_same_c-A}, the wall reduce factor $\lambda$ of the BL equation is consistently greater than that of the KP equation across various initial angles and velocities. The disparity between these coefficients diminishes as the initial angle decreases or the speed approaches $1$. Analogously, the variation of the Mach stem speed $c_w$ in figure~\ref{fig:RV_same_c-c} exhibits a similar trend. However, diverging from the observations in figure~\ref{fig:same_c-A}, it is discernible from figure~\ref{fig:RV_same_c-A} that the critical angle $\theta_{\text{cr}}^{(\text{BL})}$ is less than $\theta_{\text{cr}}^{(\text{KP})}$, and this discrepancy intensifies with an increase in speed $|\omega_0|$.

\section{Upstream-advancing Waves and Wakes}\label{sec:wake}
In the ocean, one intriguing phenomenon is surface/internal wave generation due to a uniformly running stream disturbed by a static surface pressure or bottom topography, or equivalently, owing to the relativity of motion, a constantly moving disturbance at the top/bottom of the still fluid body, resulting in a series of upstream-advancing waves. Most studies for this problem are based on horizontally one-dimensional \citep{lee1989} or (two-dimensional) anisotropic \citep{katsis1987,lee1990} models. In particular, \citet{katsis1987} explored the dynamics of two-dimensional disturbances within a horizontally infinite fluid medium, using a delta function as a forcing term for the KP equation. Subsequently, \citet{lee1990} advanced the study by examining the nonlinear wave phenomena generated when such disturbances move at speeds close to resonance in a horizontal, unbounded domain, employing the fKP equation~\eqref{eq:KP_topo} as the theoretical starting point. However, it is not appreciated for more general cases, for instance, when the transverse variation is too strong to ignore the influence of isotropism.

In the above two cases (sections \ref{sec:mach_exp} and \ref{sec:mach_ref}), we concentrate on the most active region of wave-wave interactions and compare the results between the KP and BL equations. The windowing scheme is applied in the numerical simulations to make the wave propagation in the transverse direction inapparent. However, the $(2+1)$-dimensional wave motion should be considered in an unbounded fluid, especially for the realistic ocean situation. In this section, we consider waves generated by a constantly moving topography for the standard fBL equation~\eqref{eq:BL-topo}, and the following Gaussian function is employed to model the topography:
\begin{equation}
b(x,y,t)=B(x-c_bt,y)=b_0e^{-\left( \frac{x-c_bt}{\delta_x} \right)^2-\left(\frac{y}{\delta_y}\right)^2}\,,
\end{equation}
where $c_b$ and $b_0$ denote the speed and height of the topography, respectively. Specifically, the lengths of the computational domain in the $x$- and $y$-directions are chosen as $L_x=1000$ and $L_y=2000$, and $\delta_x=L_x/200$ and $\delta_y=L_y/200$ are used in the subsequent computations. The numerical outcomes of the fBL equation for the wave induced by the moving topography are acquired through the computational methodologies delineated in section~\ref{sec:formulation}. Based on the value between the speed $c_b$ of the moving topography and the characteristic speed of the BL equation, it yields three cases: subcritical, critical, and supercritical. The details will be discussed subsequently. 

\subsection{Subcritical speed}
Our concern in this subsection is to investigate the solution behavior of~\eqref{eq:BL-topo} when a localized forcing topography moves with a subcritical speed ($c_b<1$). Thus, when the wave propagates a sufficient distance from the topography, the impact of the topography becomes negligible and can, therefore, be disregarded. Consequently, we are able to conduct a detailed analysis of the Whitham equations~\eqref{eq:modu_yt}. Here, following the approach of \citet{lee1990}, we aim to derive a similarity solution under the assumption that $k=1$ and $a$ is independent of $y$, leading to
\begin{equation}
\begin{gathered}
q=\pm \frac{y}{\sqrt{f^2-y^2}}\,,\quad a=f^2_t-1\,,\quad\eta=x\pm \sqrt{f^2-y^2}-f+C\,,
\end{gathered}\label{eq:sol_wake}
\end{equation}
where $C$ is an integration constant, and the function $f(t)$ is subject to the constraint
\begin{equation}
f(t)=\frac{2a(1+a)^{3/2}(5+3a)}{(12a^2+25a+15)a_t}\,.
\end{equation}
\begin{figure}
\centering
\begin{minipage}{2in}
\centering
\includegraphics[height=2in]{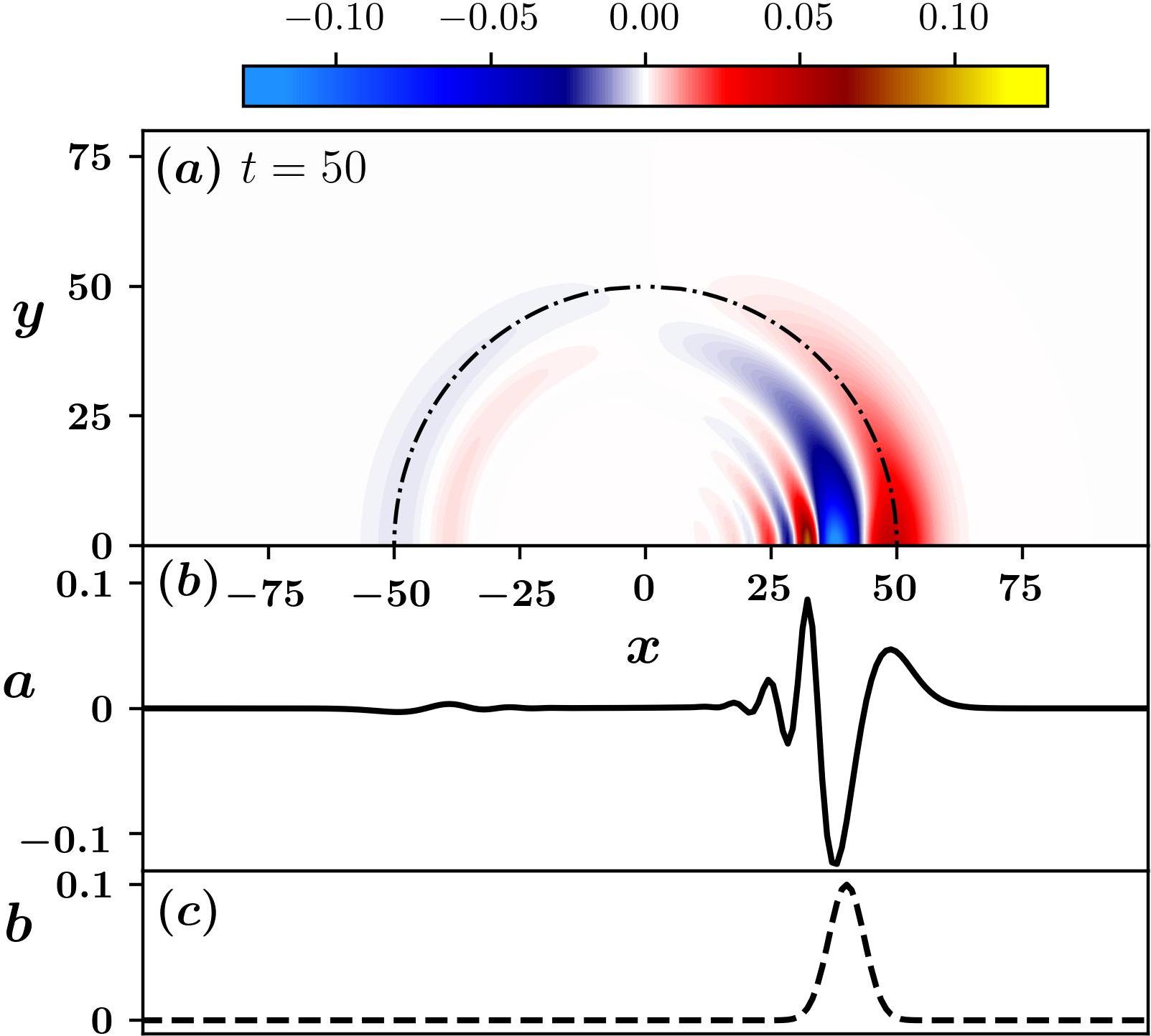}
\end{minipage}
\hspace{1cm}
\begin{minipage}{2in}
\centering
\includegraphics[height=2in]{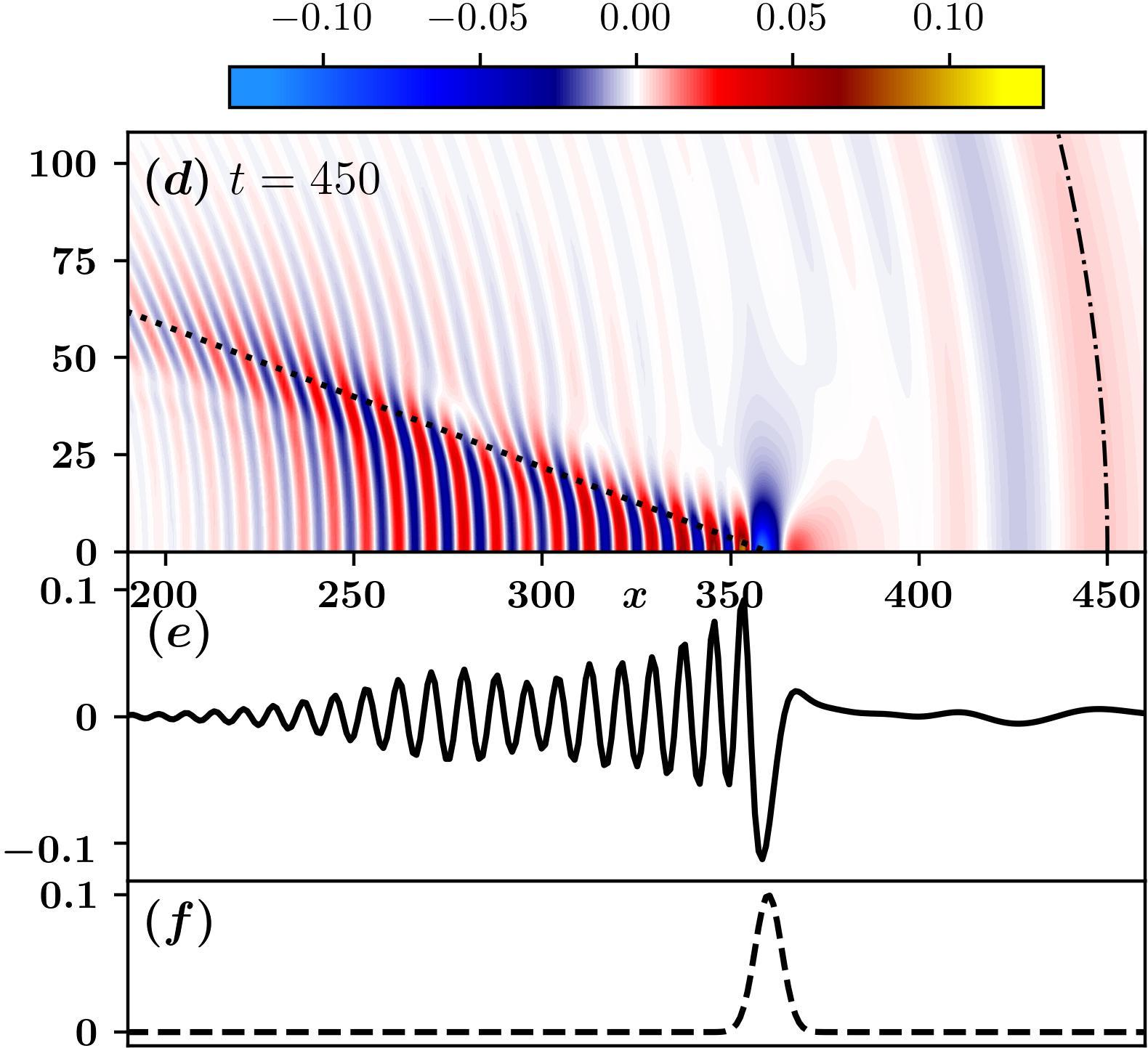}
\end{minipage}
\caption{Evolution of the wave generated by the moving topography with $b_0=0.1$ and $c_b=0.8$ at:  $t=50$ (a--c) and $t=450$ (d--f). (a, d) contour plots of the wave, where the black dash-dotted circular arcs with a radius of $t$ and a center at the origin $(0,0)$ denote the predicted locations of the start-up waves and the black dotted line denotes the linear theory prediction of the wake angle $19.96^\circ$; (b, e) $y$-cross-sections of the wave; (c, f) positions of the topography at the corresponding times.}\label{fig:topo_0.8}
\end{figure}

From the above results, one can observe that the wave's shape is a circle with the center at $(f(t)-C,0)$ and radius $f(t)$, while the wave propagates at speed $\partial_tf(t)$. Furthermore, we assume $a=0$ and $t=0$, which leads to $f(t)=\pm t$, and a start-up wave initiated by abrupt topography movement is a circle with the center at $(0,0)$ and radius $t$. Figure~\ref{fig:topo_0.8} shows the evolution of the wave generated by the moving topography with $b_0=0.1$ and $c_b=0.8$ at $t=50$ and $t=450$. It is depicted in figure~\ref{fig:topo_0.8} that the movement of the topography generates a series of oscillation waves - the so-called transverse waves, the maximum amplitude of which is of the same order as $b_0$ and gradually diminishes to zero in the $y$-direction. Consequently, the variation of $a$ in the $y$-direction is negligible, which guarantees the validity of dropping the $y$-dependence in theoretical analyses. As depicted in figures~\ref{fig:topo_0.8}a and~\ref{fig:topo_0.8}d, the black dash-dotted curves represent the predicted positions of the start-up wave based on the abovementioned discussions. It can be observed that the numerical results are in accordance with the theoretical predictions.

In figure~\ref{fig:topo_0.8}d, it can be seen that the moving topography induces a sequence of trailing waves. By focusing on wave wakes, an approximate analysis can be conducted using the linear theory. Linearizing the BL equation (with a flat bottom) yields 
\begin{equation}
\xi_{tt}-\Delta\xi_{tt}-\Delta\xi=0\,,\label{eq:BL-lin}
\end{equation}
which leads to the linear dispersion relation
\begin{equation}
-\omega^2+\boldsymbol{k}^2-\omega^2\boldsymbol{k}^2=0\,.
\end{equation}
Thus, considering the speed $c_b$ of the moving topography, the dispersion relation in the moving frame is derived as follows:
\begin{equation}
G(k,l)=c_b|\boldsymbol{k}|\cos\phi-\frac{|\boldsymbol{k}|}{\sqrt{1+\boldsymbol{k}^2}}=0\,,
\end{equation}
where $\phi$ is the inclination of the normal (direction of $\boldsymbol{k}$) to the line of the waves. Hence, the wake angle $\alpha$ denoting the slope of a ray through the topography is \citep[see][]{whitham1999}
\begin{equation}
\tan \alpha=\frac{\partial_lG(k,l)}{\partial_kG(k,l)}=\frac{c_b^2 \tan\phi}{\left(\tan^2\phi+1\right)^2-c_b^2}\,.\label{eq:alpha}
\end{equation}

Clearly, $\alpha \to 0$ as $\phi\to0$ or $\phi\to\pi/2$. Hence, the maximum value of $\alpha$ occurs when $\phi \in (0, \pi/2)$, which can be determined as
\begin{equation}
\alpha_m=\arctan\frac{3 \sqrt{3} c_b^2 \sqrt{\sqrt{4-3 c_b^2}-1}}{4 \left(2-3c_b^2+\sqrt{4-3 c_b^2}\right)},\quad \text{at}\quad \phi_m=\arctan\sqrt{\frac{\sqrt{4-3c_b^2}-1}{3}}\,.\label{eq:wave_angle}
\end{equation}

When $c_b=0$, it follows that $\alpha_m=0$. The wake angle gradually expands as $c_b$ increases, reaching $90^\circ$ as $c_b$ approaches unity. The dependence of $\alpha_m$ on $c_b$ is a remarkable wave phenomenon in shallow water, completely different from the deep-water situation, where the famous Kelvin wake angle of $19.47^\circ$ is retained regardless of the speed of the moving object on the free surface. Figure~\ref{fig:topo_0.8}d presents a comparison between the theoretical predictions and numerical results for the wake angle, depicted by the black dashed lines. It is observable that the trailing waves are all confined within the actualized region, which further confirms the validity of the linear theory.

\subsection{Critical speed}
\begin{figure}
\centering
\begin{minipage}{2in}
\centering
\includegraphics[height=2in]{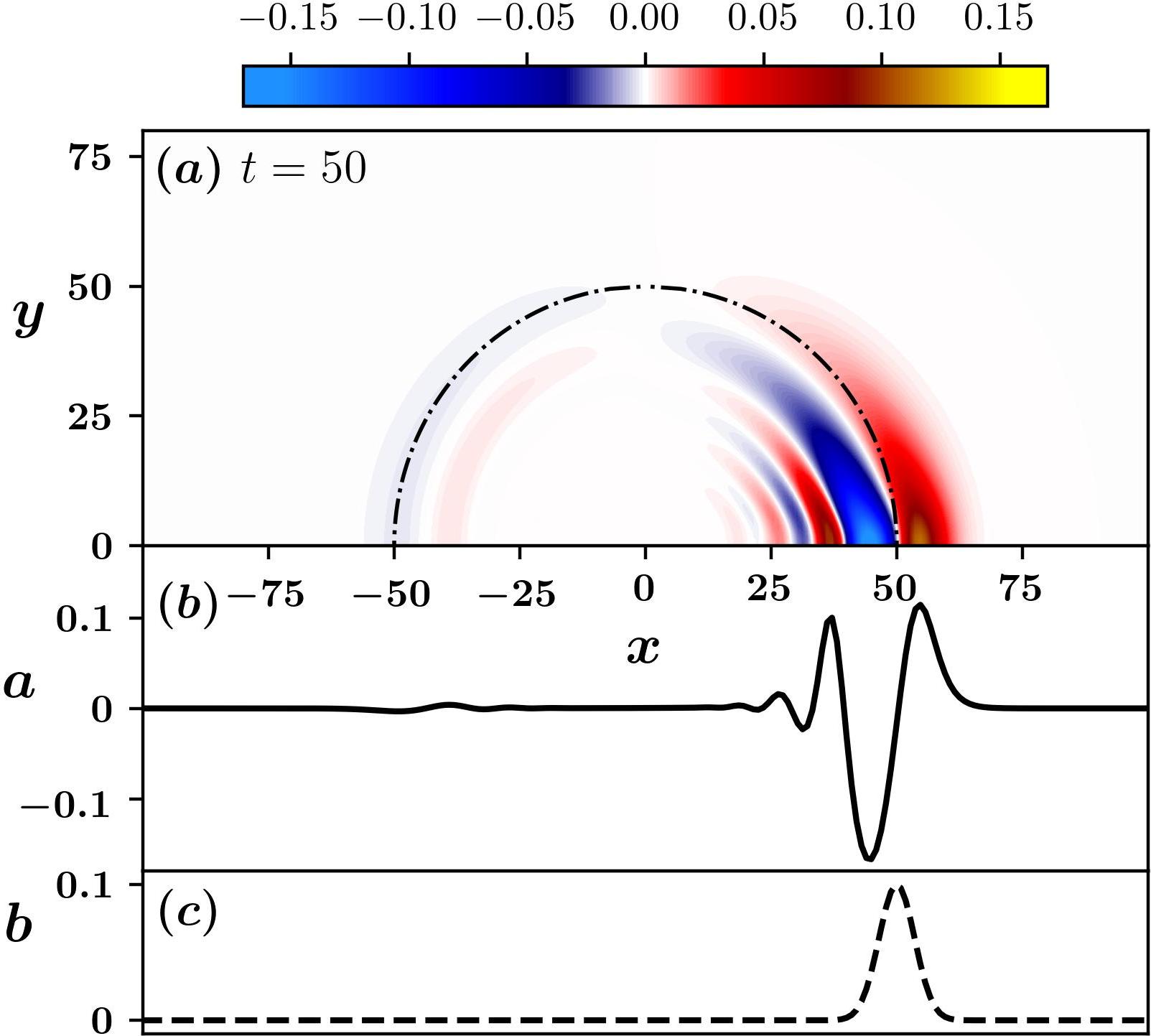}
\end{minipage}
\hspace{1cm}
\begin{minipage}{2in}
\centering
\includegraphics[height=2in]{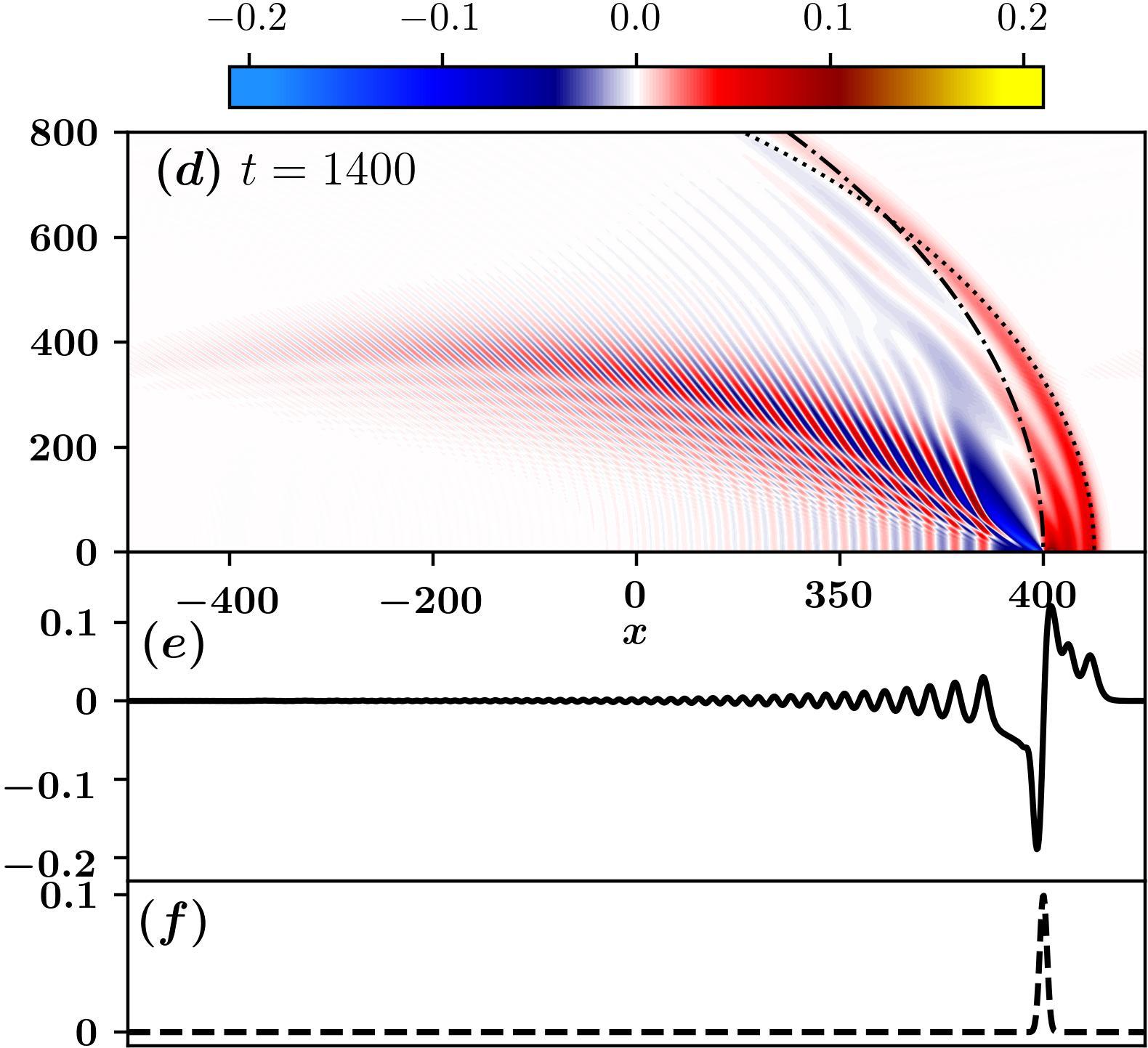}
\end{minipage}
\caption{Evolution of the wave generated by the moving topography with $b_0=0.1$ and $c_b=1$ at: $t=50$ (a--c) and $t=1400$ (d--f). (a, d) contour plots of the wave, where the black dash-dotted circular arcs with a radius of $t$ and a center at the origin $(0,0)$ denote the predicted locations of the start-up waves and the black dotted circular arc coarsely fitted as $(x-350)^2+y^2=1100^2$ denotes the locations of the first precursor wave; (b, e) $y$-cross-sections of the wave; (c, f) positions of the topography at the corresponding times.}\label{fig:topo_1}
\end{figure}

In the case of moving topography with critical speed ($c_b=1$), the wave generated by topography will propagate at the same speed as the topography, which is depicted in figure~\ref{fig:topo_1}. In figures~\ref{fig:topo_1}a and \ref{fig:topo_1}d, it can be observed that the start-up wave is ``propelled'' a certain distance by the topography due to the consistency in speed between the two, leading to the formation of what is known as the upstream-advancing waves, thereby enabling it to surpass the boundary of the theoretical circle (the black dash-dotted curves). When the start-up wave is counter to the direction of topography movement, it remains unaffected by the influence of the topography. As depicted in figures~\ref{fig:topo_1}b and \ref{fig:topo_1}e, it is evident that the ``propulsion'' by the topography occurs in the form of waves propagating forward. As time evolves, the distance over which the upstream-advancing waves are propelled increases, and the number of precursor waves also augments. During the propulsion process, the amplitude of the first precursor wave undergoes continuous variation; hence, the impact of $a_t$ in equation~\eqref{eq:sol_wake} cannot be disregarded. The simplification by assuming $a=0$ for approximation does not adequately predict the behavior. However, by disregarding the local wave characteristics near the topography and instead considering the far-field waveform, it can still be approximated by circular functions (see the black dotted curves in figure~\ref{fig:topo_1}d). Considering the wake induced by the topography, shown in figure~\ref{fig:topo_1}d, the transverse wave has significantly diminished in intensity, while the divergence wave has gained prominence in contrast to figure~\ref{fig:topo_0.8}d. From equation~\eqref{eq:wave_angle}, we can deduce that as $c_b\to1$, the wake angle approaches $\pi/2$, implying that the wake will no longer be confined within a specific angle but will gradually diffuse across the entire plane behind the topography, shown in figure~\ref{fig:topo_1}d. This diffusion, however, is a slow process, necessitating extended computational time for its evolution. 

\begin{figure}
\centering
\begin{minipage}{1.7in}
\centering
\includegraphics[height=1.5in]{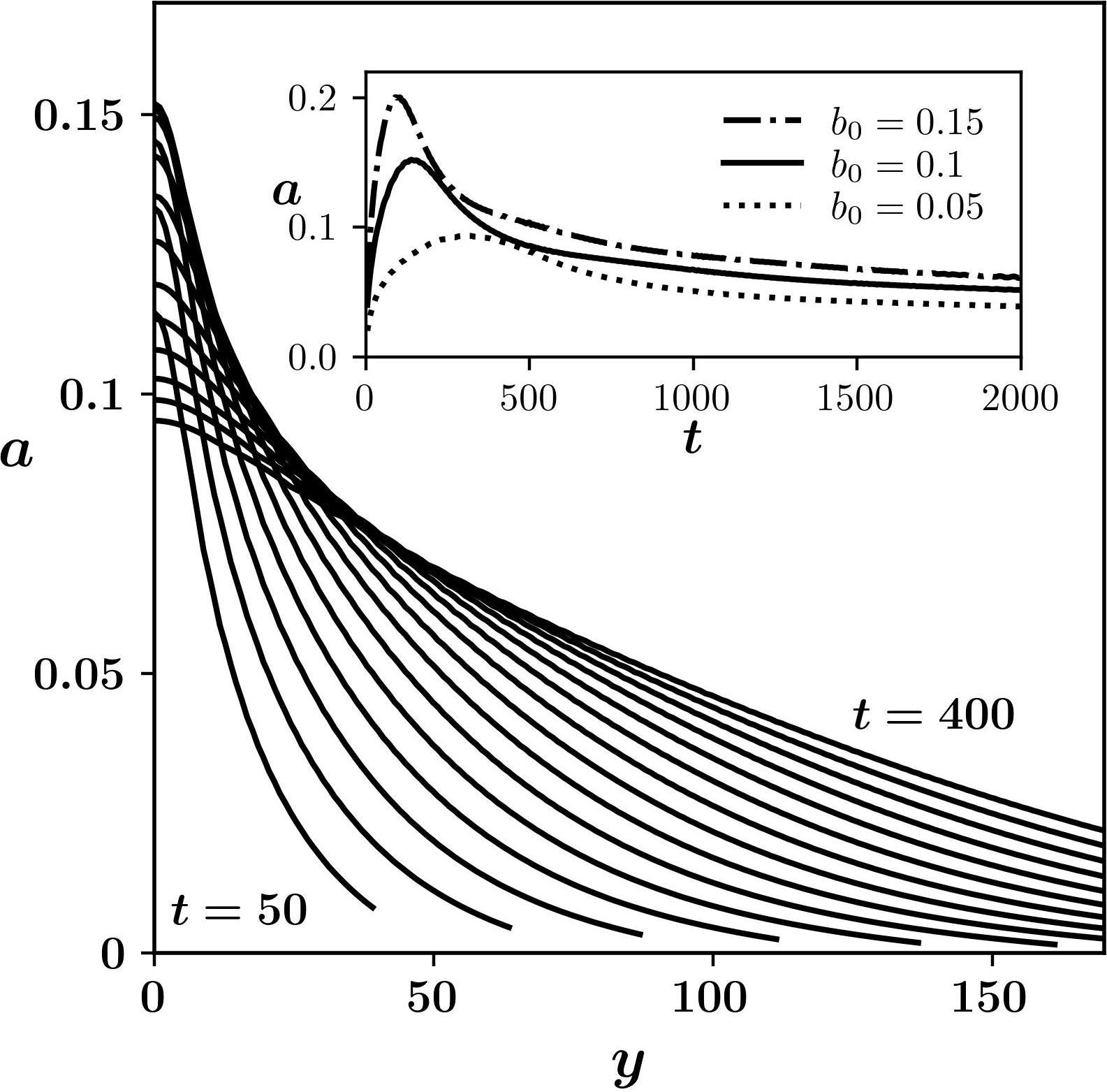}
\subcaption{}
\end{minipage}
\centering
\begin{minipage}{1.7in}
\centering
\includegraphics[height=1.5in]{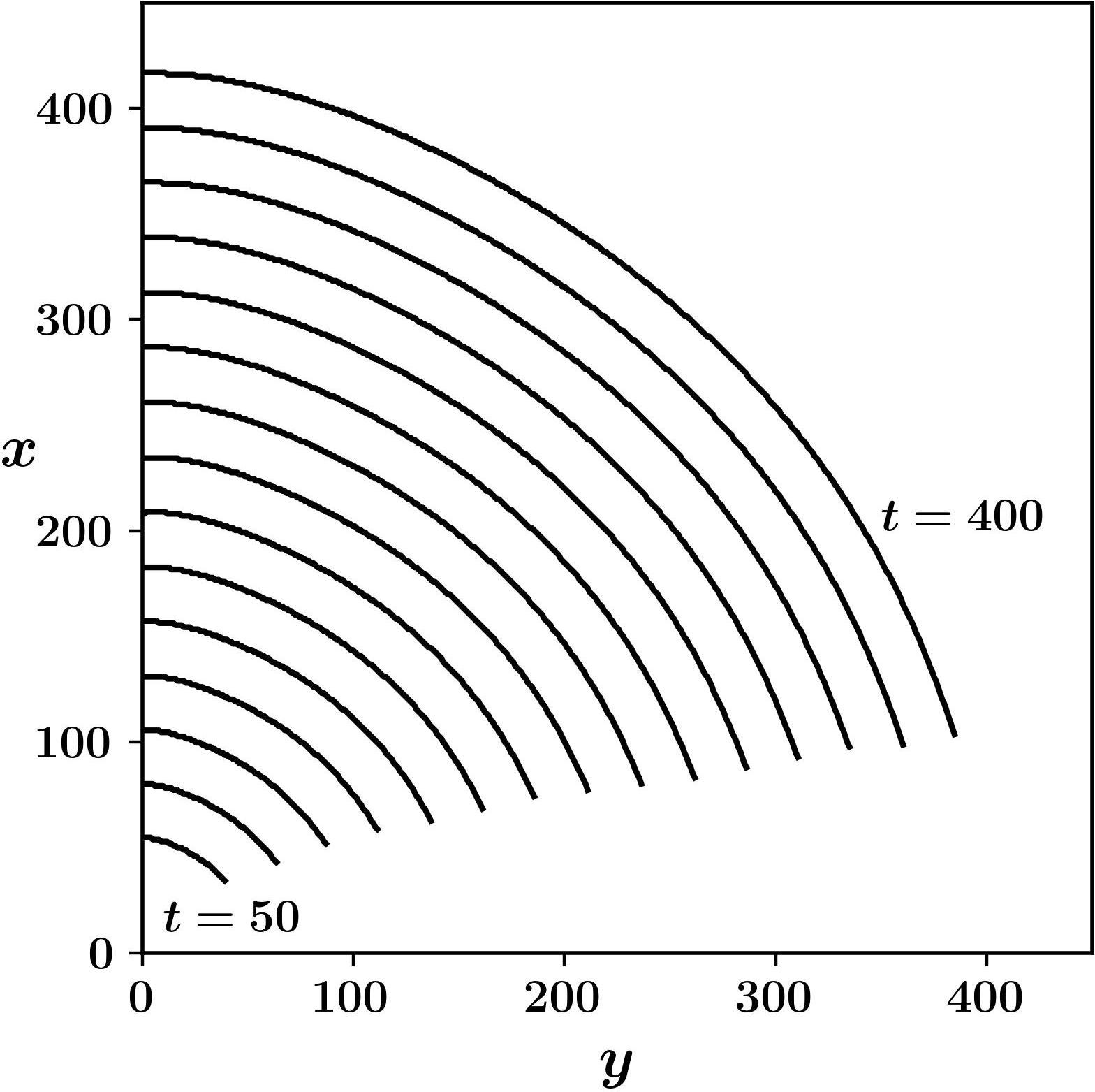}
\subcaption{}
\end{minipage}
\centering
\begin{minipage}{1.7in}
\centering
\includegraphics[height=1.5in]{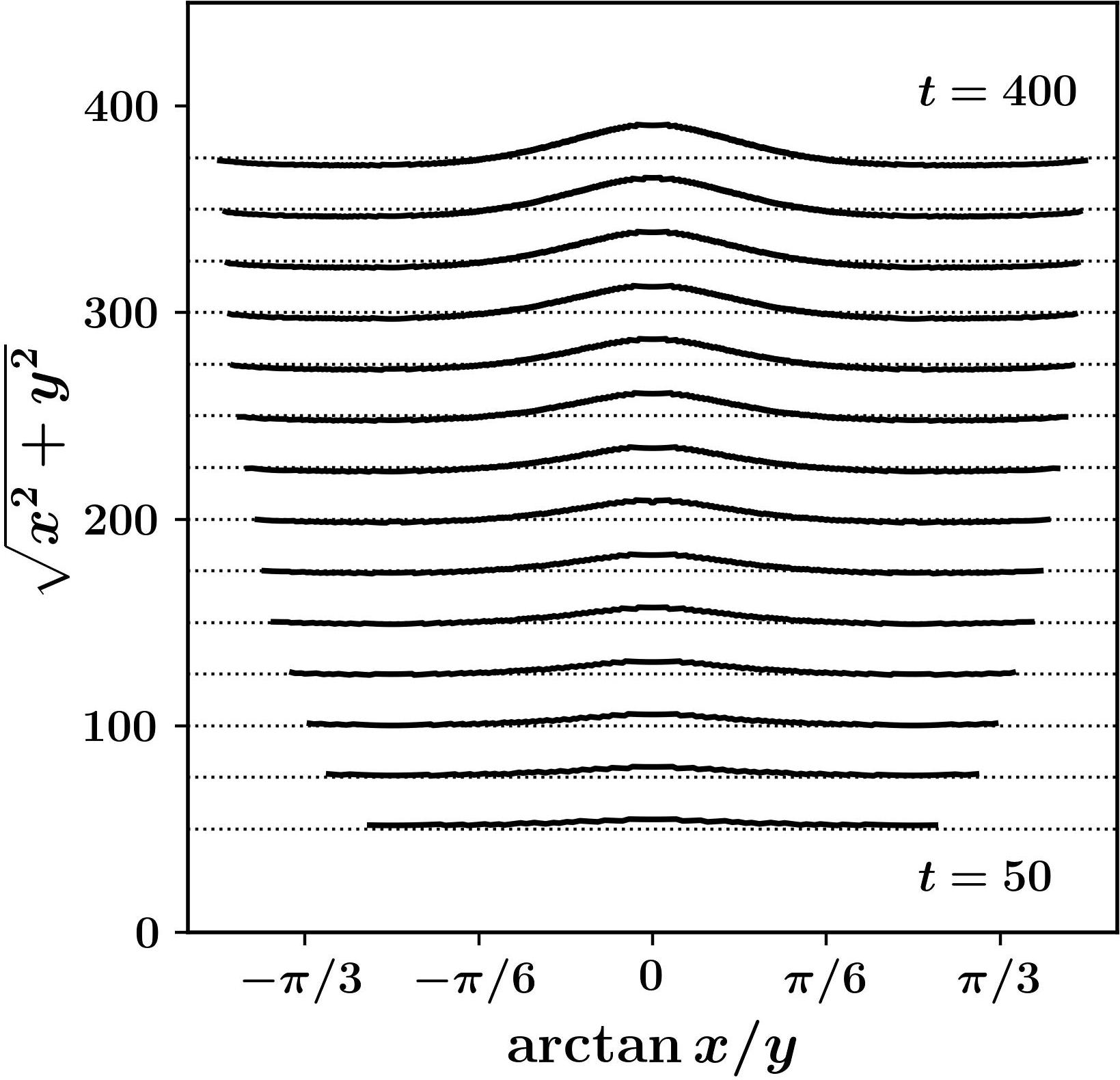}
\subcaption{}
\end{minipage}
\caption{Evolution of the first precursor wave with $c_b=1$, $b_0=0.1$ and the time interval $\Delta t=25$: (a) variation of its amplitude, where the subplot shows the amplitude evolution at the plane $y=0$ over time with the amplitude of the topography $b_0=0.05$, $0.1$, and $0.15$; (b) evolution of its crestlines in the $x-y$ plane; (c) evolution of the distance between the crestlines and the origin point $(0,0)$.}\label{fig:topo_1_ar}
\end{figure}

To delve deeper into the dynamics of the first precursor wave, we examine the morphological changes of its crestlines and the amplitude variations along these lines, considering a specific topography amplitude of $b_0=0.1$. Figure~\ref{fig:topo_1_ar}a illustrates the temporal progression of the first precursor wave's amplitude, starting from $t=50$ and displaying the results at the time interval of $\Delta t=25$ until $t=400$. This figure includes a subplot detailing the amplitude's evolution at the plane $y=0$ over time for various topography amplitudes, specifically $b_0=0.05$, $0.1$, and $0.15$. The amplitude evolution is marked by a rapid increase near the initial time, then reaching the maximum value, after which it undergoes a temperate decay. Concurrently with this decay, energy redistribution occurs, manifesting as a lateral diffusion that leads to a progressive leveling of the crestlines. Comparing the amplitudes of the first precursor waves for the three distinct topography amplitudes, one observes that an increase in the topography amplitude leads to a corresponding increase in the growth rate and peak amplitude of the precursor wave. Additionally, the decay of the wave is accelerated, and the amplitude, once it reaches a steady state, is also notably higher. The spatial evolution of the crestlines of the first precursor wave in the $x-y$ plane is shown in figure~\ref{fig:topo_1_ar}b, where the crestlines evolve in a manner that resembles a circular arc emanating from the origin $(0,0)$ as time elapses. The evolution of the distance between the crestlines and the origin is depicted in figure~\ref{fig:topo_1_ar}c. It can be observed that the distance is almost equal at any given moment, with only a slight protrusion near $y=0$, and this protrusion becomes increasingly evident as time progresses, which indicates that the peak line is essentially an arc and the main reason for the protrusion is the ``propulsion'' effect of the topography mentioned above, which causes the center of the circle to move forward. While the above findings are derived from numerical experiments, the precise patterns of how amplitude and peak lines move with the topography still require further theoretical investigations.

\subsection{Supercritical speed}
\begin{figure}
\centering
\begin{minipage}{2in}
\centering
\includegraphics[height=2in]{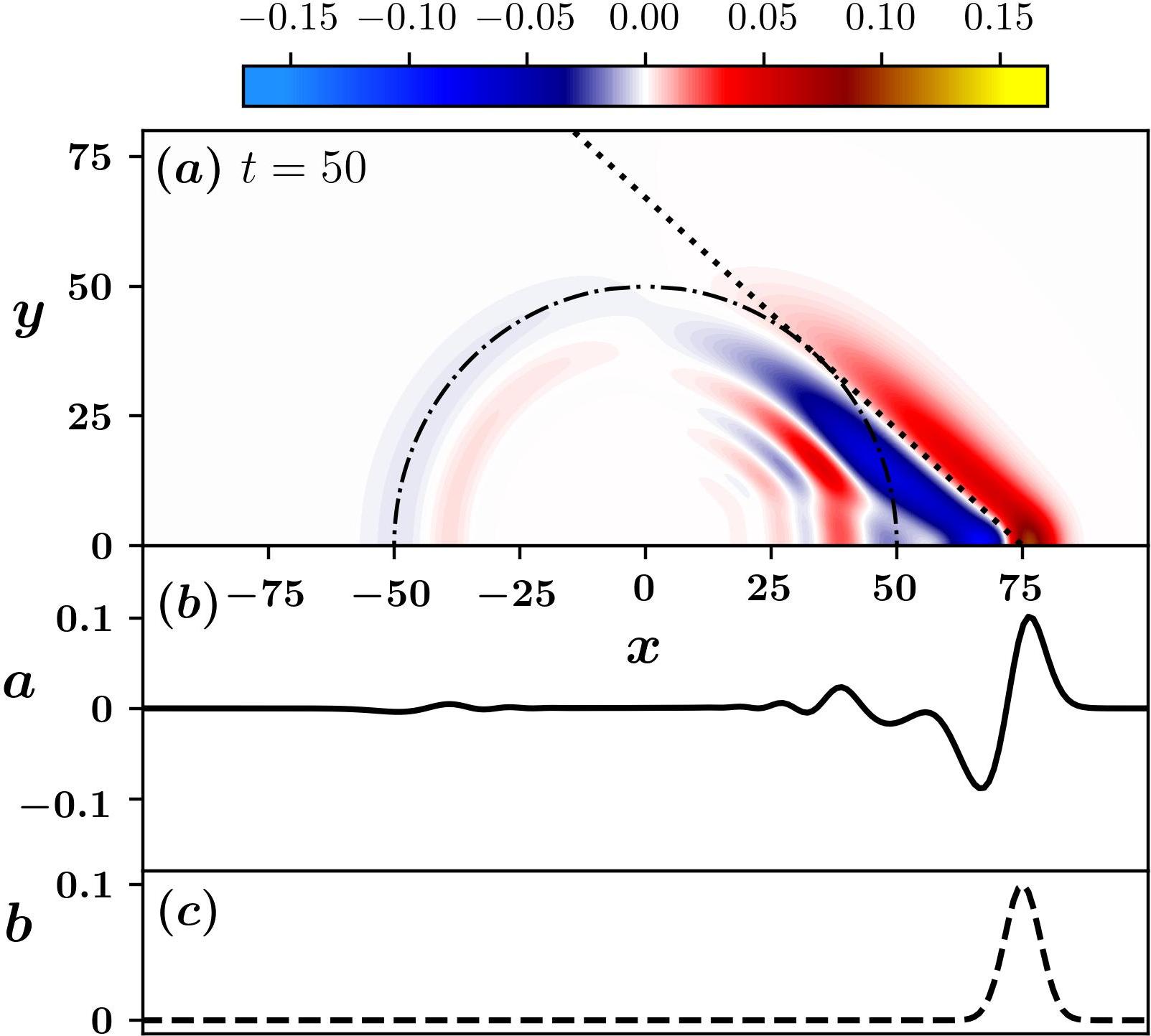}
\end{minipage}
\hspace{1cm}
\begin{minipage}{2in}
\centering
\includegraphics[height=2in]{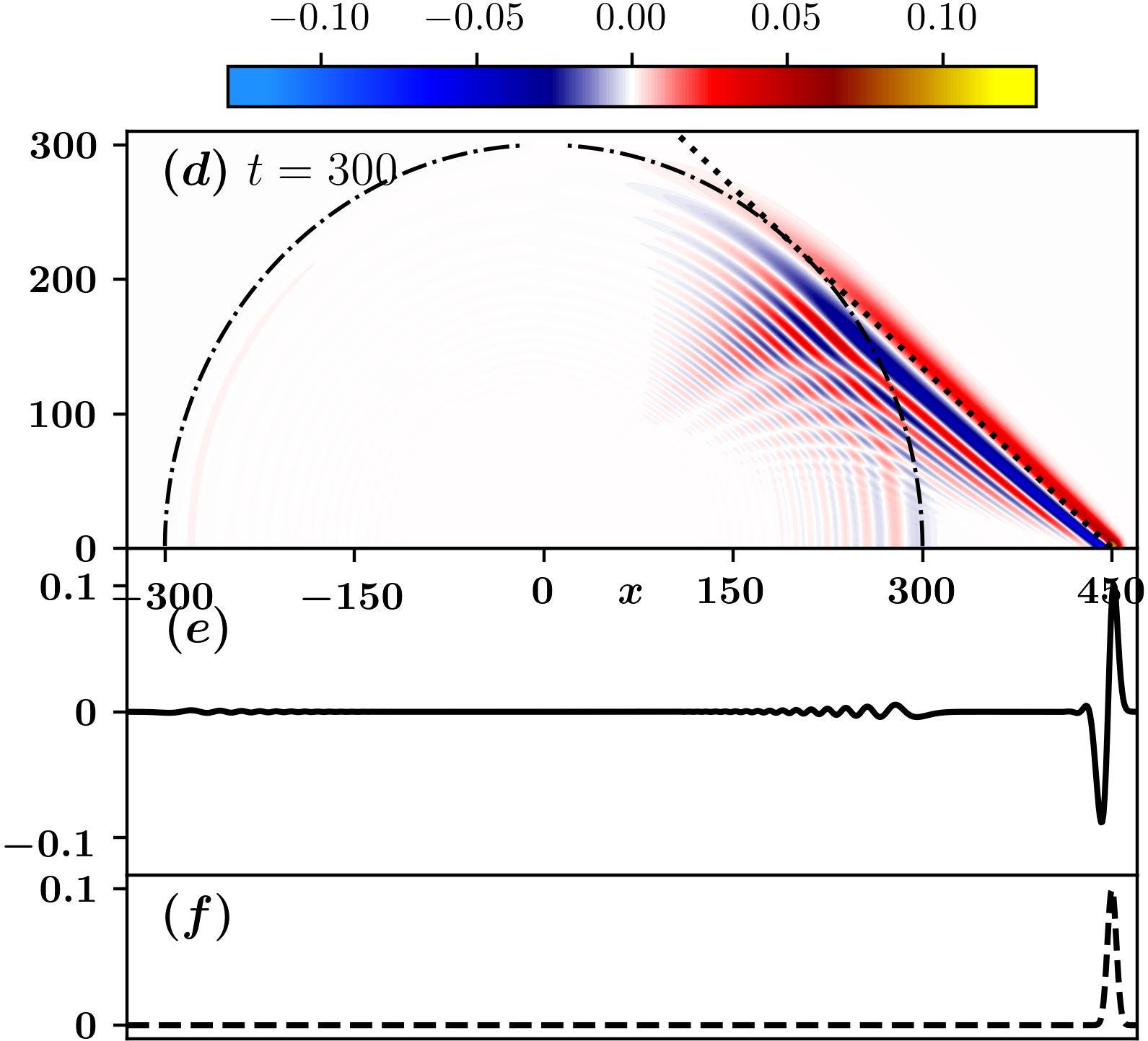}
\end{minipage}
\caption{Evolution of the wave generated by the moving topography with $b_0=0.1$ and $c_b=1.5$ at: $t=50$ (a--c) and $t=300$ (d--f). (a, d) contour plots of the wave, where the black dash-dotted circular arcs with a radius of $t$ and a center at the origin $(0,0)$ denote the predicted locations of the start-up waves and the black dotted lines denote the linear predictions of wake angles $41.81^\circ$; (b, e) $y$-cross-sections of the wave; (c, f) the positions of the topography at the corresponding times.}\label{fig:topo_1.5}
\end{figure}

When the topography moves with the supercritical speed ($c_b>1$), the upstream waves disappear; instead, a wave emerges ahead of the topography, moving at a similar speed. This wave, under the ``propelling'' influence of the topography, gradually narrows, and its amplitude slowly increases until it reaches a steady state, as shown in figures~\ref{fig:topo_1.5}b and \ref{fig:topo_1.5}e. In this case, the transverse waves vanish entirely, leaving behind only the distinct strip-like divergence waves, which can be observed in figure~\ref{fig:topo_1.5}d. For the wake angle, the condition $c_b>1$ leads to the emergence of complex values in \eqref{eq:wave_angle}, rendering the formula ineffective. Considering the situation from an alternative perspective, the anterior portion of the strip-like divergence wave, influenced by the topography, advances at a speed of $c_b$, while the rear part of the strip-like divergence wave, which is distant from the topography, is confined to the characteristic wave speed of the fluid, and can only propagate outwards at the unit speed, which can also be discerned from figures~\ref{fig:topo_1.5}a and \ref{fig:topo_1.5}d. Therefore, the strip-like divergence wave is confined within a specific angle, which starkly contrasts the wake angle for the topography moving at a subcritical speed. The wake angle $\alpha$ in the supercritical case can be obtained as $\alpha=\arcsin\left(1/c_b\right)$, which (the black dotted line) exhibits a commendable consistency with the numerical simulation shown in figures~\ref{fig:topo_1.5}a and~\ref{fig:topo_1.5}d.

\begin{figure}
\centering
\begin{minipage}{1.7in}
\centering
\includegraphics[height=1.5in]{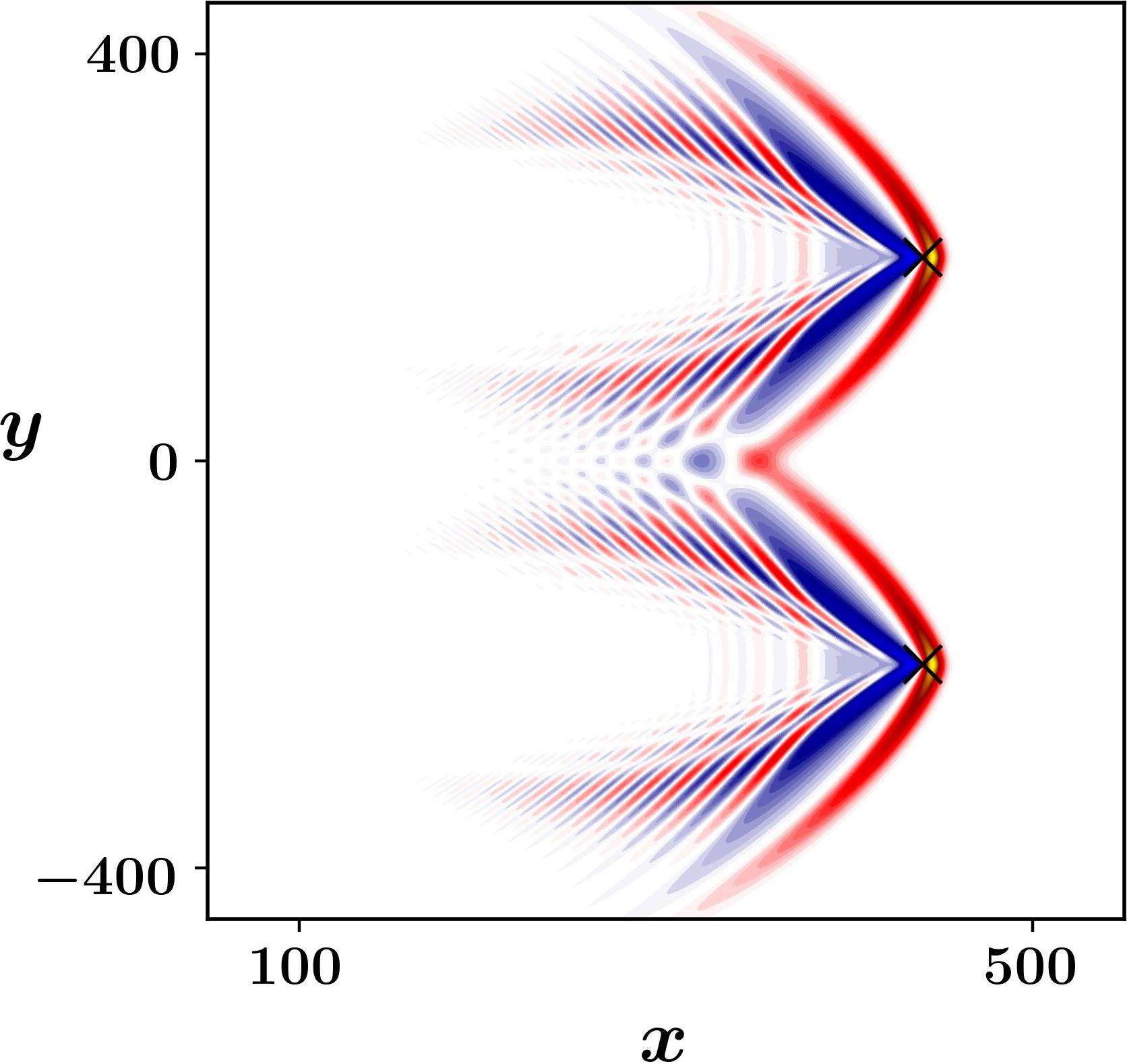}
\subcaption{$t=400$}
\end{minipage}
\centering
\begin{minipage}{1.7in}
\centering
\includegraphics[height=1.5in]{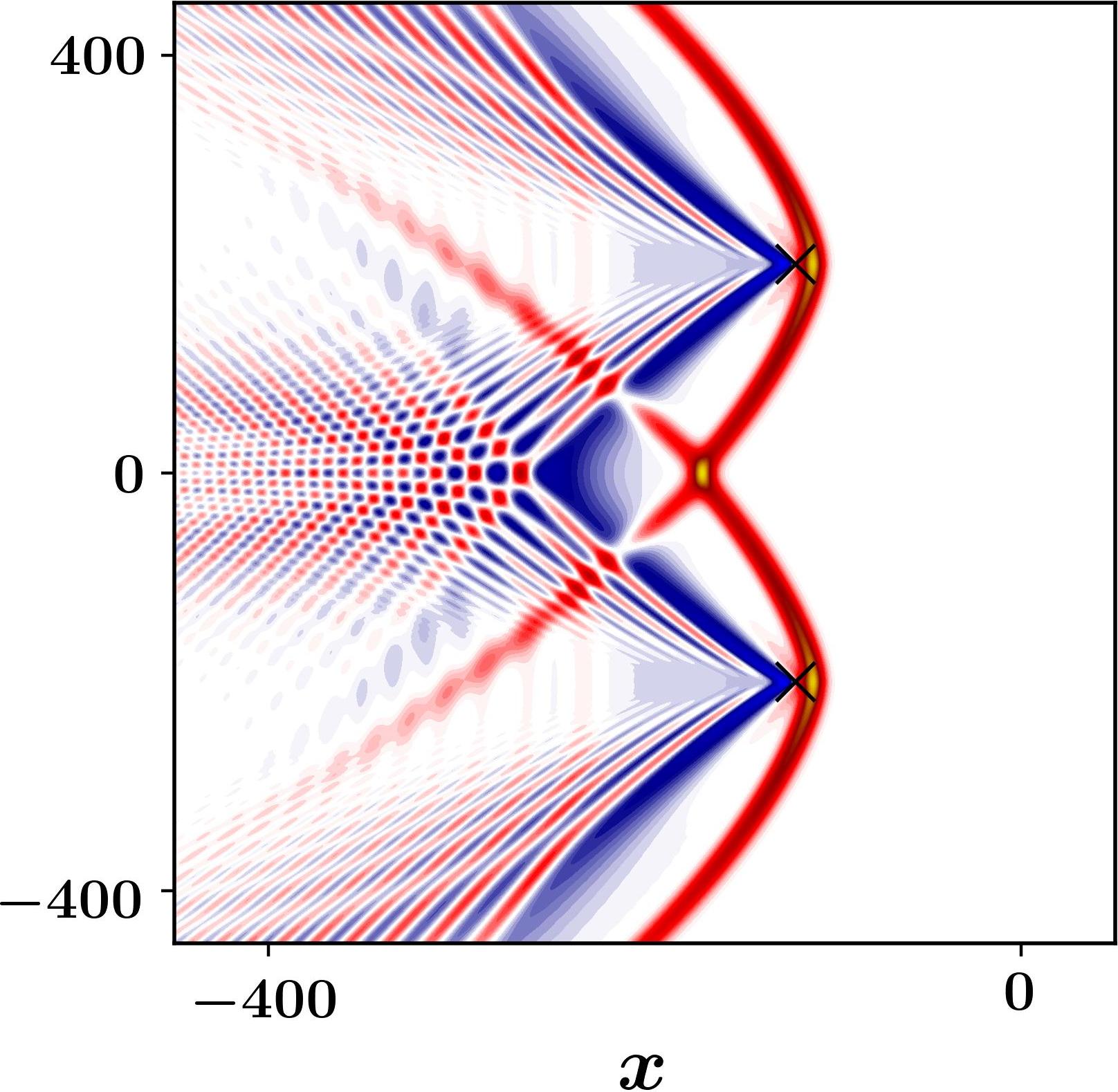}
\subcaption{$t=800$}
\end{minipage}
\centering
\begin{minipage}{1.7in}
\centering
\includegraphics[height=1.5in]{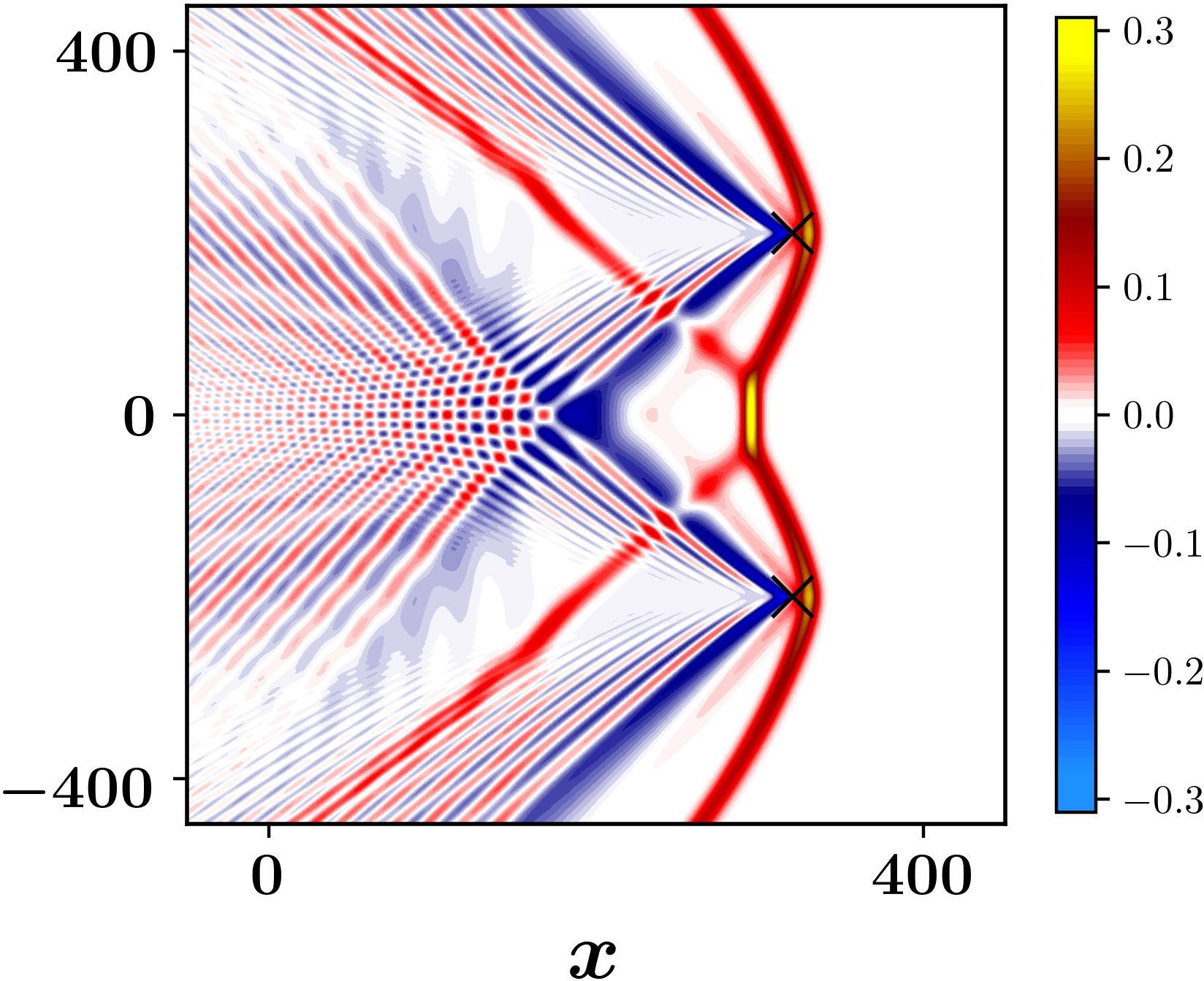}
\subcaption{$t=1200$}
\end{minipage}
\caption{Evolution of the wave generated by the moving topography~\eqref{eq:topo2} with $b_0=0.1$, $c_b=1.1$, and $l=400$ at $t=400$ (a), $800$ (b), and $1200$ (c), where the black cross `$\times$' denotes the positions of the topography peaks.}\label{fig:Topo2_1-3D}
\end{figure}

In a realistic situation, the bottom of the ocean may be complicated, with multiple topographies. When a homogeneous or density-stratified flow passes through such a region, the interactions of wave wakes may happen. To investigate these interactions, we use two Gaussian functions to describe the bottom configuration, \textit{i.e.},  
\begin{equation}
b(x,y,t)=B(x-c_bt,y-l/2)+B(x-c_bt,y+l/2)\,,\label{eq:topo2}
\end{equation}
where $l$ measures the separation distance between two locally confined topographies in the $y$-direction. The numerical results for the wakes generated by the two topographies moving at the same speed are depicted in figure~\ref{fig:Topo2_1-3D}. When $t=400$, the wakes emanating from the topographies commence to interact with each other. Up to $t=800$, the wake pattern exhibits a typical reflection characterized by the $y=0$ plane acting as the plane of symmetry. As time passes, the increase of the inner angle between the wakes, which leads to a gradual increase in amplitude, fulfills the conditions conducive to the onset of a Mach reflection. This eventually results in the formation of a Mach stem between the two wakes, as demonstrated in figure~\ref{fig:Topo2_1-3D}c.

\begin{figure}
\centering
\begin{minipage}{1.7in}
\centering
\includegraphics[height=1.5in]{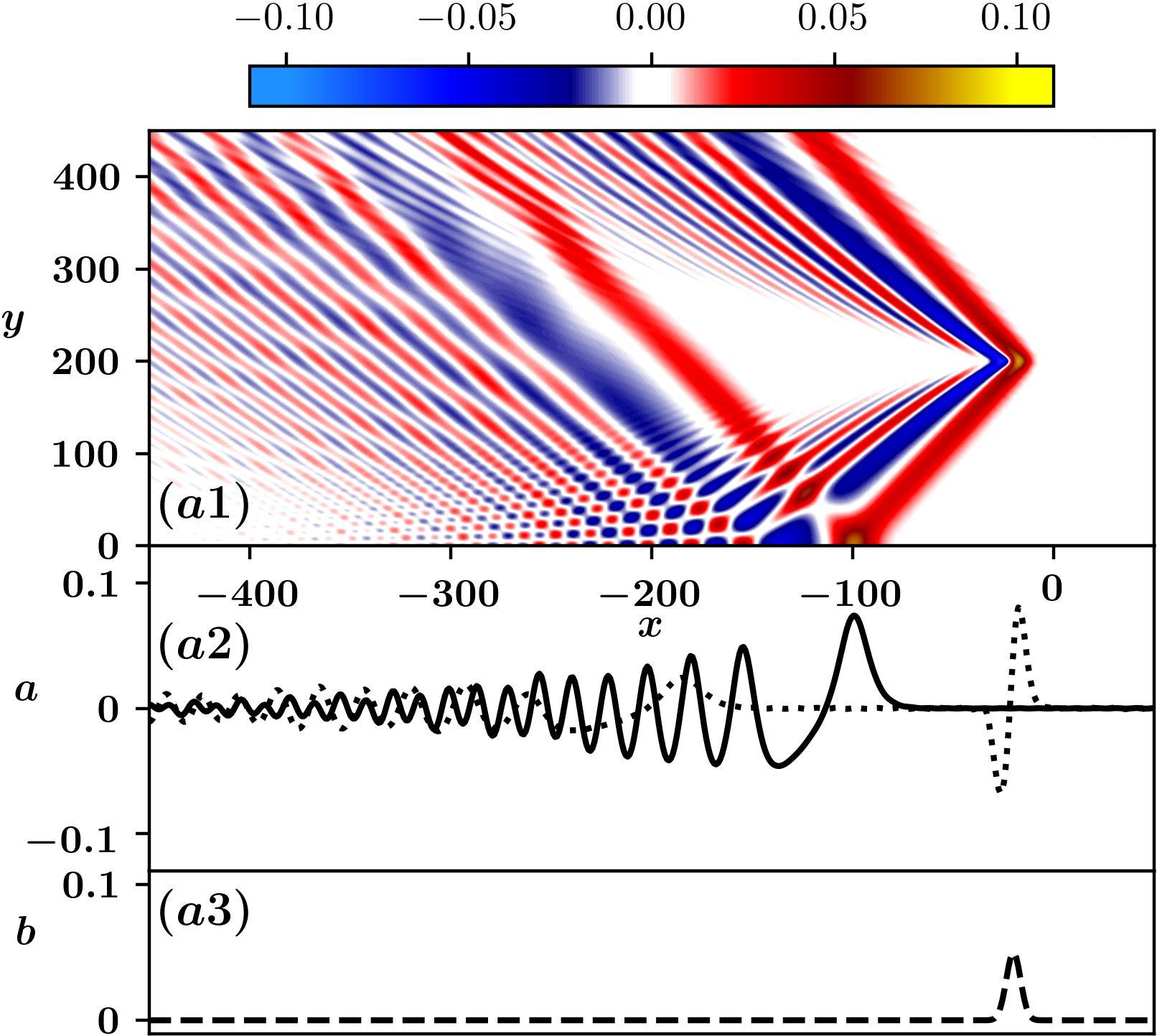}
\subcaption{$b_0=0.05$, $c_b=1.1$}
\end{minipage}
\centering
\begin{minipage}{1.7in}
\centering
\includegraphics[height=1.5in]{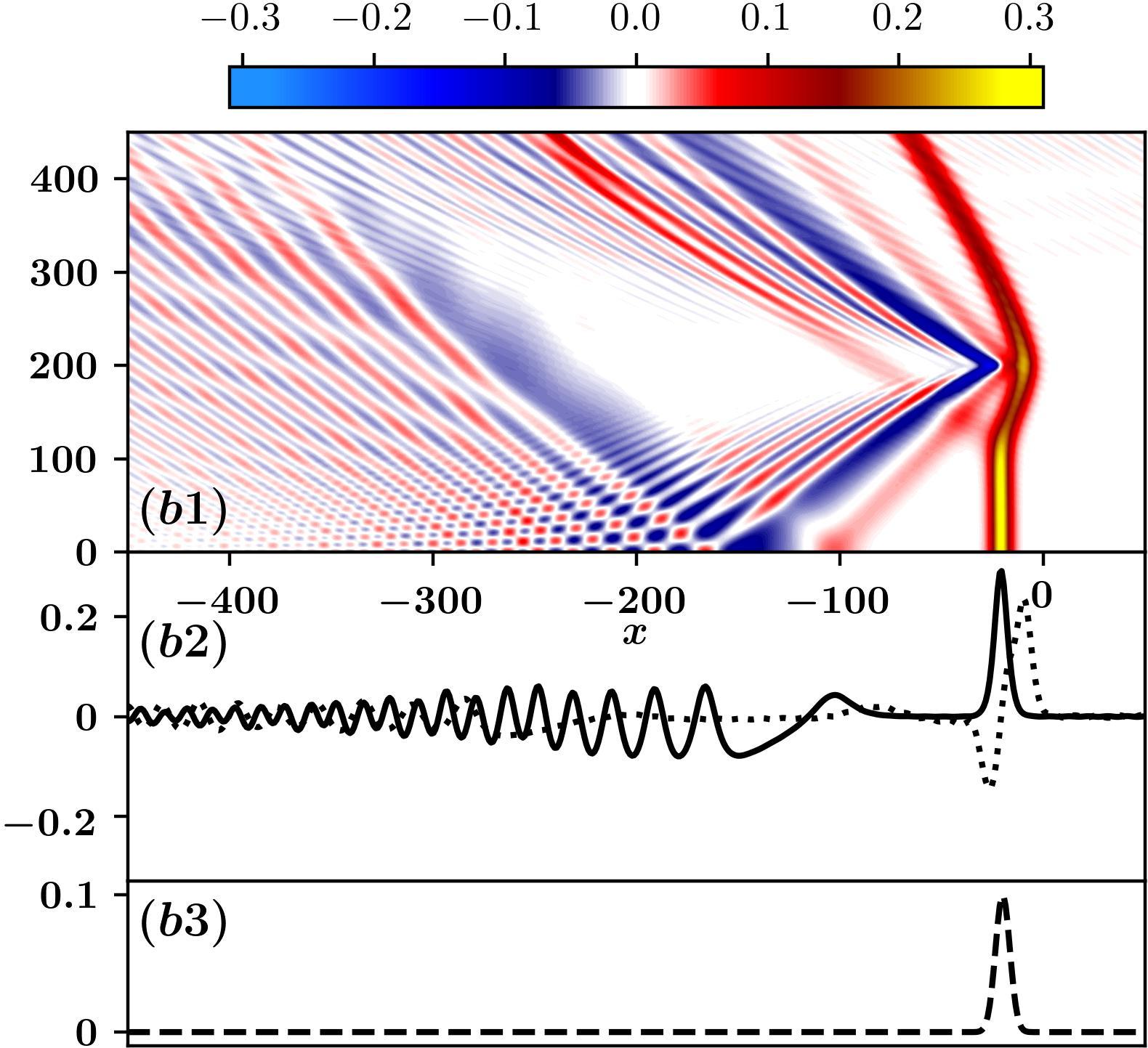}
\subcaption{$b_0=0.1$, $c_b=1.1$}
\end{minipage}
\centering
\begin{minipage}{1.7in}
\centering
\includegraphics[height=1.5in]{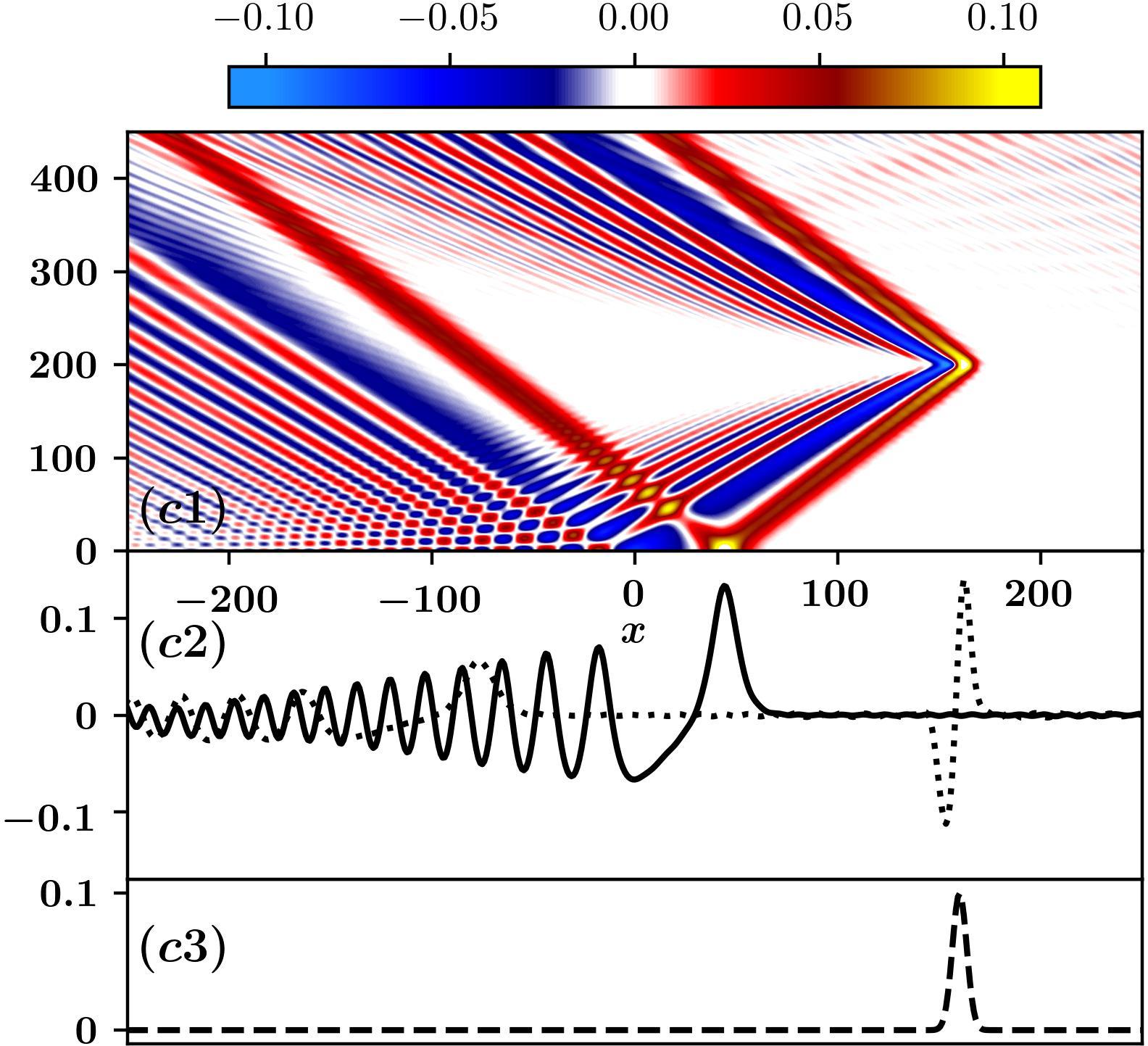}
\subcaption{$b_0=0.1$, $c_b=1.2$}
\end{minipage}
\caption{(a1, b1, c1) evolution of the wave generated by the moving topography~\eqref{eq:topo2} with different amplitudes and speeds at $t=1800$ and $l=400$. (a2, b2, c2) the black solid and dotted curves denote the wave profiles at $y=0$ and $y=200$, respectively. (a3, b3, c3) the dashed curves denote the positions of the topography at the corresponding parameters.}\label{fig:Topo2_2-3D}
\end{figure}

We conducted numerical simulations to explore whether altering topography parameters can lead to Mach reflection through wake interactions. The results for various topography speeds and amplitudes are illustrated in figure~\ref{fig:Topo2_2-3D}. Our findings indicate that Mach reflection does not occur when the amplitude of the topography is small or when the speed of the topography is large, which corresponds to a smaller angle between the two wakes. In these tests, only regular reflection is observed. Further investigations are needed to determine the specific conditions under which the Mach stem phenomenon occurs, particularly regarding the amplitude and velocity of the topography.

Note that we fix the distance between two Gaussian topographies in figures~\ref{fig:Topo2_1-3D} and ~\ref{fig:Topo2_2-3D}, setting $l=400$. If $l$ is increased, the interaction angle may change, potentially leading to a weak interaction. To investigate the condition under which the Mach stem appears, we conduct numerical experiments by varying $l$ while keeping $b_0=0.1$ and $c_b=1.1$ (the same data as figure~\ref{fig:Topo2_1-3D}). In summary, the Mach stem is generated when $l\leq1000$ with the generation time $t\in (0, 3001)$. Additionally, increasing $l$ results in longer generation times. For instance, we observe the Mach stem for $l=100$, $200$, and $1000$ when $t\in(0, T]$, where $T=490$, $1200$, and $3000.5$, respectively. However, the situation for $l>1000$ remains to be explored, as it requires significantly more time for numerical simulations, and a thorough analysis will be conducted in the future.

\section{Concluding Remarks}\label{sec:con}
Mach expansion and Mach reflection are two significant phenomena in aerodynamics that share an analogy with the dynamics of oblique solitons as they propagate and evolve in oceanic environments. This study focuses on the forced Benney--Luke (fBL) equation, which is a weakly nonlinear, isotropic, and bi-directional model that describes the motion of both surface and internal waves influenced by moving topography in oceanic settings. In the absence of topographical influences, the fBL equation simplifies to the classic Benney--Luke (BL) equation. To obtain numerical results relevant to both the BL and fBL equations, we have employed the standard pseudo-spectral method, incorporating the windowing scheme and integrating factors. We have also established the relationship between the forced Kadomtsev--Petviashvili (fKP) and fBL equations, which facilitates a more comprehensive analysis of their differences in Mach expansions and Mach reflections of oblique solitons in the context of flat topography. Using Whitham's theory, we have derived the modulation equations~\eqref{eq:modu4} for the BL equation related to soliton dynamics, and the complete spectrum of the Whitham modulation theory is discussed in Appendix~\ref{appA}. 

For the bent soliton initial data (corresponding to the initial angle $\theta_0<0$), the Mach expansion has been observed in the BL equation for relatively small magnitudes of the inclined angle $|\theta_0|$: the propagation of bent soliton results in the gradual extension of a perpendicular soliton at the bend point, referred to as the Mach stem. The amplitude of the Mach stem diminishes progressively as the angle $|\theta_0|$ increases towards a critical threshold, leading to the eventual dissolution of this structure. Concurrently, a regular expansion occurs, which manifests a circular wavefront. The conclusions drawn about the BL equation are contrasted with those regarding the KP equation, as depicted in figure~\ref{fig:same_c}. Notably, the wall reduction factor for the BL equation is consistently greater than that for the KP equation, and this discrepancy diminishes as the initial angle $|\theta_0|$ or speed $|\omega_0|$ increases. Conversely, the speed of the Mach stem is found to be in good agreement with various values of the initial speed $\omega_0$, with discrepancies only emerging when the initial angle $\theta_0$ approaches the critical angle $\theta_{\text{cr}}$ that distinguishes Mach expansions from regular expansions.

While for the reverse bent soliton initial data, \textit{i.e.,} $\theta_0>0$, a notable distinction arises: a Mach stem forms with an amplitude that surpasses that of the initial soliton, and subsequently, a smaller soliton emanates from the bend point in the direction of $\theta<0$. These three solitons converge at a single point, forming a tripartite structure referred to as the three-wave resonance soliton, akin to the Mach reflection. Utilizing both classical and modified Rankine--Hugoniot (RH) conditions, the amplitude and slope of the Mach stem $(a_w,q_w)$ and the reflected soliton $(a_i,q_i)$ can be numerically determined, along with the velocity of the triple-wave resonance point $(s_x,s_y)$. The growth rate of the length of Mach stem, $s_y$, is inversely related to the initial angle $\theta_0$. When $s_y$ reduces to zero as $\theta_0$ increases, the Mach stem fails to form; instead, an X-shaped soliton appears, indicative of regular reflection, where the reflected soliton gradually transitions into one with an identical amplitude $a_0$ but an opposite slope $-q_0$. Comparative analyses on Mach reflections between the BL and KP equations are presented in figure~\ref{fig:RV_same_c}, illustrating that the critical angle $\theta_{\text{cr}}^{(\text{BL})}$ for the BL equation is less than $\theta_{\text{cr}}^{(\text{KP})}$ for the KP equation, and this discrepancy becomes more pronounced as the speed $|\omega_0|$ increases.

Building upon the modulation equations of the BL equation, we have conducted a further investigation into the evolution of waves generated by uniformly moving topography. One of the most notable findings is that the waves in the far field exhibit a circular pattern, diverging from the parabolic wavefronts in the KP equation typically associated with its anisotropic nature. Considering the difference between the topography speed $c_b$ and the characteristic speed of shallow water, we have studied wave patterns and dynamics across the subcritical, critical, and supercritical regimes. In the subcritical regime, the topography induces a start-up wave at the instant of movement commencement, diffusing radially at a characteristic speed and forming a circular boundary. The wake angle can be determined using the linear theory, with its relationship to the topography speed expressed by equation~\eqref{eq:wave_angle}. When the topography attains the critical velocity, the wave is ``propelled'' by the topography, leading to the development of an upstream-advancing wave and resulting in an arc-shaped wavefront. When the topographic speed exceeds the critical value, transverse waves vanish within the wake, leaving only strip-like dispersive waves confined to an angle defined by $\alpha=\arcsin(1/c_b)$. When two topographies, symmetrically distributed along the $x$-axis, move at identical supercritical speeds, their generated wakes can interact with one another. Specifically, depending on the selected amplitudes and speeds of the terrains, Mach reflection may occur between the two wakes; otherwise, regular reflection will manifest.

The conclusions drawn above rely on the modulation theory of the BL equation. A more thorough and extensive exploration is necessary to understand the upstream-advancing wave dynamics, particularly regarding topographies moving at a critical speed. Additionally, the precise conditions that lead to the formation of a Mach stem during the interaction of two wakes -- including the amplitude and speed of each Gaussian topography, as well as the separation distance between them -- merit further investigation.

\backsection[Acknowledgements]{This work was partially supported by the National Natural Science Foundation of China grants (Nos. 12325207 and 12101590) and the Strategic Priority Research Program of the Chinese Academy of Sciences (Nos. XDB0640000 and XDB0640200).} 

\backsection[Declaration of Interests]{The authors report no conflict of interest.}

\backsection[Author ORCIDs]{Zhan Wang, https://orcid.org/0000-0003-4393-2118; Xu-Dan Luo, https://orcid.org/0000-0002-3833-5530.}

\appendix
\section{Derivation of the general Whitham system for the BL equation}\label{appA}
In the appendix, we derive the full Whitham modulation equations for the BL equation~\eqref{eq:BL-topo-flat}. Starting from equation~\eqref{E: q expansion}, we substitute the expansion into equation~\eqref{eq:e_BL} and collect terms according to the power of $\varepsilon$. To leading order $O(1/\varepsilon^{2})$, we have
\begin{equation}
-\omega^{2}\left(k^{2}+l^{2}\right)\xi^{(0)}_{\eta\eta\eta\eta}+\left(\omega^{2}-k^{2}-l^{2}\right)\xi^{(0)}_{\eta\eta}-3\omega\left(k^{2}+l^{2}\right)\xi^{(0)}_{\eta}\xi^{(0)}_{\eta\eta}=0\,.
\label{E: leading order}
\end{equation}
Integrating equation~\eqref{E: leading order} over the real line yields
\begin{equation}
\label{E: leading order'}
-\omega^{2}\left(k^{2}+l^{2}\right)\xi^{(0)}_{\eta\eta\eta}+\left(\omega^{2}-k^{2}-l^{2}\right)\xi^{(0)}_{\eta}-\frac{3\omega}{2}\left(k^{2}+l^{2}\right)\left(\xi^{(0)}_{\eta}\right)^{2}=c_{1}\,,
\end{equation}
where $c_1$ is an integration constant. Multiplying expression~\eqref{E: leading order'} by $2\xi^{(0)}_{\eta\eta}$ and integrating again, one obtains
\begin{equation}
\omega^{2}\left(k^{2}+l^{2}\right)\left(\xi^{(0)}_{\eta\eta}\right)^{2}=-\omega\left(k^{2}+l^{2}\right)\left(\xi^{(0)}_{\eta}\right)^{3}+\left(\omega^{2}-k^{2}-l^{2}\right)\left(\xi^{(0)}_{\eta}\right)^{2}-2c_{1}\xi^{(0)}_{\eta}-c_{2}\,,
\label{E: cubic}
\end{equation}
where $c_2$ is also an integration constant. The above differential equation admits an elliptic function solution $\xi^{(0)}_{\eta}$, which satisfies 
\begin{equation}
\label{E: f}
\left(\xi^{(0)}_{\eta\eta}\right)^{2}=\frac{1}{\omega}\left(\lambda_{1}-\xi^{(0)}_{\eta}\right)\left(\lambda_{2}-\xi^{(0)}_{\eta}\right)\left(\lambda_{3}-\xi^{(0)}_{\eta}\right):=f\left(\xi^{(0)}_{\eta}\right)\,,
\end{equation}
where $\lambda_{1}\leq \lambda_{2}\leq \lambda_{3}$, assumed real, are the roots of the cubic polynomial on the right-hand side of equation~\eqref{E: cubic}. The normalization of the elliptic function with period one gives the following cnoidal wave solution:
\begin{equation}
\label{E: q^{(0)}'}
\xi^{(0)}_{\eta}=A+B\text{cn}^{2}\left(2K_{m}(\eta-\eta_{*});m\right)\,,
\end{equation}
where $\text{cn}(\cdot~; m)$ is one of the Jacobi elliptic functions, $m$ is the elliptic parameter, $K_{m}$ is the complete elliptic integral of the first kind, and $\eta_{*}$ is an integration constant. A direct calculation yields
\begin{equation}
\xi^{(0)}_{\eta\eta}=-4BK_{m}\;\text{cn}(2K_{m}(\eta-\eta_{*});m)\;\text{sn}(2K_{m}(\eta-\eta_{*});m)\;\text{dn}(2K_{m}(\eta-\eta_{*});m)
\end{equation}
and
\begin{equation}
\begin{split}
\xi^{(0)}_{\eta\eta\eta\eta}=&64BK_{m}^{3}\;\text{cn}(2K_{m}(\eta-\eta_{*});m)\;\text{sn}(2K_{m}(\eta-\eta_{*});m)\;\text{dn}(2K_{m}(\eta-\eta_{*});m)\\
&\cdot\left[1-2m+3m\;\text{cn}^{2}(2K_{m}(\eta-\eta_{*});m)\right]\,.
\end{split}
\end{equation}
The solution~\eqref{E: q^{(0)}'} can be easily verified by substituting the above into equation~\eqref{E: leading order}. Note that $\xi^{(0)}_{\eta}$ is the well-known exact cnoidal wave solution to the KP equation, which is consistent with the result discussed in \cite{ablowitz2006}. In what follows, we will find $A$ and $B$ regarding physical variables.

{\bf Case 1.} If $\omega>0$, then it is clear that the real oscillating solution corresponds to the motion of $\xi^{(0)}_{\eta}$ between two zeros where $f\left(\xi^{(0)}_{\eta}\right)\geq0$, that is, in the interval $\lambda_{2}\leq \xi^{(0)}_{\eta} \leq \lambda_{3}$. In this case, 
\begin{equation*}
A=\lambda_{2}\,,\quad B=\lambda_{3}-\lambda_{2}\,,\quad m=\frac{\lambda_{3}-\lambda_{2}}{\lambda_{3}-\lambda_{1}}\,,\quad\omega=\frac{\lambda_{3}-\lambda_{1}}{16K_{m}^{2}}\,.
\end{equation*}
It follows from \eqref{E: cubic}--\eqref{E: f} that $\lambda_{1}+\lambda_{2}+\lambda_{3}=\frac{\omega^{2}-k^{2}-l^{2}}{\omega(k^{2}+l^{2})}$. Thus,
\begin{equation*}
A=\frac{\omega^{2}-k^{2}-l^{2}}{3\omega(k^{2}+l^{2})}+\frac{16}{3}\omega(1-2m)K_{m}^{2}\,,\quad B=16m\omega K_{m}^{2}\,.
\end{equation*}

{\bf Case 2.} If $\omega<0$, then $f\left(\xi^{(0)}_{\eta}\right)\geq0$ whenever $\lambda_{1}\leq \xi^{(0)}_{\eta} \leq \lambda_{2}$. In this case, 
\begin{equation*}
A=\lambda_{2}\,,\quad B=\lambda_{1}-\lambda_{2}\,,\quad m=\frac{\lambda_{2}-\lambda_{1}}{\lambda_{3}-\lambda_{1}}\,,\quad\omega=-\frac{\lambda_{3}-\lambda_{1}}{16K_{m}^{2}}\,. 
\end{equation*}
Recalling $\lambda_{1}+\lambda_{2}+\lambda_{3}=\frac{\omega^{2}-k^{2}-l^{2}}{\omega(k^{2}+l^{2})}$, one deduces
\begin{equation*}
A=\frac{\omega^{2}-k^{2}-l^{2}}{3\omega(k^{2}+l^{2})}+\frac{16}{3}\omega(1-2m)K_{m}^{2}\,,\quad B=16m\omega K_{m}^{2}\,.
\end{equation*}
In summary, for any nonzero $\omega$, we obtain 
\begin{equation}
A=\frac{\omega^{2}-k^{2}-l^{2}}{3\omega(k^{2}+l^{2})}+\frac{B}{3}\left(\frac{1}{m}-2\right)=\frac{\omega^{2}-k^{2}-l^{2}}{3\omega(k^{2}+l^{2})}+\frac{16}{3}\omega(1-2m)K_{m}^{2}
\end{equation} 
and 
\begin{equation}
B=16m\omega K_{m}^{2}\,.
\end{equation}

By integrating expression~\eqref{E: q^{(0)}'} with respect to $\eta$, one derives the following traveling wave solution to the BL equation~\eqref{eq:e_BL}:
\begin{equation}
\label{E: BL general solution}
\xi^{(0)}=A\eta+\frac{B}{2m K_{m}}\left[E(2K_{m}(\eta-\eta_{*}))-2K_{m}(1-m)(\eta-\eta_{*})\right]\,,
\end{equation}
where $E(2K_{m}(\eta-\eta_{*})):=E(\phi,m)$ is the incomplete elliptic integral of the second kind, and $\phi=\text{am}(2K_{m}(\eta-\eta_{*}))$ denotes the corresponding Jacobi amplitude. Specifically, solution~\eqref{E: BL general solution} is an unbounded travelling wave if $A\not\equiv0$.

We proceed with the derivation by examining the next-to-leading order terms, namely $O(1/\varepsilon)$; after a tedious calculation, one obtains
\begin{equation}
\label{E: next order}
\omega^{2}\left(k^{2}+l^{2}\right)\xi^{(1)}_{\eta\eta\eta\eta}+\left(k^{2}+l^{2}-\omega^{2}\right)\xi^{(1)}_{\eta\eta}+3\omega\left(k^{2}+l^{2}\right)\left(\xi^{(0)}_{\eta}\xi^{(1)}_{\eta}\right)_{\eta}=g\,,
\end{equation}
where
\begin{equation}
\label{E: g}
\begin{split}
g=&-\left(\omega_{T}+k_{X}+l_{Y}\right)\xi^{(0)}_{\eta}-2\omega \xi^{(0)}_{\eta T}
-2k\xi^{(0)}_{\eta X}-2l \xi^{(0)}_{\eta Y}-\left(\omega^{2}k_{X}+\omega^{2}l_{Y}-l^{2}\omega_{T}\right.\\
&\left.-k^{2}\omega_{T}+4\omega k \omega_{X}+4\omega l \omega_{Y}\right)\xi^{(0)}_{\eta\eta\eta}+2\omega\left(k^{2}+l^{2}\right)\xi^{(0)}_{\eta\eta\eta T}-2\omega^{2}k\xi^{(0)}_{\eta\eta\eta X}\\
&-2\omega^{2}l\xi^{(0)}_{\eta\eta\eta Y}-\left(2k\omega_{X}+2l\omega_{Y}+\omega k_{X}+\omega l_{Y}\right)\left(\xi^{(0)}_{\eta}\right)^{2}+\left(k^{2}+l^{2}\right)\xi^{(0)}_{\eta\eta}\xi^{(0)}_{T}\\
&-2\omega k \xi^{(0)}_{\eta\eta}\xi^{(0)}_{X}-2\omega l \xi^{(0)}_{\eta\eta}\xi^{(0)}_{Y}+2\left(k^{2}+l^{2}\right)\xi^{(0)}_{\eta}\xi^{(0)}_{\eta T}-4\omega k\xi^{(0)}_{\eta}\xi^{(0)}_{\eta X}\\
&-4\omega l\xi^{(0)}_{\eta}\xi^{(0)}_{\eta Y}\,.
\end{split}
\end{equation}

Recall that the solution $\xi^{(0)}_{\eta}$ of~\eqref{E: leading order} is periodic with period one, namely,  $\xi_{\eta}(\eta+1)=\xi_{\eta}(\eta)$. Integrating expression~\eqref{E: next order} with respect to $\eta$ over a period, one needs the periodicity condition
\begin{equation}
\label{periodicity condition}
\int_{0}^{1}g\;\text{d}\eta=0
\end{equation}
to avoid the secular terms, giving rise to a modulation equation. Multiplying expression~\eqref{E: next order} by $\xi^{(0)}_{\eta}$ and integrating over $[0,1]$ gives the solvability condition
\begin{equation}
\label{solvability condition}
\int_{0}^{1}\xi^{(0)}_{\eta}g\;\text{d}\eta=0\,,
\end{equation}
which provides another modulation equation. Overall, we have four Whitham modulation equations, including the first two compatibility conditions in~\eqref{eq:whitham1}, the periodicity condition~\eqref{periodicity condition}, and the solvability condition~\eqref{solvability condition}. By substituting expression~\eqref{E: g} into~\eqref{periodicity condition} and~\eqref{solvability condition}, we have
\begin{equation}
\label{periodicity condition'}
\begin{split}
0=&\;2\omega \partial_Tg_{1}+2k\partial_Xg_{1}+2l \partial_Yg_{1}+\left(\omega_{T}+k_{X}+l_{Y}\right)g_{1}-\frac{1}{2}\left(k^{2}+l^{2}\right)\partial_T g_{2}\\
&\;+k\omega \partial_X g_{2}+l\omega \partial_Y g_{2}+\left(2k\omega_{X}+2l\omega_{Y}+\omega k_{X}+\omega l_{Y}\right)g_{2}+(A\\
&\;+B\alpha)\left[2k\omega A_{X}+2l\omega A_{Y}-(k^{2}+l^{2})A_{T}-\frac{1-m}{m}(2k\omega+2l\omega-k^{2}-l^{2})\right]\,,
\end{split}
\end{equation}
\begin{equation}
\label{solvability condition'}
\begin{split}
0=&\;k\partial_X g_{2}+l\partial_Y g_{2}+\omega\partial_T g_{2}+\left(\omega_{T}+k_{X}+l_{Y}\right)g_{2}
+\left(2k\omega_{X}+2l\omega_{Y}+\omega k_{X}+\omega l_{Y}\right)g_{3}\\
&\;+\omega k \partial_X g_{3}+\omega l\partial_Y g_{3}-\frac{1}{2}\left(k^{2}+l^{2}\right)\partial_T g_{3}
-\omega^{2}k\partial_X g_{4}-\omega^{2}l\partial_Y g_{4}+\omega\left(k^{2}+l^{2}\right)\partial_T g_{4}\\
&\;-\left(4\omega k \omega_{X}+4\omega l \omega_{Y}+\omega^{2}k_{X}+\omega^{2}l_{Y}-l^{2}\omega_{T}-k^{2}\omega_{T}\right)g_{4}+\left(A^{2}+2AB\alpha\right.\\
&\;\left.+B^{2}\alpha^{2}\right)\left[2\omega k \left(A_{X}-\frac{1-m}{m}\right)+2\omega l \left(A_{Y}-\frac{1-m}{m}\right)-(k^{2}+l^{2})\left(A_{T}-\frac{1-m}{m}\right)\right]\,,
\end{split}
\end{equation}
where $\alpha=\text{cn}^{2}(-2K_{m}\eta_{*};m)$, $g_{j}=\int_{0}^{1}\left(\xi^{(0)}_{\eta}\right)^{j}\;\text{d}\eta$ for $j=1,2,3$, and $g_{4}=\int_{0}^{1}\left(\xi^{(0)}_{\eta\eta}\right)^{2}\;\text{d}\eta$. Using solution~\eqref{E: q^{(0)}'} and the properties of the Jacobi elliptic functions \citep[see][formulas 312, 361]{byrd2013}, one obtains 
\begin{equation}
\begin{aligned}
g_{1}=&A+\frac{B}{m K_{m}}[E_{m}-(1-m)K_{m}]\,,\\
g_{2}=&A^{2}+\frac{2AB}{m K_{m}}[E_{m}-(1-m)K_{m}]+\frac{B^{2}}{3m^{2}K_{m}}[(2-3m)(1-m)K_{m}+2(2m-1)E_{m}]\,,\\
g_{3}=&A^{3}+\frac{3A^{2}B}{m K_{m}}[E_{m}-(1-m)K_{m}]+\frac{AB^{2}}{m^{2}K_{m}}[(2-3m)(1-m)K_{m}+2(2m-1)E_{m}]\\
&+\frac{B^{3}}{5mK_{m}}\bigg\{\frac{4(2m-1)}{3m^{2}}\big[(2-3m)(1-m)K_{m}+2(2m-1)E_{m}\big]+\frac{3(1-m)}{m}\big[E_{m}\\
&-(1-m)K_{m}\big]\bigg\}\,,\\
g_{4}=&\frac{16B^{2}K_{m}}{15m^{2}}[(1-m)(m-2)K_{m}+2(m^{2}-m+1)E_{m}]\,,
\end{aligned}
\end{equation}
where $E_{m}$ is the complete elliptic integral of the second kind. 

Hence, there are four Whitham modulation equations: the first two compatibility conditions in~\eqref{eq:whitham1}, along with~\eqref{periodicity condition'} and~\eqref{solvability condition'}, for the four dependent variables: $k$, $l$, $\omega$, and $m$. Furthermore, to delve into the soliton dynamics of the BL equation, the traveling wave solution~\eqref{E: BL general solution} can be specialized to a solitary wave (kink wave) by taking $m\to1$. We claim that a solitary wave solution implies $\omega^{2}-k^{2}-l^{2}>0$. Indeed, a soliton solution yields $A\equiv0$; otherwise, it is unbounded. The following analysis is based on different signs of $\omega$.

{\bf Case 1.} If $\omega>0$, $m \to 1$ is equivalent to $\lambda_{1} \to \lambda_{2}$. It then follows that $\omega=\frac{\lambda_{3}-\lambda_{1}}{16K_{m}^{2}}$, and $A=\frac{\omega^{2}-k^{2}-l^{2}}{3\omega(k^{2}+l^{2})}-\frac{\lambda_{3}-\lambda_{1}}{3}=0$ implies $B=\lambda_{3}-\lambda_{1}=\frac{\omega^{2}-k^{2}-l^{2}}{\omega(k^{2}+l^{2})}>0$.

{\bf Case 2.} If $\omega<0$, $m \to 1$ means $\lambda_{2} \to \lambda_{3}$. Hence, $\omega=-\frac{\lambda_{3}-\lambda_{1}}{16K_{m}^{2}}$, and $A=\frac{\omega^{2}-k^{2}-l^{2}}{3\omega(k^{2}+l^{2})}+\frac{\lambda_{3}-\lambda_{1}}{3}=0$ leads to $B=\lambda_{1}-\lambda_{3}=\frac{\omega^{2}-k^{2}-l^{2}}{\omega(k^{2}+l^{2})}<0$.

Therefore, $\omega^{2}-k^{2}-l^{2}>0$ holds for any nonzero $\omega$ with $|\omega|=\frac{\lambda_{3}-\lambda_{1}}{16K_{m}^{2}}$.

Meanwhile, passing to this limit in the phase $\eta$, one obtains $2K_{m}\eta\to \frac{\sqrt{\lambda_{3}-\lambda_{1}}}{2\epsilon\sqrt{|\omega|}}(kX+lY-\omega T)$. Hence, solution~\eqref{E: BL general solution} reduces to a kink wave, \textit{i.e.,}
\begin{equation}
\xi^{(0)}(X,Y,T;\varepsilon)=2\sqrt{a}~\text{sgn} (\omega)\tanh\left[\frac{\sqrt{a}}{2\varepsilon|\omega|}(kX+lY-\omega T)\right],
\end{equation}
where $a=\frac{\omega^{2}-k^{2}-l^{2}}{k^{2}+l^{2}}>0$, and the integration constant $\eta_{*}$ in~\eqref{E: BL general solution} is taken to be $0$. The corresponding soliton slope is denoted by $q=l/k$. Without loss of generality, we assume the wavenumber $k>0$. Then, the sign of velocity is uniquely determined by the sign of frequency $\omega$. In addition, returning to the original independent variables associated with equation~\eqref{eq:BL-topo-flat}, the general four modulation equations can be reduced to the following three modulation equations in terms of three physical variables: $k$, $q$, and $a$:
\begin{equation}
k_{t}+\text{sgn} (\omega)\left(k \sqrt{a+1} \sqrt {1+q^{2}}\right)_{x}=0\,,
\label{eq:mod_so_1}
\end{equation}
\begin{equation}
q_{t}+\text{sgn}(\omega)\left[ q_{x}\left(\sqrt{a+1} \sqrt {1+q^{2}}\right)-q\left(\sqrt{a+1} \sqrt {1+q^{2}}\right)_{x}+\left(\sqrt{a+1} \sqrt {1+q^{2}}\right)_{y}\right]=0\,,
\label{eq:mod_so_2}
\end{equation}
\begin{equation}
\text{sgn} (\omega)\left[\frac{(4a+5)a\sqrt{a}}{a+1}\right]_{t}+\left[\frac{(3a+5)a\sqrt{a}}{\sqrt{a+1} \sqrt {1+q^{2}}}\right]_{x}+\left[\frac{(3a+5)qa\sqrt{a}}{\sqrt{a+1} \sqrt {1+q^{2}}}\right]_{y}=0\,.\label{eq:mod_so_3}
\end{equation}
In fact, \eqref{eq:mod_so_2} is derived from the last two compatibility conditions in \eqref{eq:whitham1}, and the third relation holds automatically if the phase $\eta(X,Y,T)$ is smooth, as mentioned in section~\ref{sec:modu}. Hence, \eqref{eq:mod_so_1}--\eqref{eq:mod_so_2} only rely on the first two compatibility conditions in \eqref{eq:whitham1}. Upon contemplating the propagation of waves in the leftward direction, \textit{i.e.}, $\omega<0$, the modulation equations~\eqref{eq:mod_so_1}--\eqref{eq:mod_so_3}, as well as $k_y-(qk)_x=0$, can be simplified to the form of equations~\eqref{eq:modu4_1}--\eqref{eq:modu4_4} as delineated in the main text.


\begin{thebibliography}{60}
\expandafter\ifx\csname natexlab\endcsname\relax\def\natexlab#1{#1}\fi
\def\au#1{#1} \def\ed#1{#1} \def\yr#1{#1}\def\at#1{#1}\def\jt#1{\textit{#1}}
\def\bt#1{#1}\def\bvol#1{\textbf{#1}} \def\vol#1{#1} \def\pg#1{#1}
\def\publ#1{#1}\def\arxiv#1{#1}\def\org#1{#1}\def\st#1{\textit{#1}}

\bibitem[Ablowitz \& Baldwin\/(2012)]{ablowitz2012}
{\sc \au{Ablowitz, M. J.} \& \au{Baldwin, D. E.}} \yr{2012} \at{Nonlinear shallow ocean-wave soliton interactions on flat beaches}.
\jt{Phys. Rev. E} \bvol{86}~(3), \pg{036305}.

\bibitem[Ablowitz {\em et al.\/}(2017)]{ablowitz2017}
{\sc \au{Ablowitz, M. J.}, \au{Biondini, G.} \& \au{Wang, Q.}} \yr{2017}
\at{Whitham modulation theory for the Kadomtsev--Petviashvili equation}.
\jt{Proc. R. Soc. A} \bvol{473}~(2204), \pg{20160695}.

\bibitem[Ablowitz \& Clarkson\/(1991)]{ablowitz1991}
{\sc \au{Ablowitz, M. J.} \& \au{Clarkson, P. A.}} \yr{1991} {\em Solitons, Nonlinear Evolution Equations and Inverse Scattering\/}. 
{\em London Mathematical Society Lecture Note Series} 149.  \publ{Cambridge; New York: Cambridge University Press}.

\bibitem[Ablowitz \& Curtis\/(2011)]{ablowitz2011}
{\sc \au{Ablowitz, M. J.} \& \au{Curtis, C. W.}} \yr{2011} \at{On the evolution of perturbations to solutions of the Kadomtsev--Petviashvilli equation using the Benney--Luke equation}.  
\jt{J. Phys. A} \bvol{44}~(19), \pg{195202}.

\bibitem[Ablowitz \& Curtis\/(2013)]{ablowitz2013}
{\sc \au{Ablowitz, M. J.} \& \au{Curtis, C. W.}} \yr{2013}
\at{Conservation laws and non-decaying solutions for the Benney--Luke equation}.  
\jt{Proc. R. Soc. A} \bvol{469}~(2152), \pg{20120690}.

\bibitem[Ablowitz {\em et al.\/}(2006)]{ablowitz2006}
{\sc \au{Ablowitz, M. J.}, \au{Fokas, A. S.} \& \au{Musslimani, Z. H.}}
\yr{2006}  \at{On a new non-local formulation of water waves}.  \jt{J. Fluid Mech.} \bvol{562}, \pg{313}.

%
\bibitem[Bazhenova {\em et al.\/}(1975)]{bazhenova1975}
{\sc \au{Bazhenova, T. V.}, \au{Gvozdeva, L. G.}, \au{Komarov, V. S.} \& \au{Sukhov, B. G.}} \yr{1975}  
\at{Diffraction of strong shock waves by convex corners}. 
\jt{Fluid Dynam.} \bvol{8}~(4), \pg{611--619}.

\bibitem[Benney \& Luke\/(1964)]{benney1964}
{\sc \au{Benney, D. J.} \& \au{Luke, J. C.}} \yr{1964} \at{On the interactions of permanent waves of finite amplitude}.
\jt{J. Math. Phys.} \bvol{43}~(1-4), \pg{309--313}.

\bibitem[Biondini\/(2007)]{biondini2007}
{\sc \au{Biondini, G.}} \yr{2007} \at{Line soliton interactions of the Kadomtsev--Petviashvili equation}.  
\jt{Phys. Rev. Lett.} \bvol{99}~(6), \pg{064103}.

\bibitem[Biondini \& Chakravarty\/(2006)]{biondini2006}
{\sc \au{Biondini, G.} \& \au{Chakravarty, S.}} \yr{2006} \at{Soliton solutions of the Kadomtsev--Petviashvili II equation}.  
\jt{Journal of Mathematical Physics}  \bvol{47}~(3),  \pg{033514}.

\bibitem[Biondini {\em et al.\/}(2020)]{biondini2020}
{\sc \au{Biondini, G.}, \au{Hoefer, M. A.} \& \au{Moro, A.}} \yr{2020}
\at{Integrability, exact reductions and special solutions of the KP--Whitham equations}.  
\jt{Nonlinearity}  \bvol{33}~(8), \pg{4114--4132}.

\bibitem[Biondini {\em et al.\/}(2009)]{biondini2009}
{\sc \au{Biondini, G.}, \au{Maruno, K.-I.}, \au{Oikawa, M.} \& \au{Tsuji, H.}}
\yr{2009}  \at{Soliton interactions of the Kadomtsev--Petviashvili equation and generation of large-amplitude water waves}.
\jt{Stud. Appl. Math.} \bvol{122}~(4), \pg{377--394}.

\bibitem[Bokhove \& Kalogirou\/(2016)]{bokhove2016}
{\sc \au{Bokhove, O.} \& \au{Kalogirou, A.}} \yr{2016} \at{Variational water wave modelling: from continuum to experiment}. 
In {\em Lectures on the Theory of Water Waves}, \bt{1st edn. (ed. \ed{Bridges, T. J., Groves, M. D. \& Nicholls, D. P.})},  
\pg{pp. 226--260}. \publ{Cambridge University Press}.

\bibitem[Byrd \& Friedman\/(2013)]{byrd2013}
{\sc \au{Byrd, P. F.} \& \au{Friedman, M. D.}} \yr{2013} {\em Handbook of Elliptic Integrals for Engineers and Scientists\/}, 2nd edn.
\publ{Berlin, Germany: Springer}.

\bibitem[Cornish\/(1910)]{cornish1910}
{\sc \au{Cornish, V.}} \yr{1910} {\em Waves of the Sea and Other Water Waves\/}.  \publ{London, T. Fisher Unwin}.

\bibitem[Curtis \& Shen\/(2015)]{curtis2015}
{\sc \au{Curtis, C. W.} and \au{Shen, S. S. P.}} \yr{2015}
\at{Three-dimensional surface water waves governed by the forced Benney--Luke equation}.  \jt{Stud. Appl. Math.}
\bvol{135}~(4), \pg{447--465}.

%
\bibitem[Funakoshi\/(1980)]{funakoshi1980}
{\sc \au{Funakoshi, M.}} \yr{1980} \at{Reflection of obliquely incident solitary waves}. \jt{J. Phys. Soc. Jpn.}
\bvol{49}~(6), \pg{2371--2379}.

\bibitem[Gidel {\em et al.\/}(2017)]{gidel2017}
{\sc \au{Gidel, F.}, \au{Bokhove, O.} \& \au{Kalogirou, A.}}
\yr{2017} \at{Variational modelling of extreme waves through oblique interaction of solitary waves: Application to Mach reflection}.
\jt{Nonlinear Proc. Geoph.} \bvol{24}~(1), \pg{43--60}.

\bibitem[Gilmore {\em et al.\/}(1950)]{gilmore1950}
{\sc \au{Gilmore, F. R.}, \au{Plesset, M. S.} \& \au{Crossley, H. E.}}
\yr{1950} \at{The analogy between hydraulic jumps in liquids and shock waves in gases}.  \jt{J. Appl. Phys.}
\bvol{21}~(3), \pg{243--249}.

\bibitem[Grava {\em et al.\/}(2018)]{grava2018}
{\sc \au{Grava, T.}, \au{Klein, C.} \& \au{Pitton, G.}} \yr{2018}
\at{Numerical study of the Kadomtsev--Petviashvili equation and dispersive shock waves}.  
\jt{Proc. Roy. Soc. A} \bvol{474}~(2210), \pg{20170458}.

%
\bibitem[Kadomtsev \& Petviashvili\/(1970)]{kadomtsev1970}
{\sc \au{Kadomtsev, B. B.} \& \au{Petviashvili, V. I.}} \yr{1970}  \at{On the stability of solitary waves in weakly dispersing media}.
\jt{Sov. Phys. Dokl.} \bvol{15}, \pg{539--541}.

\bibitem[Kao \& Kodama\/(2012)]{kao2012}
{\sc \au{Kao, C.-Y.} \& \au{Kodama, Y.}} \yr{2012}  \at{Numerical study of the KP equation for non-periodic waves}.  
\jt{Math. Comput. Simulat.} \bvol{82}~(7), \pg{1185--1218}.

\bibitem[Katsis \& Akylas\/(1987)]{katsis1987}
{\sc \au{Katsis, C.} and \au{Akylas, T. R.}} \yr{1987}  
\at{On the excitation of long nonlinear water waves by a moving pressure distribution. Part 2. Three-dimensional effects}.  
\jt{J. Fluid Mech.} \bvol{177}, \pg{49--65}.

\bibitem[Kodama\/(2004)]{kodama2004}
{\sc \au{Kodama, Y.}} \yr{2004} \at{Young diagrams and {\emph{N}}-soliton solutions of the KP equation}.  
\jt{J. Phys. A} \bvol{37}~(46), \pg{11169--11190}.

\bibitem[Kodama\/(2010)]{kodama2010}
{\sc \au{Kodama, Y.}} \yr{2010} \at{KP solitons in shallow water}.
\jt{J. Phys. A} \bvol{43}~(43), \pg{434004}.

\bibitem[Kodama\/(2018)]{kodama2018}
{\sc \au{Kodama, Y.}} \yr{2018} {\em Solitons in Two-Dimensional Shallow Water\/}.  
\publ{Philadelphia, PA: {Society for Industrial and Applied Mathematics}}.

\bibitem[Kodama {\em et al.\/}(2009)]{kodama2009}
{\sc \au{Kodama, Y.}, \au{Oikawa, M.} \& \au{Tsuji, H.}} \yr{2009} \at{Soliton solutions of the KP equation with V-shape initial waves}.
\jt{J. Phys. A} \bvol{42}~(31), \pg{312001}.

\bibitem[Kodama \& Yeh\/(2016)]{kodama2016}
{\sc \au{Kodama, Y.} \& \au{Yeh, H.}} \yr{2016} \at{The KP theory and Mach reflection}.  
\jt{J. Fluid Mech.} \bvol{800}, \pg{766--786}.

\bibitem[Krehl \& van der Geest\/(1991)]{krehl1991}
{\sc \au{Krehl, P.} \& \au{van der Geest, M.}} \yr{1991}  \at{The discovery of the Mach reflection effect and its demonstration in an auditorium}.
\jt{Shock Waves} \bvol{1}~(1), \pg{3--15}.

\bibitem[Lee \& Grimshaw\/(1990)]{lee1990}
{\sc \au{Lee, S.-J.} \& \au{Grimshaw, R. H. J.}} \yr{1990}
\at{Upstream-advancing waves generated by three-dimensional moving disturbances}.  
\jt{Phys. Fluids} \bvol{2}~(2), \pg{194--201}.

\bibitem[Lee {\em et al.\/}(1989)]{lee1989}
{\sc \au{Lee, S.-J.}, \au{Yates, G. T.} \& \au{Wu, T. Y.}}
\yr{1989} \at{Experiments and analyses of upstream-advancing solitary waves generated by moving disturbances}.  
\jt{J. Fluid Mech.} \bvol{199}, \pg{569--593}.

\bibitem[Li {\em et al.\/}(2011)]{li2011}
{\sc \au{Li, W.}, \au{Yeh, H.} \& \au{Kodama, Y.}} \yr{2011} \at{On the Mach reflection of a solitary wave: Revisited}.  
\jt{J. Fluid Mech.} \bvol{672}, \pg{326--357}.

%
\bibitem[Mach \& Wosyka\/(1875)]{mach1875}
{\sc \au{Mach, E.} \& \au{Wosyka, J.}} \yr{1875} \at{Ueber einige mechanische Wirkungen des elektrischen Funkens}.  
\jt{Ann. Phys. - Berlin} \bvol{232}~(11), \pg{407--416}.

\bibitem[Melville\/(1980)]{melville1980}
{\sc \au{Melville, W. K.}} \yr{1980} \at{On the Mach reflexion of a solitary wave}.  
\jt{J. Fluid Mech.} \bvol{98}~(2), \pg{285--297}.

\bibitem[Miles\/(1977{\natexlab{{\em a\/}}})]{miles1977}
{\sc \au{Miles, J. W.}} \yr{1977{\natexlab{{\em a\/}}}} \at{Obliquely interacting solitary waves}.  
\jt{J. Fluid Mech.} \bvol{79}~(1), \pg{157--169}.

\bibitem[Miles\/(1977{\natexlab{{\em b\/}}})]{miles1977a}
{\sc \au{Miles, J. W.}} \yr{1977{\natexlab{{\em b\/}}}} \at{Resonantly interacting solitary waves}.  
\jt{J. Fluid Mech.} \bvol{79}~(1), \pg{171--179}.

\bibitem[Milewski\/(1998)]{milewski1998}
{\sc \au{Milewski, P. A.}} \yr{1998} \at{A formulation for water waves over topography}.  
\jt{Stud. Appl. Math.} \bvol{100}~(1), \pg{95--106}.

\bibitem[Milewski \& Tabak\/(1999)]{milewski1999a}
{\sc \au{Milewski, P. A.} and \au{Tabak, E. G.}} \yr{1999}  
\at{A pseudospectral procedure for the solution of nonlinear wave equations with examples from free-surface flows}.  
\jt{SIAM J. Sci. Comput.} \bvol{21}~(3), \pg{1102--1114}.

\bibitem[Neu\/(2015)]{neu2015}
{\sc \au{Neu, J. C.}} \yr{2015} {\em Singular Perturbation in the Physical Sciences\/}. {\em Graduate Studies in Mathematics\/} volume 167.
\publ{Providence, Rhode Island: American Mathematical Society}.

%
%
\bibitem[Perroud\/(1957)]{perroud1957}
{\sc \au{Perroud, P. H.}} \yr{1957} \at{The solitary wave reflection along a straight vertical wall at oblique incidence}. PhD thesis, University of California, Berkeley.

\bibitem[Reichenbach\/(1983)]{reichenbach1983}
{\sc \au{Reichenbach, H.}} \yr{1983} \at{Contributions of Ernst Mach to fluid mechanics}.  
\jt{Annu. Rev. Fluid Mech.}  \bvol{15}~(1), \pg{1--29}.

\bibitem[Ryskamp {\em et al.\/}(2021)]{ryskamp2021}
{\sc \au{Ryskamp, S.}, \au{Maiden, M. D.}, \au{Biondini, G.} \& \au{Hoefer, M. A.}} \yr{2021}  
\at{Evolution of truncated and bent gravity wave solitons: The Mach expansion problem}.  
\jt{J. Fluid Mech.} \bvol{909}, \pg{A24}.

\bibitem[Ryskamp {\em et al.\/}(2022)]{ryskamp2022}
{\sc \au{Ryskamp, S. J.}, \au{Hoefer, M. A.} \& \au{Biondini, G.}}
\yr{2022} \at{Modulation theory for soliton resonance and Mach reflection}.  
\jt{Proc. Roy. Soc. A} \bvol{478}~(2259), \pg{20210823}.

\bibitem[Schlatter {\em et al.\/}(2005)]{schlatter2005}
{\sc \au{Schlatter, P.}, \au{Adams, N. A.} \& \au{Kleiser, L.}} \yr{2005}  
\at{A windowing method for periodic inflow/outflow boundary treatment of non-periodic flows}.  
\jt{J. Comput. Phys.}  \bvol{206}~(2), \pg{505--535}.

\bibitem[Shu\/(1998)]{shu1998}
{\sc \au{Shu, C.-W.}} \yr{1998}  
\at{Essentially non-oscillatory and weighted essentially non-oscillatory schemes for hyperbolic conservation laws}.  
\bt{In {\em Advanced Numerical Approximation of Nonlinear Hyperbolic Equations\/} (ed. \ed{Cockburn, B., Shu, C.-W., Johnson, C., Tadmor, E. \& Quarteroni, A.})}, \pg{pp. 325--432}. \publ{Berlin, Heidelberg: Springer}.

\bibitem[Tanaka\/(1993)]{tanaka1993}
{\sc \au{Tanaka, M.}} \yr{1993}  \at{Mach reflection of a large-amplitude solitary wave}.  
\jt{J. Fluid Mech.} \bvol{248}, \pg{637--661}.

\bibitem[Tsuji \& Oikawa\/(2007)]{tsuji2007}
{\sc \au{Tsuji, H.} and \au{Oikawa, M.}} \yr{2007}  
\at{Oblique interaction of solitons in an extended Kadomtsev--Petviashvili equation}.  
\jt{J. Phys. Soc. Jpn.} \bvol{76}~(8), \pg{084401}.

\bibitem[Wang \& Pawlowicz\/(2012)]{wang2012a}
{\sc \au{Wang, C.} \& \au{Pawlowicz, R.}} \yr{2012}  
\at{Oblique wave-wave interactions of nonlinear near-surface internal waves in the Strait of Georgia}.  
\jt{J. Geophys. Res. Oceans} \bvol{117}~(C6), \pg{2012JC008022}.

\bibitem[Whitham\/(1957)]{whitham1957}
{\sc \au{Whitham, G. B.}} \yr{1957}  
\at{A new approach to problems of shock dynamics Part I Two-dimensional problems}.  
\jt{J. Fluid Mech.} \bvol{2}~(2), \pg{145--171}.

\bibitem[Whitham\/(1959)]{whitham1959}
{\sc \au{Whitham, G. B.}} \yr{1959} \at{A new approach to problems of shock dynamics Part 2. Three-dimensional problems}.  
\jt{J. Fluid Mech.} \bvol{5}~(3), \pg{369--386}.

\bibitem[Whitham\/(1999)]{whitham1999}
{\sc \au{Whitham, G. B.}} \yr{1999} {\em Linear and Nonlinear Waves\/}. {\em Pure and Applied Mathematics\/} 1237.  
\publ{New York Chichester Weinheim: Wiley}.

\bibitem[Yuan {\em et al.\/}(2018)]{yuan2018l}
{\sc \au{Yuan, C.}, \au{Grimshaw, R. H. J.}, \au{Johnson, E.} \& \au{Chen, X.}} \yr{2018}  
\at{The propagation of internal solitary waves over variable topography in a horizontally two-dimensional framework}.  
\jt{J. Phys. Oceanogr.} \bvol{48}~(2), \pg{283--300}.

\bibitem[Yuan \& Wang\/(2022)]{yuan2022}
{\sc \au{Yuan, C.} \& \au{Wang, Z.}} \yr{2022}  
\at{On diffraction and oblique interactions of horizontally two-dimensional internal solitary waves}.  
\jt{J. Fluid Mech.} \bvol{936}, \pg{A20}.

\bibitem[Yuan {\em et al.\/}(2020)]{yuan2020}
{\sc \au{Yuan, C.}, \au{Wang, Z.} \& \au{Chen, X.}} \yr{2020}  
\at{The derivation of an isotropic model for internal waves and its application to wave generation}.  
\jt{Ocean Model.}  \bvol{153}, \pg{101663}.

\end{thebibliography}


\end{document}